\documentclass{aa}  

\usepackage{graphicx}
\usepackage{txfonts}
\begin{document}

   \title{Re-exploring molecular complexity with ALMA (ReMoCA): Interstellar detection of urea}

   \subtitle{}

   \author{A. Belloche \inst{1}
          \and R.~T. Garrod \inst{2}
          \and H.~S.~P.~M{\"u}ller \inst{3}
          \and K.~M. Menten \inst{1}
          \and I. Medvedev \inst{4}
          \and J. Thomas \inst{4}
          \and Z. Kisiel \inst{5}
          }

   \institute{Max-Planck-Institut f\"{u}r Radioastronomie, Auf dem H\"{u}gel 69, 53121 Bonn, Germany\\ \email{belloche@mpifr-bonn.mpg.de}
         \and Departments of Chemistry and Astronomy, University of Virginia, Charlottesville, VA 22904, USA        
         \and I. Physikalisches Institut, Universit{\"a}t zu K{\"o}ln, Z{\"u}lpicher Str. 77, 50937 K{\"o}ln, Germany
         \and Department of Physics, Wright State University, Dayton, OH, 45435, USA
         \and Institute of Physics, Polish Academy of Sciences, Al. Lotnikow 32/46, 02-668 Warszawa, Poland
             }

   \date{Received March 7, 2019; accepted June 5, 2019}

 
  \abstract
   {Urea, NH$_2$C(O)NH$_2$, is a molecule of great importance in organic 
   chemistry and biology. Two searches for urea in the interstellar 
   medium have been reported in the past, but neither were conclusive.}
   {We want to take advantage of the increased sensitivity and angular 
   resolution provided by the Atacama Large Millimeter/submillimeter Array
   (ALMA) to search for urea toward the hot molecular cores embedded in the 
   high-mass-star-forming region Sgr~B2(N).}
   {We used the new spectral line survey named ReMoCA (Re-exploring Molecular 
   Complexity with ALMA) that was performed toward Sgr~B2(N) with 
   ALMA in its observing cycle 4 between 84~GHz and 114~GHz. The spectra were 
   analyzed under 
   the local thermodynamic equilibrium approximation. We constructed a full
   synthetic spectrum that includes all the molecules identified so far. We used
   new spectroscopic predictions for urea in its vibrational ground state and 
   first vibrationally excited state to search for this complex organic 
   molecule in the ReMoCA data set. We employed the gas-grain chemical kinetics 
   model MAGICKAL to interpret the astronomical observations.
   }
   {We report the secure detection of urea toward the hot core Sgr~B2(N1) at a 
   position called N1S slightly offset from the continuum peak, which avoids 
   obscuration by the dust. The identification of urea relies on nine 
   clearly detected transitions. We derive a column density of 
   $2.7\times10^{16}$~cm$^{-2}$ for urea, two orders of magnitude lower than the
   column density of formamide, and one order of magnitude below that of 
   methyl isocyanate, acetamide, and N-methylformamide. The latter molecule is
   reliably identified toward N1S with 60 clearly detected lines, confirming 
   an earlier claim of its tentative interstellar detection. We report the 
   first interstellar detections of NH$_2$CH$^{18}$O and $^{15}$NH$_2$CHO.
   We also report the 
   nondetection of urea toward the secondary hot core Sgr~B2(N2) with an 
   abundance relative to the other four species at least one order of magnitude 
   lower than toward the main hot core. Our chemical model roughly reproduces 
   the relative abundances of formamide, methyl isocyanate, acetamide, and 
   N-methylformamide, but it overproduces urea by at least one order of 
   magnitude.}
   {Urea is clearly detected in one of the hot cores. Comparing the full 
   chemical composition of Sgr~B2(N1S) and Sgr~B2(N2) may help understand why 
   urea is at least one order of magnitude less abundant in the latter source.}

   \maketitle
%

\section{Introduction}
\label{s:introduction}

The chemical composition of astronomical sources, in particular star-forming 
regions, can be a good probe of their physical properties, their evolutionary
stage, and their history. This knowledge can even help us establish connections 
with the chemical composition of small bodies like comets and asteroids in our 
solar system, and thereby possibly learn about the conditions that were 
prevailing when life appeared on Earth. With the advent of broadband receivers 
at observational facilities, spectral line surveys covering wide frequency 
ranges with sufficient spectral resolution have become steadily more efficient 
and turned into ideal tools to derive the chemical composition of (many) 
astronomical sources in a systematic way \citep[e.g.,][]{Bergner17,Bonfand17}. 
In particular, such surveys are essential for the robust identification of 
complex organic molecules (COMs) in the interstellar medium (ISM), 
defined as molecules containing carbon with six atoms or more 
\citep[][]{Herbst09}. More than 200 interstellar 
molecules have been identified in the ISM since the advent of radio astronomy 
\citep[e.g.,][]{McGuire18}, with an approximately constant detection rate 
of about four new molecules every year since the early 1970s. While new 
molecules, including ions and radicals, with few atoms are still being 
discovered, much effort has been put into searching for more complex species 
with the aim of exploring the degree of complexity that interstellar chemistry 
can achieve and understanding the chemical processes that lead to this 
complexity.

Complex organic molecules have been found in diverse types of 
astronomical environments, but the first detections of these species were most 
of the time achieved in star-forming regions. The hot ($\sim$100--200~K) and 
dense ($10^6$--$10^8$~cm$^{-3}$) molecular regions around new born 
(proto)stars, called hot cores and hot corinos around high-mass and low-mass 
objects, respectively, have indeed turned out to be excellent targets to 
probe the complexity of interstellar chemistry. Among these sources, the hot 
cores embedded in the giant molecular cloud Sagittarius~B2 (Sgr~B2) close to 
the Galactic center, despite their large distance from us 
\citep[8.3~kpc,][]{Reid14}, are prime targets because their high column 
densities 
\citep[10$^{24}$--10$^{25}$~cm$^{-2}$ at arcsecond scale,][]{Qin11,Bonfand17} 
facilitate the detection of low-abundance molecules. 

Building on a spectral line survey performed with the 30\,m single-dish
telescope of the Institut de Radioastronomie Millim\'etrique (hereafter
IRAM 30\,m telescope) toward Sgr~B2(N) and Sgr~B2(M) \citep[][]{Belloche13}, 
which led to 
the detection of several new COMs \citep[][]{Belloche08,Belloche09}, we have 
taken advantage of the advent of the Atacama Large Millimeter/submillimeter 
Array (ALMA) to perform the EMoCA survey (Exploring Molecular Complexity with 
ALMA) toward Sgr~B(N) in order to probe further the chemical composition of 
the hot cores embedded within it \citep[][]{Belloche16}. The high angular 
resolution achieved with ALMA for this survey was essential to make 
significant progress given that the single-dish spectra were already close to 
the spectral confusion limit. With a half-power beam width (HPBW) of 
$\sim$1.6$\arcsec$, 
the EMoCA survey revealed that the rotational lines detected toward the 
northern hot core of Sgr~B2(N), which we call Sgr~B2(N2), had widths as narrow 
as $\sim$5~km~s$^{-1}$, whereas the main hot core, Sgr~B2(N1), located 
$\sim$5$\arcsec$ to the south of Sgr~B2(N2), still had broader linewidths 
($\sim$7~km~s$^{-1}$) and showed many lines with prominent wings. While the 
resulting spectral confusion prevented any significant progress in determining 
the chemical composition of Sgr~B(N1) compared to what we achieved with the 
single-dish telescope, the narrow linewidths of Sgr~B2(N2) and its separation 
from Sgr~B2(N1) allowed us to identify two new COMs toward Sgr~B2(N2) in the 
EMoCA survey: the branched alkyl molecule \textit{iso}-propyl cyanide 
\citep[\textit{i}-C$_3$H$_7$CN,][]{Belloche14} and, tentatively, 
N-methylformamide \citep[CH$_3$NHCHO,][]{Belloche17}. To make further progress 
on the chemical composition of Sgr~B2(N1), we performed a new spectral line 
survey with ALMA at even higher angular resolution and sensitivity. In keeping 
with the EMoCA survey, we call this new survey ReMoCA, which stands for 
\hbox{Re-exploring} Molecular Complexity with ALMA. As an initial outcome of 
ReMoCA, we present the results of our search for urea toward Sgr~B2.

Urea (NH$_2$C(O)NH$_2$), also called carbonyl diamide or carbamide, is a 
complex organic molecule in the astrochemical sense. It assumes a special role 
in the history of modern science. In 1828, Friedrich W{\"o}hler synthesized 
this organic molecule, which is associated with biological sources, out of 
inorganic compounds \citep[][]{Woehler1828}. This is 
widely seen as the start of modern organic chemistry, effectively rendering 
obsolete the need for a special vital force associated with living 
organisms only, which was the mainstream thinking at that time. Urea is 
thermodynamically speaking the most stable molecule in the CH$_4$N$_2$O family 
\citep[][]{Fourre16}. 
None of the CH$_4$N$_2$O isomers have been securely detected in the ISM so far, 
but urea has been identified in meteorites \citep[e.g.,][]{Hayatsu75} and
laboratory experiments conducted on interstellar ice analogs have shown that 
urea can be formed under ultraviolet irradiation \citep[e.g.,][]{Nuevo10}. 
Both results suggest that urea may exist in the ISM and motivate astronomers 
to search for it. Two searches for urea have been reported in 
the literature so far. \citet{Raunier04} compared the infrared spectra of the 
young stellar object NGC~7538~IRS9 recorded with the Short Wavelength 
Spectrometer (SWS) onboard the Infrared Space Observatory (ISO) to their
laboratory spectra of ices containing ammonia, water, and the products of the 
ultraviolet irradiation of isocyanic acid that leads to the formation of urea, 
among other species. The similarity between the shapes of the broad 
bands detected with ISO and the laboratory spectra led them to claim a 
tentative identification of urea (and formamide) in NGC~7538~IRS9. Because of 
the nonuniqueness of IR band assignments, the astrochemical community 
has not retained this tentative identification.

More recently, \citet{Remijan14} reported on a multitelescope campaign 
performed to search for urea in the millimeter-wavelength range. The team
targeted Sgr~B2(N) with five observational facilities, both interferometers 
and single-dish telescopes. They covered urea transitions between 100~GHz and 
250~GHz and found possible evidence for transitions of this molecule 
at 102~GHz and 232~GHz with four spectral lines assigned to urea using a 
local-thermodynamic-equilibrium (LTE) synthetic spectrum of its rotational 
emission. Other transitions predicted by their model were consistent with the 
observed spectra but could not be assigned due to blends with other species. 
Due to the small number of assigned transitions and the complexity of 
the hot-core spectrum, which they did not attempt to model in its entirety, 
\citet{Remijan14} refrained from claiming an interstellar detection of
urea.

We report below a detection of urea with ALMA using the ReMoCA survey. 
Section~\ref{s:obs_red} presents the observational details and explains how
the ALMA data were reduced. Section~\ref{s:spectro} describes how the 
spectroscopic predictions of urea were obtained from laboratory measurements. 
Section~\ref{s:results} reports our astronomical results on urea as well as 
several other related molecules, in particular N-methylformamide. In 
Sect.~\ref{s:chemistry}, we put the astronomical results into a broader 
astrochemical context using numerical simulations of interstellar chemistry. 
The results are discussed in Sect.~\ref{s:discussion} and we give the 
conclusion of our findings in Sect.~\ref{s:conclusions}.

\section{Observations and data reduction}
\label{s:obs_red}

\begin{table*}
\caption{Observational setups of the ReMoCA survey.}
\label{t:setup}
\centering
\begin{tabular}{cccccccccccc}
\hline
\hline
\noalign{\smallskip}
Setup & \multicolumn{2}{c}{Frequency range} & Date of & $t_{\rm start}$\tablefootmark{a} & $N_{\rm a}$\tablefootmark{b} & Baseline  & $t_{\rm int}$\tablefootmark{d} & pwv\tablefootmark{e} & \multicolumn{3}{c}{Calibrators\tablefootmark{f}} \\
\cline{2-3} \cline{10-12}
\noalign{\smallskip}
      & LSB & USB  & observation & (UTC) &  & range\tablefootmark{c} & & & B & A & P \\
      & {\small (GHz)} & {\small (GHz)} & {\small yyyy-mm-dd} & {\small hh:mm} & & {\small (m)} & {\small (min)} & {\small (mm)} & & & \\
\hline
\noalign{\smallskip}
S1 & 84.1--87.8  & 96.1--99.8   & 2016-10-29 & 22:19 & 40 & 19--1100 & 47 & 2.3 & 1 & 1 & 2 \\
   &             &              & 2016-10-30 & 20:33 & 42 & 17--1100 & 47 & 1.2 & 1 & 3 & 2 \\
   &             &              & 2016-11-03 & 20:10 & 42 & 19--1100 & 47 & 1.4 & 1 & 3 & 2 \\
   &             &              & 2016-11-04 & 20:11 & 40 & 21--1100 & 47 & 2.0 & 1 & 3 & 2 \\
S2 & 87.8--91.5  & 99.8--103.5  & 2016-11-04 & 21:39 & 40 & 21--1100 & 47 & 2.2 & 1 & 3 & 2 \\
   &             &              & 2016-11-06 & 15:52 & 40 & 19--1100 & 47 & 2.0 & 4 & 3 & 2 \\
   &             &              & 2017-05-10 & 06:14 & 43 & 17--1100 & 47 & 2.7 & 1 & 1 & 2 \\
   &             &              & 2017-07-01 & 06:05 & 36 & 21--2200 & 47 & 1.0 & 1 & 3 & 2 \\
S3 & 91.4--95.1  & 103.4--107.1 & 2017-05-22 & 04:50 & 43 & 15--1100 & 47 & 2.8 & 1 & 3 & 2 \\
   &             &              & 2017-05-22 & 06:12 & 43 & 15--1100 & 47 & 4.0 & 1 & 1 & 2 \\
   &             &              & 2017-05-22 & 07:23 & 43 & 15--1100 & 47 & 3.6 & 1 & 3 & 2 \\
   &             &              & 2017-05-22 & 08:40 & 43 & 15--1100 & 47 & 3.8 & 1 & 3 & 2 \\
S4 & 95.1--98.8  & 107.1--110.8 & 2017-07-15 & 01:25 & 42 & 19--1500 & 48 & 0.8 & 1 & 3 & 2 \\
   &             &              & 2017-07-23 & 01:40 & 46 & 17--3600 & 48 & 3.0 & 1 & 3 & 2 \\
   &             &              & 2017-07-23 & 02:59 & 46 & 17--3600 & 48 & 2.8 & 1 & 3 & 2 \\
   &             &              & 2017-07-23 & 04:18 & 46 & 17--3600 & 48 & 2.1 & 1 & 3 & 2 \\
S5 & 98.7--102.4 & 110.7--114.4 & 2017-07-23 & 05:36 & 46 & 17--3600 & 50 & 1.6 & 1 & 1 & 2 \\
   &             &              & 2017-07-24 & 02:27 & 44 & 31--3600 & 50 & 0.5 & 1 & 3 & 2 \\
   &             &              & 2017-07-24 & 03:47 & 44 & 31--3600 & 50 & 0.3 & 1 & 3 & 2 \\
   &             &              & 2017-07-24 & 05:06 & 44 & 31--3600 & 50 & 0.3 & 1 & 1 & 2 \\
\hline
\end{tabular}
\tablefoot{
\tablefoottext{a}{Start time of observation.}
\tablefoottext{b}{Number of ALMA 12\,m antennas.}
\tablefoottext{c}{Minimum and maximum projected baseline separations.}
\tablefoottext{d}{On-source integration time.}
\tablefoottext{e}{Precipitable water vapour content of the atmosphere.}
\tablefoottext{f}{Bandpass (B), amplitude (A), and phase (P) calibrators. The 
calibrators are: 1: J1924-2914, 2: J1744-3116, 3: J1733-1304, 4: J1517-2422.}
}
\end{table*}

\subsection{Observations}
\label{ss:obs}
The observations of the ReMoCA survey were conducted in a very similar way as
for the EMoCA survey \citep[][]{Belloche16}. We performed a complete spectral
line survey toward Sgr~B2(N) with ALMA between 84.1~GHz and 114.4~GHz. The 
phase center was set at ($\alpha, \delta$)$_{\rm J2000}$=
($17^{\rm h}47^{\rm m}19{\fs}87, -28^\circ22'16{\farcs}0$), a position that is 
located halfway between the hot cores Sgr~B2(N1) and (N2) that are separated 
by $4.9\arcsec$ in the north-south direction \citep[][]{Belloche16}. The size 
(HPBW) of the primary beam of the 12\,m antennas varies between $69''$ at 
84~GHz and $51''$ at 114~GHz \citep{Remijan15}.

Details about the observational setup are given in Table~\ref{t:setup}. The 
survey was divided into five spectral setups that were observed independently
from each other. Each setup was observed in four observing blocks of 
47--50~min on-source integration time each. Each setup was observed in only
one polarization and delivered four spectral windows of bandwidths 1875 MHz 
each, two per sideband. The sideband separation is 12~GHz. Each pair of 
adjacent spectral windows has an overlap of about 50~MHz. The observations were
performed with a channel spacing of 244.141~kHz and the spectra were 
smoothed to a resolution of 488.3 kHz (1.7 to 1.3~km~s$^{-1}$). All 
observations were performed during ALMA's observing cycle 4.

\subsection{Data reduction}
\label{ss:reduction}

The data was calibrated and imaged with the Common Astronomy Software 
Applications package (CASA), version 4.2.0 (r28322) 
for setups S1 to S4 and version 4.2.1 (r29047) for setup S5. We used 
the procedures provided by the Joint ALMA Observatory to apply the 
bandpass, amplitude, and phase calibrations. In addition, three or four 
iterations of self-calibration were performed using a strong spectral line 
detected toward Sgr~B2(N1) in each setup. The deconvolution was performed 
with the \textit{csclean} imager mode and a \textit{Briggs} weighting scheme 
with a \textit{robust} parameter of 0.5. The pixel size was set to 
0.12$\arcsec$ for setups S1--S3 and 0.06$\arcsec$ for setups S4--S5, with
1000 $\times$ 1000 pixels and 1890 $\times$ 1890 pixels, respectively.

\begin{table}
\caption{Beam sizes and noise levels.}
\label{t:beam_noise}
\centering
\begin{tabular}{cccccccc}
\hline
\hline
\noalign{\smallskip}
S.\tablefootmark{a} & \hspace*{-3ex} W.\tablefootmark{b} & \hspace*{-3ex} Freq. range\tablefootmark{c} & \multicolumn{2}{c}{Synthesized beam} & \hspace*{-2ex} & \multicolumn{2}{c}{rms\tablefootmark{e}} \\
\cline{4-5} \cline{7-8}
\noalign{\smallskip}
      & & \hspace*{-3ex} {\small (MHz)} & $HPBW$ & \hspace*{-2ex} P.A.\tablefootmark{d} & \hspace*{-2ex} & \hspace*{-2ex} {\small (mJy/} & \hspace*{-2ex} {\small (K)} \\
      & &     & {\small ($''\times''$)} & \hspace*{-2ex} {\small ($^\circ$)}    & \hspace*{-2ex} & \hspace*{-2ex} {\small beam)} & \\
\hline
\noalign{\smallskip}
S1 & \hspace*{-3ex} 0 &  \hspace*{-3ex} 84\,112 -- 85\,990  & $0.86 \times 0.67$ & \hspace*{-2ex} $-84$ & \hspace*{-2ex} & \hspace*{-2ex} 0.98 & \hspace*{-2ex} 0.29 \\
   & \hspace*{-3ex} 1 &  \hspace*{-3ex} 85\,938 -- 87\,815  & $0.90 \times 0.66$ & \hspace*{-2ex} $-87$ & \hspace*{-2ex} & \hspace*{-2ex} 0.94 & \hspace*{-2ex} 0.26 \\
   & \hspace*{-3ex} 2 &  \hspace*{-3ex} 96\,116 -- 97\,993  & $0.74 \times 0.59$ & \hspace*{-2ex} $-84$ & \hspace*{-2ex} & \hspace*{-2ex} 0.92 & \hspace*{-2ex} 0.27 \\
   & \hspace*{-3ex} 3 &  \hspace*{-3ex} 97\,941 -- 99\,818  & $0.72 \times 0.57$ & \hspace*{-2ex} $-86$ & \hspace*{-2ex} & \hspace*{-2ex} 0.94 & \hspace*{-2ex} 0.29 \\
S2 & \hspace*{-3ex} 0 &  \hspace*{-3ex} 87\,763 -- 89\,640  & $0.85 \times 0.62$ & \hspace*{-2ex}  $87$ & \hspace*{-2ex} & \hspace*{-2ex} 1.02 & \hspace*{-2ex} 0.30 \\
   & \hspace*{-3ex} 1 &  \hspace*{-3ex} 89\,588 -- 91\,465  & $0.93 \times 0.61$ & \hspace*{-2ex}  $83$ & \hspace*{-2ex} & \hspace*{-2ex} 1.05 & \hspace*{-2ex} 0.28 \\
   & \hspace*{-3ex} 2 &  \hspace*{-3ex} 99\,766 -- 101\,643 & $0.72 \times 0.60$ & \hspace*{-2ex}  $83$ & \hspace*{-2ex} & \hspace*{-2ex} 0.82 & \hspace*{-2ex} 0.23 \\
   & \hspace*{-3ex} 3 & \hspace*{-3ex} 101\,591 -- 103\,468 & $0.72 \times 0.62$ & \hspace*{-2ex}  $66$ & \hspace*{-2ex} & \hspace*{-2ex} 0.74 & \hspace*{-2ex} 0.19 \\
S3 & \hspace*{-3ex} 0 &  \hspace*{-3ex} 91\,403 -- 93\,280  & $0.70 \times 0.62$ & \hspace*{-2ex}  $89$ & \hspace*{-2ex} & \hspace*{-2ex} 0.85 & \hspace*{-2ex} 0.28 \\
   & \hspace*{-3ex} 1 &  \hspace*{-3ex} 93\,228 -- 95\,105  & $0.69 \times 0.59$ & \hspace*{-2ex} $-86$ & \hspace*{-2ex} & \hspace*{-2ex} 0.81 & \hspace*{-2ex} 0.27 \\
   & \hspace*{-3ex} 2 & \hspace*{-3ex} 103\,405 -- 105\,282 & $0.63 \times 0.53$ & \hspace*{-2ex} $-85$ & \hspace*{-2ex} & \hspace*{-2ex} 0.87 & \hspace*{-2ex} 0.29 \\
   & \hspace*{-3ex} 3 & \hspace*{-3ex} 105\,230 -- 107\,107 & $0.61 \times 0.52$ & \hspace*{-2ex} $-86$ & \hspace*{-2ex} & \hspace*{-2ex} 0.92 & \hspace*{-2ex} 0.31 \\
S4 & \hspace*{-3ex} 0 &  \hspace*{-3ex} 95\,062 -- 96\,939  & $0.57 \times 0.46$ & \hspace*{-2ex} $-53$ & \hspace*{-2ex} & \hspace*{-2ex} 0.39 & \hspace*{-2ex} 0.20 \\
   & \hspace*{-3ex} 1 &  \hspace*{-3ex} 96\,887 -- 98\,764  & $0.56 \times 0.45$ & \hspace*{-2ex} $-53$ & \hspace*{-2ex} & \hspace*{-2ex} 0.39 & \hspace*{-2ex} 0.20 \\
   & \hspace*{-3ex} 2 & \hspace*{-3ex} 107\,064 -- 108\,942 & $0.51 \times 0.41$ & \hspace*{-2ex} $-54$ & \hspace*{-2ex} & \hspace*{-2ex} 0.43 & \hspace*{-2ex} 0.22 \\
   & \hspace*{-3ex} 3 & \hspace*{-3ex} 108\,890 -- 110\,767 & $0.50 \times 0.40$ & \hspace*{-2ex} $-54$ & \hspace*{-2ex} & \hspace*{-2ex} 0.45 & \hspace*{-2ex} 0.23 \\
S5 & \hspace*{-3ex} 0 &  \hspace*{-3ex} 98\,714 -- 100\,591 & $0.43 \times 0.30$ & \hspace*{-2ex} $-78$ & \hspace*{-2ex} & \hspace*{-2ex} 0.69 & \hspace*{-2ex} 0.66 \\
   & \hspace*{-3ex} 1 & \hspace*{-3ex} 100\,539 -- 102\,417 & $0.42 \times 0.29$ & \hspace*{-2ex} $-78$ & \hspace*{-2ex} & \hspace*{-2ex} 0.68 & \hspace*{-2ex} 0.66 \\
   & \hspace*{-3ex} 2 & \hspace*{-3ex} 110\,717 -- 112\,594 & $0.39 \times 0.26$ & \hspace*{-2ex} $-77$ & \hspace*{-2ex} & \hspace*{-2ex} 0.81 & \hspace*{-2ex} 0.78 \\
   & \hspace*{-3ex} 3 & \hspace*{-3ex} 112\,542 -- 114\,419 & $0.38 \times 0.25$ & \hspace*{-2ex} $-77$ & \hspace*{-2ex} & \hspace*{-2ex} 0.95 & \hspace*{-2ex} 0.95 \\
\hline
\end{tabular}
\tablefoot{
\tablefoottext{a}{Setup.}
\tablefoottext{b}{Spectral window.}
\tablefoottext{c}{The frequencies correspond to rest frequencies at a systemic 
velocity of 62~km~s$^{-1}$.}
\tablefoottext{d}{Position angle East from North.}
\tablefoottext{e}{Median rms noise level measured in the channel maps of the 
continuum-removed data cubes.}
}
\end{table}

Given the high number of spectral lines detected toward the hot cores, their
different systemic velocities, and the velocity gradients across some of them,
it is impossible to split the line and continuum emission in the Fourier 
domain. Therefore we performed the splitting in the image plane, in an 
automatic way using the GILDAS/CLASS 
software\footnote{\label{fn:gildas}See http://www.iram.fr/IRAMFR/GILDAS.}.
Because of the presence of numerous absorption lines produced by the Sgr~B2 
molecular cloud itself and the diffuse and translucent clouds along the line of 
sight \citep[see, e.g.,][]{Thiel19a}, selecting the channels suitable for 
fitting the level of the continuum emission in an automatic way toward the hot 
core positions is not straightforward. We selected four positions that are free 
of hot-core emission but are associated with a strong continuum emission, 
among others the ultra-compact \ion{H}{II} (UC\ion{H}{II}) regions K4 and K1, 
in order to find the channels that contain signal in absorption (and some in 
emission as well, coming from larger scales). Then, for each pixel, we compute 
the channel intensity distribution after masking these channels. Provided all 
channels with absorption have been properly masked and the baseline does not 
contain any artifacts due to inaccuracies of the bandpass calibration or the 
imaging process, we expect the part of the channel distribution below the 
continuum baseline to follow a Gaussian pattern. We estimated the width of 
this lower part on pixels that appeared free from line emission and used this 
width to define the level of the baseline in each pixel. The beam sizes and 
position angles as well as the median noise levels measured in the channel 
maps of the continuum-subtracted datacubes are reported in 
Table~\ref{t:beam_noise}. The beam sizes are 2.5--4 times smaller than the 
ones achieved with the EMoCA survey. The sensitivity in flux density of the 
ReMoCA survey is a factor of three times better than that of the EMoCA survey.

In practice, the spectral baselines in some areas of the field of view suffer 
from artifacts that we have not yet fully understood and that may be
due to inaccuracies of the bandpass calibration or to missing information 
related to the incomplete sampling of the UV plane. Therefore, the 
splitting of the line and continuum emission performed so far is still 
preliminary and we plan to improve it in the future. Nevertheless, it is 
sufficient to start analyzing the continuum-subtracted spectra of the 
hot cores detected in the field of view.

\section{Spectroscopy}
\label{s:spectro}

The rotational spectrum of urea has received only modest attention in the 
spectroscopic community. Until recently the investigations were limited to 
frequencies below 50~GHz and to the ground vibrational state. 
\citet{urea_rot_1975} studied the microwave spectrum and determined 
its sizeable dipole moment $\mu=\mu_b=3.83$~D as well as $^{14}$N quadrupole 
coupling parameters. \citet{urea_HFS_1986} and \citet{urea_HFS_1996} 
studied the $^{14}$N hyperfine structure (HFS) in greater detail 
and determined unsplit rest frequencies with greater accuracy. 
\citet{1997JMoSt.413..405G} studied the rotational spectra of several 
isotopologs to determine the structure of urea, in particular the exact shape 
of the molecule. They also mentioned the identification of transitions 
pertaining to two excited vibrational states whose vibrational energy 
they estimated from relative intensities to be $\sim$155~cm$^{-1}$ 
($\sim$225~K) and $\sim$365~cm$^{-1}$ ($\sim$525~K), which they 
interpreted as the NH$_2$ wagging fundamental and first overtone 
states, respectively. Additional investigations of the isotopologs containing 
one and two $^{15}$N were also carried out \citep{15N-urea_rot_2004}. 

\citet{Remijan14} reported fairly extensive transition rest frequencies 
in the range of 59~GHz to 233~GHz in their account of searching for urea 
toward Sgr~B2(N). At about the same time, we started an investigation into the 
millimeter and submillimeter spectrum of urea 
\citep{mss_2013,isms_2014,Kisiel14}.
Our laboratory spectra of urea were collected with a system reported in 
\citet{Fosnight13}. The sample was thermally evolved into vapor by heating 
solid urea to 120$^\circ$C. A steady flow of urea through a two-meter long 
absorption cell heated to 120$^\circ$C facilitated the acquisition of 
continuous spectra in the 210--270 GHz and 300--500 GHz bands. Two commercial 
heterodyne systems produced by Virginia Diodes, Inc.\footnote{Available from 
http://www.virginiadiodes.com/.} covering these two bands were used in the 
study. The heterodyne systems were driven by a custom microwave synthesizer 
reported in \citet{Medvedev10}. The spectra were recorded using a frequency 
modulation technique. The second-derivative (2f) spectra were digitized from 
the output of a lock-in amplifier. The 2f spectra were then normalized by the 
baseline power, recorded independently. 

\begin{figure}
\centerline{\resizebox{1.0\hsize}{!}{\includegraphics[angle=0]{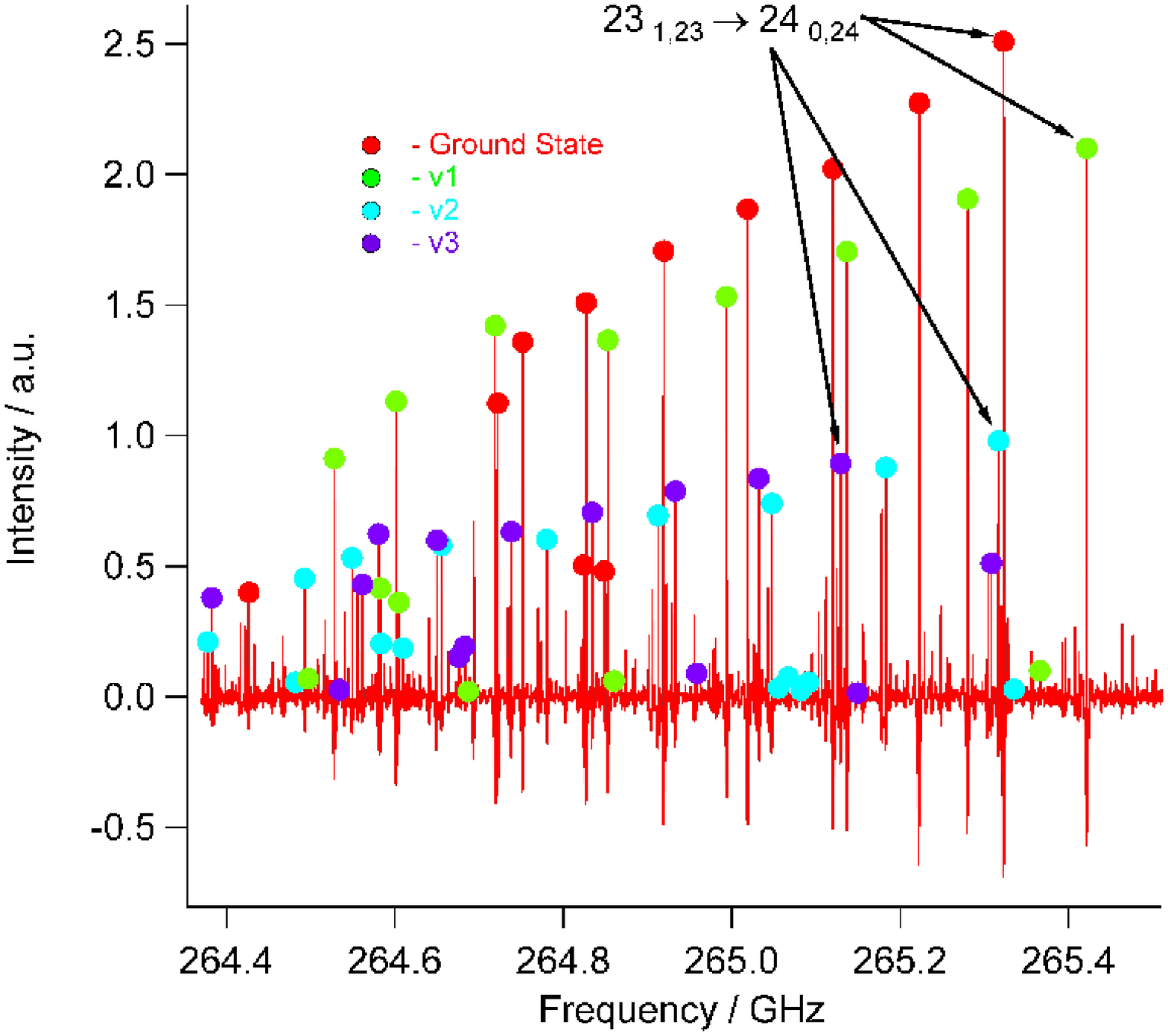}}}
\caption{Representative laboratory spectrum of urea used for spectral
assignment of the four strongest vibrational states. Red, green, cyan, and 
purple markers represent the assigned transitions of ground, first, second, and
third excited states, respectively. Quantum numbers for one transition with 
matching rotational quantum numbers across all four states are marked with 
arrows. These types of transitions were used for the calculation of the 
vibrational energies through spectral intensity ratios.}
\label{f:lab}
\end{figure}

Besides transitions within the ground vibrational state of urea, we identified 
numerous transitions within excited vibrational states of which three have 
been analyzed thus far along with the ground vibrational state. Over one 
thousand transitions with rotational quantum numbers matching across 
vibrational states were identified at frequencies ranging from 207 to 500~GHz. 
Their quantum numbers $J$ and $K_{\rm a}$ range from 9 to 66 and from 0 to 31,
respectively. Figure~\ref{f:lab} shows a representative 
spectrum of urea with markers representing our spectroscopic assignments of 
the ground state and the three strongest vibrational states.  
The power-normalized spectra allowed us to use relative spectral intensities 
to calculate  relative vibrational energies of the three excited vibrational 
states that we assigned. The relative energy of the first excited state was 
determined to be at $47 \pm 44$~cm$^{-1}$. The nature of the three excited 
vibrational states is not entirely clear, but quantum-chemical 
calculations by \citet{Inostroza12} suggest that these may be the three lowest 
modes associated with the two large-amplitude NH$_2$ wagging modes.

A detailed account of the analysis of this data set will be reported elsewhere
(Medvedev et al., in prep.). In short, all fits were performed with the SPFIT
/SPCAT package \citep[][]{Pickett91} using Watson's A-reduced asymmetric top 
Hamiltonian in the oblate top representation III$^{\rm l}$ \citep[][]{Watson77}. 
For the ground and first excited vibrational states, isolated single state fits 
were sufficient. The second and third vibrational states were found to be 
mutually perturbed and required a Coriolis-type coupled fit. Each of the three 
fits reproduced measured frequencies to a standard deviation of less than 
50~kHz, and allowed us to predict the rest frequencies with a precision of 
better than 10~kHz. Both ground and first excited states were fitted with 12 
effective Hamiltonian parameters which included all quartic and four sextic 
parameters ($H_J$, $H_{JK}$, $H_{KJ}$, and $H_K$).

\section{Astronomical results}
\label{s:results}

\subsection{Continuum emission}
\label{ss:cont}

\begin{figure*}
\centerline{\resizebox{0.7\hsize}{!}{\includegraphics[angle=0]{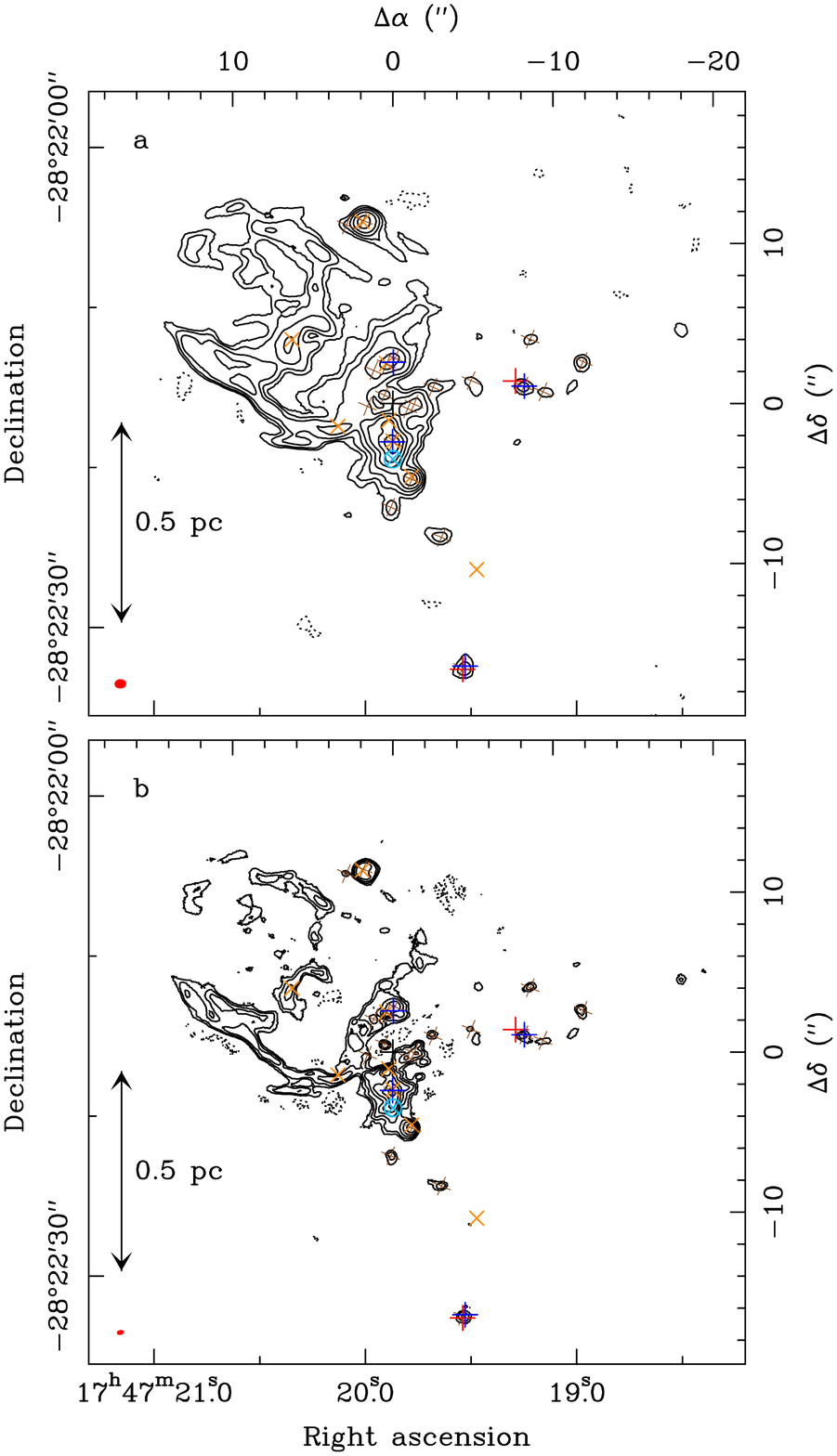}}}
\caption{Continuum map at 99.2~GHz extracted from setups S1 (\textbf{a}) and 
S5 (\textbf{b}), respectively. The contours start at 4$\sigma$ (\textbf{a}) and 
5$\sigma$ (\textbf{b}) with $\sigma$ the rms noise level equal to
0.95~mJy/beam and 0.42~mJy/beam, respectively, and then increase by a factor of
two. Negative contours are shown as dotted lines. In each panel, the red ellipse
represents the beam (HPBW). The black cross marks the phase center, the dark 
blue crosses the hot cores N1-N5 reported by \citet{Bonfand17}, the light blue
cross enclosed in a circle the position N1S, the brown crosses the 1~mm 
continuum sources detected 
by \citet{SanchezMonge17} with ALMA, the red crosses Class II methanol masers 
\citep[][]{Caswell96}, and the orange crosses the HII regions reported by 
\citet{Gaume95}. The names of some of these sources are indicated in 
Fig.~\ref{f:linecounts}. The maps are not corrected for primary-beam 
attenuation.}
\label{f:contmap}
\end{figure*}

Figure~\ref{f:contmap} shows two maps of continuum emission obtained with
the ReMoCA survey at 99.2 GHz. The top and bottom maps are extracted from 
setups S1 and S5 with beam sizes of $\sim$0.64$\arcsec$ and 
$\sim$0.36$\arcsec$, respectively. Although the splitting of the line and 
continuum emissions is still preliminary (see Sect.~\ref{ss:reduction}), the 
similarities between both maps give us confidence in their reliability. Many 
faint sources match well the positions of continuum sources detected by 
\citet{SanchezMonge17} with ALMA at higher frequencies (211--275~GHz) with a 
similar angular resolution ($\sim$0.4$\arcsec$). The ReMoCA continuum maps are 
also very similar to the map shown in Fig.~3 of \citet{Ginsburg18}, obtained 
with ALMA at 3~mm with a similar angular resolution ($\sim$0.5$\arcsec$). 

The northeastern part of the maps shown in Fig.~\ref{f:contmap} is dominated 
by free-free emission from the \ion{H}{ii} regions K4, K5, and K6, well 
detected at 22~GHz with the Very Large Array \citep[see Fig.~6 of][]{Gaume95} 
but not in the 1.3~mm maps of \citet{SanchezMonge17} in which dust emission 
dominates. The extended emission detected with ReMoCA around the hot core 
Sgr~B2(N1) does not have any counterpart in the VLA map, suggesting that it is 
dominated by dust emission as seen in the 1.3~mm maps. Compact emission is 
clearly detected in the ReMoCA maps toward the UC\ion{H}{ii} region K1 while 
it is barely seen at 1.3~mm, suggesting that it is dominated by free-free 
emission in the ReMoCA maps. Free-free emission emitted by the UC\ion{H}{ii} 
regions K2 (coincident with N1) and K3 certainly also contributes to the 
continuum emission detected with ReMoCA.

\subsection{Map of spectral line density}

\begin{figure*}
\centerline{\resizebox{1.0\hsize}{!}{\includegraphics[angle=0]{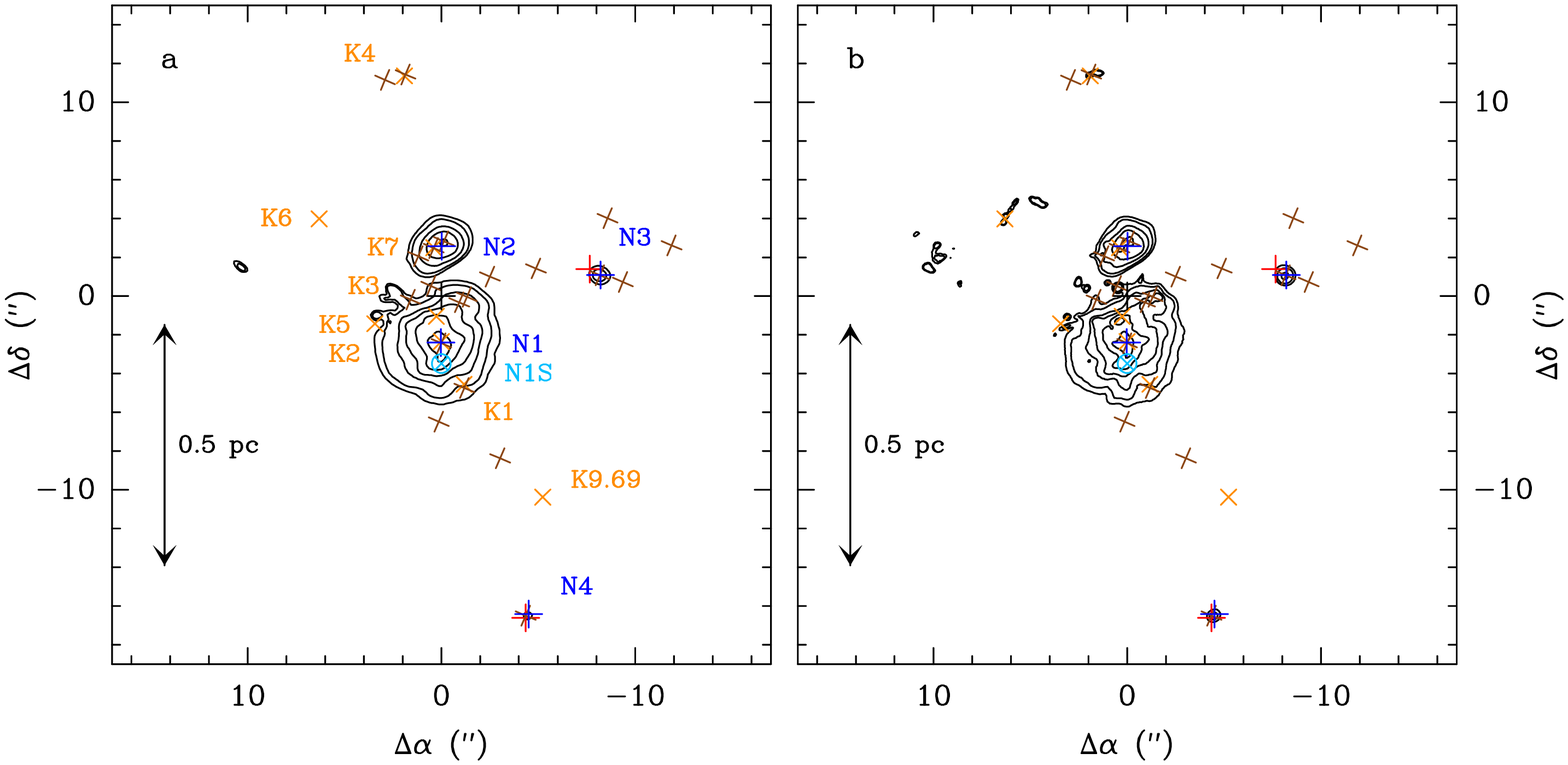}}}
\caption{Contour maps of the number of channels with continuum-subtracted flux
density above the 10$\sigma$ level. \textbf{a)} Only setups S1, S2, and S3,
which have beam sizes on the order of 0.6\arcsec--0.7\arcsec, were taken into 
account. The contours are 1000, 2000, 5000, 10\,000 counts, and then increase 
by steps of 10\,000 counts. \textbf{b)} Only setups S4 and S5, which have beam 
sizes of approximately 0.4\arcsec--0.5\arcsec, were taken into account. The 
contours are 700, 1400, 3500, 7000 counts, and then increase by steps of 7000 
counts. The symbols in both panels are the same as in Fig.~\ref{f:contmap}. 
The lowest contours in some parts of the maps, for instance toward K4 or K6, 
are affected by inaccuracies in the baseline subtraction and may not trace 
true line emission.}
\label{f:linecounts}
\end{figure*}

Figure~\ref{f:linecounts} shows maps of channel counts, that is the number of
channels with continuum-subtracted flux density above a threshold of 
10$\sigma$ 
\citep[for a similar map based on the EMoCA survey, see][]{Bonfand17}. 
More sensitive maps with a lower threshold will be computed
in the future when the splitting of the line and continuum emissions will have
been finalized. Channels in the frequency ranges covered several times were 
counted only once. Because of the different angular resolutions of the five 
spectral setups, we computed two maps: the first one (Fig.~\ref{f:linecounts}a) 
using setups S1, S2, and S3 only, the second one (Fig.~\ref{f:linecounts}b)
using setups S4 and S5 only. Several positions are affected by inaccuracies in 
the baseline subtraction, which in some cases leads to an overestimation of the 
number of channels with true line emission. This is in particular the case 
around the UC\ion{H}{ii} regions K4 and K6. Despite this issue, we recover the 
peaks detected by \citet{Bonfand17} toward N3 and N4 that led to the discovery 
of these hot cores. There may be other new, compact sources in the ReMoCA 
channel count maps, but a closer inspection of the line-continuum splitting 
will be required before we can draw any firm conclusions about these faint 
structures.

\subsection{Position selection}
\label{ss:position}

At the high angular resolution achieved with ReMoCA, the spectrum toward the 
continuum peak position of Sgr~B2(N1) is severely affected by the optical 
depth of the continuum emission that masks most of the line emission. The 
situation is similar to that encountered for the protostellar binary 
IRAS~16293-2422 in the PILS survey \citep[][]{Jorgensen16}. As in the case of
IRAS~16293-2422, we decided to analyze a position slightly offset from the
dust continuum peak. We selected the position at 
($\alpha, \delta$)$_{\rm J2000}$=
($17^{\rm h}47^{\rm m}19{\fs}870, -28^\circ22'19{\farcs}48$), about 1$\arcsec$ to 
the south of Sgr B2(N1). We call this position Sgr~B2(N1S). It is shown as a
light blue cross enclosed in a circle in Figs.~\ref{f:contmap} and 
\ref{f:linecounts}. This position is sufficiently close to N1 to allow us to 
derive the chemical composition of
the hot core, but still far enough to suffer less from the opacity of the
dust emission. In our search for a suitable position around N1, we paid 
attention to the widths of the detected spectral lines. We selected N1S 
because of its narrow linewidths, on the order of 5~km~s$^{-1}$, which is as 
narrow as the width of Sgr B2(N2)'s spectral lines in the EMoCA survey. This  
position does not show prominent linewings as was the case for Sgr~B2(N1) at 
poorer angular resolution in the EMoCA survey. The narrow linewidths and the
absence of strong linewing emission reduce the level of spectral confusion. 
The ReMoCA survey thus allows us to investigate the molecular composition of 
N1, the main hot core of Sgr~B2(N), in greater detail than was possible before 
with the EMoCA survey.

We followed the same method as \citet{Belloche16} to model the spectrum of N1S. 
Details about the modeling can be found there. In short, we use the Weeds 
software \citep[][]{Maret11} to produce synthetic spectra of complex organic 
molecules under the LTE assumption, which is well 
justified given the high densities that characterize Sgr~B2(N)'s hot cores 
\citep[][]{Bonfand17}. We started from the list of molecules identified toward 
N2 with the EMoCA survey to build a full model of the COM emission detected 
with ReMoCA toward N1S. We assumed a source size of 2$\arcsec$ for all 
species, which is more extended than the beam. While this full model is still 
preliminary, it fits the observed spectrum well enough to allow for a search 
for new molecules.

\subsection{Detection of urea}
\label{ss:urea}

\begin{figure}
\centerline{\resizebox{1.0\hsize}{!}{\includegraphics[angle=0]{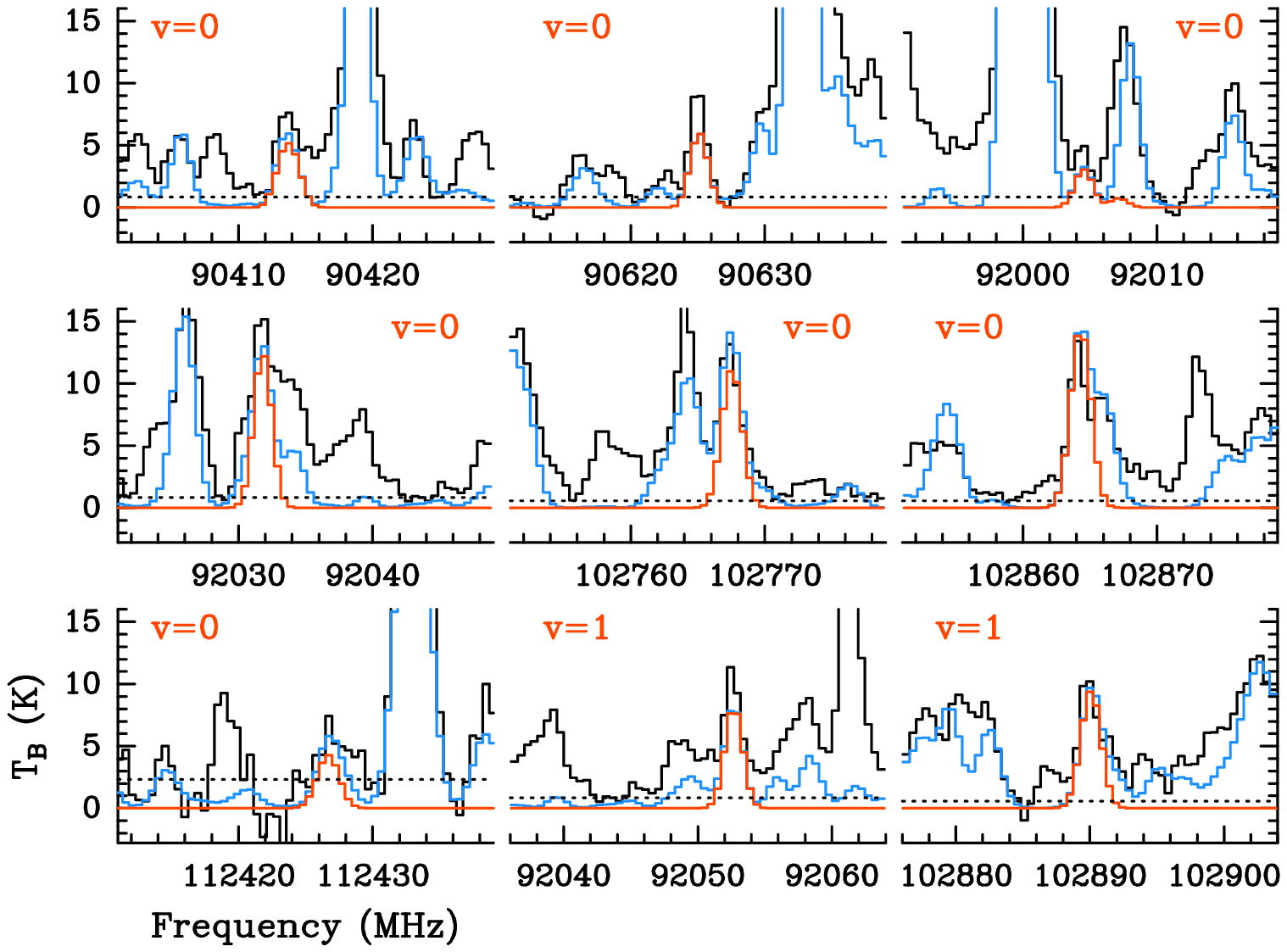}}}
\caption{Transitions of NH$_2$C(O)NH$_2$ in its vibrational ground state or 
first vibrationally excited state detected in the ReMoCA survey toward 
Sgr~B2(N1S). The best-fit LTE synthetic spectrum of NH$_2$C(O)NH$_2$ is 
displayed in red and overlaid on the observed spectrum of Sgr~B2(N1S) shown in 
black. The blue synthetic spectrum contains the contributions of all molecules 
identified in our survey so far, including NH$_2$C(O)NH$_2$. The dotted line 
indicates the $3\sigma$ noise level. 
}
\label{f:urea_det}
\end{figure}

We used the spectroscopic predictions described in Sect.~\ref{s:spectro} to
search for NH$_2$C(O)NH$_2$ toward Sgr~B2(N1S). The spectroscopic 
predictions were formatted in the same way as in the Cologne database for
molecular spectroscopy
\citep[CDMS,][]{Endres16}, using the
same partition function values as version 1 of the CDMS entry 60517 of urea.
The spectra of the transitions 
covered by the ReMoCA survey that are expected to be detectable are shown in
Fig.~\ref{f:spec_nh2conh2_ve0} for the vibrational ground state and in 
Fig.~\ref{f:spec_nh2conh2_ve1} for the first vibrationally excited state. We
assumed a temperature of 160~K to compute a synthetic spectrum for 
NH$_2$C(O)NH$_2$. The synthetic spectrum matches the ReMoCA spectrum relatively
well, as evaluated by visual inspection, but the temperature is not well 
constrained, as shown below with the population diagram analysis. Many 
transitions of 
urea are blended with emission lines from other species and cannot be 
unambiguously identified, but nine transitions, marked with green stars in 
Figs.~\ref{f:spec_nh2conh2_ve0} and \ref{f:spec_nh2conh2_ve1}, suffer from 
less contamination and can be considered as clearly detected. These detected 
transitions are presented in a concise way in Fig.~\ref{f:urea_det} and their
spectroscopic parameters are listed in Table~\ref{t:urea_det}. The parameters 
used to fit the urea spectrum are reported in Table~\ref{t:coldens}. 

\begin{table*}
 \begin{center}
 \caption{
 Spectroscopic parameters of urea transitions detected toward Sgr~B2(N1S).
}
 \label{t:urea_det}
 \vspace*{0.0ex}
 \begin{tabular}{rccccrrrrcrrrr}
 \hline\hline
  & & & & & \multicolumn{9}{c}{Quantum numbers} \\ 
 \multicolumn{1}{c}{Frequency} & \multicolumn{1}{c}{$\Delta f$\tablefootmark{a}}  & \multicolumn{1}{c}{$A_{\rm ul}$\tablefootmark{b}} & \multicolumn{1}{c}{$E_{\rm u}$\tablefootmark{c}}  & \multicolumn{1}{c}{$g_{\rm u}$\tablefootmark{d}} & \multicolumn{4}{c}{Upper level} & -- & \multicolumn{4}{c}{Lower level} \\ 
  \multicolumn{1}{c}{\small (MHz)} & \multicolumn{1}{c}{\small (kHz)} & \multicolumn{1}{c}{\small (10$^{-5}$ s$^{-1}$)} & \multicolumn{1}{c}{\small (K)} & & \multicolumn{1}{c}{$J_{\rm u}$} & \multicolumn{1}{c}{$K_{\rm a,u}$} & \multicolumn{1}{c}{$K_{\rm c,u}$} & \multicolumn{1}{c}{$\varv_{\rm u}$\tablefootmark{e}} & -- & \multicolumn{1}{c}{$J_{\rm l}$} & \multicolumn{1}{c}{$K_{\rm a,l}$} & \multicolumn{1}{c}{$K_{\rm c,l}$} & \multicolumn{1}{c}{$\varv_{\rm l}$\tablefootmark{e}}  \\ 
 \hline
90413.2891 &  5.7 &  1.62 &   107.6 &  31 & 15 &  7 &  8 & 0 & -- & 15 &  6 &  9 & 0 \\ 
90414.1278 &  5.7 &  1.62 &   107.6 &  31 & 15 &  8 &  8 & 0 & -- & 15 &  7 &  9 & 0 \\ 
90624.9667 &  5.8 &  1.52 &    92.1 &  29 & 14 &  6 &  8 & 0 & -- & 14 &  5 &  9 & 0 \\ 
90625.2268 &  5.8 &  1.52 &    92.1 &  29 & 14 &  7 &  8 & 0 & -- & 14 &  6 &  9 & 0 \\ 
92004.5312 &  4.6 &  2.31 &    17.6 &  13 &  6 &  3 &  4 & 0 & -- &  5 &  2 &  3 & 0 \\ 
92031.8092 &  6.8 &  3.79 &    20.8 &  17 &  8 &  0 &  8 & 0 & -- &  7 &  1 &  7 & 0 \\ 
92031.8092 &  6.8 &  3.79 &    20.8 &  17 &  8 &  1 &  8 & 0 & -- &  7 &  0 &  7 & 0 \\ 
102767.5263 &  6.4 &  4.55 &    24.6 &  17 &  8 &  1 &  7 & 0 & -- &  7 &  2 &  6 & 0 \\ 
102767.5465 &  6.4 &  4.55 &    24.6 &  17 &  8 &  2 &  7 & 0 & -- &  7 &  1 &  6 & 0 \\ 
102864.3049 &  7.4 &  5.37 &    25.7 &  19 &  9 &  0 &  9 & 0 & -- &  8 &  1 &  8 & 0 \\ 
102864.3049 &  7.4 &  5.37 &    25.7 &  19 &  9 &  1 &  9 & 0 & -- &  8 &  0 &  8 & 0 \\ 
112426.6894 &  7.0 &  2.07 &    82.9 &  29 & 14 &  4 & 10 & 0 & -- & 14 &  3 & 11 & 0 \\ 
112426.6895 &  7.0 &  2.07 &    82.9 &  29 & 14 &  5 & 10 & 0 & -- & 14 &  4 & 11 & 0 \\ 
\hline 
92052.6888 &  7.2 &  3.79 &    92.7 &  17 &  8 &  0 &  8 & 1 & -- &  7 &  1 &  7 & 1 \\ 
92052.6889 &  7.2 &  3.79 &    92.7 &  17 &  8 &  1 &  8 & 1 & -- &  7 &  0 &  7 & 1 \\ 
102890.0663 &  7.9 &  5.38 &    97.7 &  19 &  9 &  0 &  9 & 1 & -- &  8 &  1 &  8 & 1 \\ 
102890.0663 &  7.9 &  5.38 &    97.7 &  19 &  9 &  1 &  9 & 1 & -- &  8 &  0 &  8 & 1 \\ 
 \hline
 \end{tabular}
 \end{center}
 \vspace*{-2.5ex}
 \tablefoot{
 \tablefoottext{a}{Frequency uncertainty.}
 \tablefoottext{b}{Einstein coefficient for spontaneous emission.}
 \tablefoottext{c}{Upper-level energy.}
 \tablefoottext{d}{Upper-level degeneracy.}
 \tablefoottext{e}{$\varv = 0$ corresponds to the vibrational ground state and $\varv = 1$ to the first vibrationally excited state.}
 }
 \end{table*}

Two transitions from within the first vibrationally excited state are in 
marginal agreement with the observed spectrum when their synthetic spectrum is 
added to the contribution of all other identified molecules. The first slight
discrepancy occurs at 89.881~GHz where the full model overestimates the 
detected signal by $\sim$5$\sigma$ (see Fig.~\ref{f:spec_nh2conh2_ve1}). Given 
that the full model also 
overestimates the two lines (of other species) around $\sim$89.87~GHz, the
reason for this discrepancy is most likely an overestimation of the baseline 
level. The second discrepancy occurs at $\sim$102.645--102.650~GHz, with the 
full model that includes contributions from all identified molecules
overestimating the detected signal by $\sim$15$\sigma$ (see blue spectrum in 
Fig.~\ref{f:spec_nh2conh2_ve1}). The urea line is blended with transitions of 
several species, in particular a transition of C$_2$H$_5$CN in its 
vibrationally excited state $\varv_{20}=1$ that has a large frequency 
uncertainty of 1.1~MHz. We suspect that the frequency of this transition is
incorrect. When we remove it from the full model, the remaining discrepancy
at the frequency of the urea line is reduced to $\sim$10$\sigma$. Given that the
model also overestimates the observed line at $\sim$102.641 GHz (which is a 
blend of \textit{c}-C$_2$H$_4$O and \textit{n}-C$_3$H$_7$CN, both unambiguously 
detected) by $\sim$11$\sigma$, hinting maybe at a slight overestimation of the 
baseline level, this second discrepancy is likely not to be significant. All 
in all, these two slight discrepancies do not invalidate our identification of 
urea.

In addition to the vibrational ground state and first vibrationally excited 
state, we also looked for transitions from within the second and third 
vibrationally excited states reported in Sect.~\ref{s:spectro}. Unfortunately,
they are either predicted to be below the $3\sigma$ detection limit or blended 
with stronger lines of other species, which prevents their identification.

\begin{table*}[!ht]
 \begin{center}
 \caption{
 Parameters of our best-fit LTE model of urea, N-methylformamide, and other related molecules toward Sgr~B2(N1S).
}
 \label{t:coldens}
 \vspace*{-1.2ex}
 \begin{tabular}{lcrccccccr}
 \hline\hline
 \multicolumn{1}{c}{Molecule} & \multicolumn{1}{c}{Status\tablefootmark{a}} & \multicolumn{1}{c}{$N_{\rm det}$\tablefootmark{b}} & \multicolumn{1}{c}{Size\tablefootmark{c}} & \multicolumn{1}{c}{$T_{\mathrm{rot}}$\tablefootmark{d}} & \multicolumn{1}{c}{$N$\tablefootmark{e}} & \multicolumn{1}{c}{$F_{\rm vib}$\tablefootmark{f}} & \multicolumn{1}{c}{$\Delta V$\tablefootmark{g}} & \multicolumn{1}{c}{$V_{\mathrm{off}}$\tablefootmark{h}} & \multicolumn{1}{c}{$\frac{N_{\rm ref}}{N}$\tablefootmark{i}} \\ 
  & & & \multicolumn{1}{c}{\small ($''$)} & \multicolumn{1}{c}{\small (K)} & \multicolumn{1}{c}{\small (cm$^{-2}$)} & & \multicolumn{1}{c}{\small (km~s$^{-1}$)} & \multicolumn{1}{c}{\small (km~s$^{-1}$)} & \\ 
 \hline
 NH$_2$C(O)NH$_2$, $\varv=0$$^\star$ & d & 7 &  2.0 &  160 &  2.7 (16) & 1.86 & 5.0 & 0.0 &       1 \\ 
 \hspace*{13.2ex} $\varv=1$ & d & 2 &  2.0 &  160 &  2.7 (16) & 1.86 & 5.0 & 0.0 &       1 \\ 
\hline 
 CH$_3$NHCHO, $\varv=0$$^\star$ & d & 38 &  2.0 &  160 &  2.6 (17) & 1.19 & 4.5 & 0.5 &       1 \\ 
 \hspace*{12.5ex} $\varv_{\rm t}=1$ & d & 15 &  2.0 &  160 &  2.6 (17) & 1.19 & 4.5 & 0.5 &       1 \\ 
 \hspace*{12.5ex} $\varv_{\rm t}=2$ & d & 7 &  2.0 &  160 &  2.6 (17) & 1.19 & 4.5 & 0.5 &       1 \\ 
\hline 
 CH$_3$NCO, $\varv=0$$^\star$ & d & 49 &  2.0 &  200 &  2.5 (17) & 1.00 & 5.0 & 0.0 &       1 \\ 
 \hspace*{9.3ex} $\varv_{\rm b}=1$ & d & 2 &  2.0 &  200 &  2.5 (17) & 1.00 & 5.0 & 0.0 &       1 \\ 
\hline 
 NH$_2$CHO, $\varv=0$$^\star$ & d & 20 &  2.0 &  160 &  2.9 (18) & 1.09 & 6.0 & 0.0 &       1 \\ 
 \hspace*{9.5ex} $\varv_{12}=1$ & d & 14 &  2.0 &  160 &  2.9 (18) & 1.09 & 6.0 & 0.0 &       1 \\ 
 NH$_2$$^{13}$CHO, $\varv=0$ & d & 11 &  2.0 &  200 &  1.1 (17) & 1.17 & 5.5 & 0.0 &      28 \\ 
 \hspace*{11.0ex} $\varv_{12}=1$ & d & 1 &  2.0 &  200 &  1.1 (17) & 1.17 & 5.5 & 0.0 &      28 \\ 
 NH$_2$CH$^{18}$O, $\varv=0$ & d & 2 &  2.0 &  200 &  2.1 (16) & 1.17 & 5.5 & 0.0 &     140 \\ 
 $^{15}$NH$_2$CHO, $\varv=0$ & d & 2 &  2.0 &  200 &  9.4 (15) & 1.17 & 5.5 & 0.0 &     314 \\ 
\hline 
 CH$_3$C(O)NH$_2$, $\varv=0$$^\star$ & d & 80 &  2.0 &  160 &  4.1 (17) & 1.16 & 5.0 & 0.0 &       1 \\ 
 \hspace*{11.8ex} $\varv_{\rm t}=1$ & d & 53 &  2.0 &  160 &  4.1 (17) & 1.16 & 5.0 & 0.0 &       1 \\ 
 \hspace*{11.8ex} $\varv_{\rm t}=2$ & d & 18 &  2.0 &  160 &  4.1 (17) & 1.16 & 5.0 & 0.0 &       1 \\ 
 \hspace*{11.8ex} $\Delta\varv_{\rm t} \neq 0$ & d & 2 &  2.0 &  160 &  4.1 (17) & 1.16 & 5.0 & 0.0 &       1 \\ 
\hline 
 \end{tabular}
 \end{center}
 \vspace*{-2.5ex}
 \tablefoot{
 \tablefoottext{a}{d: detection.}
 \tablefoottext{b}{Number of detected lines \citep[conservative estimate, see Sect.~3 of][]{Belloche16}. One line of a given species may mean a group of transitions of that species that are blended together.}
 \tablefoottext{c}{Source diameter (\textit{FWHM}).}
 \tablefoottext{d}{Rotational temperature.}
 \tablefoottext{e}{Total column density of the molecule. $x$ ($y$) means $x \times 10^y$. An identical value for all listed vibrational/torsional states of a molecule means that LTE is an adequate description of the vibrational/torsional excitation.}
 \tablefoottext{f}{Correction factor that was applied to the column density to account for the contribution of vibrationally excited states, in the cases where this contribution was not included in the partition function of the spectroscopic predictions.}
 \tablefoottext{g}{Linewidth (\textit{FWHM}).}
 \tablefoottext{h}{Velocity offset with respect to the assumed systemic velocity of Sgr~B2(N1S), $V_{\mathrm{sys}} = 62$ km~s$^{-1}$.}
 \tablefoottext{i}{Column density ratio, with $N_{\rm ref}$ the column density of the previous reference species marked with a $\star$.}
 }
 \end{table*}

\begin{table}
 \begin{center}
 \caption{
 Rotational temperatures derived from population diagrams toward Sgr~B2(N1S).
}
 \label{t:popfit}
 \vspace*{0.0ex}
 \begin{tabular}{lll}
 \hline\hline
 \multicolumn{1}{c}{Molecule} & \multicolumn{1}{c}{States\tablefootmark{a}} & \multicolumn{1}{c}{$T_{\rm fit}$\tablefootmark{b}} \\ 
  & & \multicolumn{1}{c}{\small (K)} \\ 
 \hline
NH$_2$C(O)NH$_2$ & $\varv=0$, $\varv=1$ &   346 (187) \\ 
\hline 
CH$_3$NHCHO & $\varv=0$, $\varv_{\rm t}=1$, $\varv_{\rm t}=2$ &   187 (17) \\ 
\hline 
CH$_3$NCO & $\varv=0$, $\varv_{\rm b}=1$ &   197 (24) \\ 
\hline 
NH$_2$CHO & $\varv=0$, $\varv_{12}=1$ & 159.3 (5.0) \\ 
NH$_2$$^{13}$CHO & $\varv=0$, $\varv_{12}=1$ &   284 (91) \\ 
\hline 
CH$_3$CONH$_2$ & $\varv_{\rm t}=0$, $\varv_{\rm t}=1$, $\varv_{\rm t}=2$, $\Delta\varv_{\rm t} \neq 0$ &   226 (13) \\ 
\hline 
 \end{tabular}
 \end{center}
 \vspace*{-2.5ex}
 \tablefoot{
 \tablefoottext{a}{Vibrational states that were taken into account to fit the population diagram.}
 \tablefoottext{b}{The standard deviation of the fit is given in parentheses. As explained in Sect.~3 of \citet{Belloche16} and in Sect.~\ref{ss:urea} of this work, this uncertainty is purely statistical and should be viewed with caution. It may be underestimated.}
 }
 \end{table}

We also built a population diagram following the same method as 
\citet{Belloche16} and using all nine detected transitions, as well as a few 
additional ones for which we have partly identified the contaminating species 
(Fig.~\ref{f:popdiag_nh2conh2}). The formal fit to this population diagram,
after correction for the line optical depth and the contamination by the other
identified species (Fig.~\ref{f:popdiag_nh2conh2}b), yields a rotational 
temperature of $350 \pm 190$~K (see Table~\ref{t:popfit}). On top of the high
uncertainty, we think that this is not a reliable number. The first reason is 
that the population diagram method originally described by \citet{Goldsmith99} 
requires the background continuum temperature to be negligible compared to the 
excitation temperature of the transitions. This is no longer the case at the 
high angular resolution achieved by the ReMoCA survey. An obvious sign that 
this assumption is not valid is the distribution of the synthetic (red) 
datapoints in Fig.~\ref{f:popdiag_nh2conh2}b. After correction for the line 
optical depth, they should fall onto a straight line, modulo a small 
dispersion due to the finite ranges used to integrate the line intensities, but 
this is not the case: the shift by a factor of 1.5 between the synthetic 
datapoints in the range 80--110~K results from these datapoints coming from 
setups with different angular resolutions, hence different continuum 
background levels. This dependence on the background level affects the 
observed datapoints as well and cannot be accounted for by a simple linear fit 
to the population diagram. 

The second reason why the fitted rotational temperature is not reliable is 
that some transitions suffer from contamination from still unidentified 
species, especially in their linewings, which leads to an overestimation of 
their integrated intensities. This additional uncertainty is not included in 
the error bars plotted in the population diagram, which are purely statistical.
Finally, there may be some systematic uncertainties 
related to the still preliminary line-continuum splitting. Given these three 
reasons, the first one being the most severe, this population diagram 
should be viewed with caution. We rather consider it as a convenient tool to 
visualize in a concise way the match between the synthetic and observed 
spectra. We stress that the quality of this match is carefully evaluated 
by visual inspection of 
the spectra themselves which we use to optimize the model parameters.

The model that contains contributions from all the species that we identified 
toward Sgr~B2(N2) 
by analyzing the EMoCA survey over the past five years fits well the ReMoCA 
spectrum of this source. We used this model to search for urea toward 
N2. No sign of urea was found and the upper limit on its column density 
is reported in Table~\ref{t:coldens_n2}, assuming the same rotational 
temperature (180~K) and source size (0.9$\arcsec$) as derived for 
CH$_3$C(O)NH$_2$ toward N2 by \citet{Belloche17}. The upper limit lies more than
one order of magnitude below the column density derived for urea toward
N1S.

\begin{table*}[!ht]
 \begin{center}
 \caption{
 Parameters of our best-fit LTE model of selected complex organic molecules toward Sgr~B2(N2).
}
 \label{t:coldens_n2}
 \vspace*{-1.2ex}
 \begin{tabular}{lcrcccccc}
 \hline\hline
 \multicolumn{1}{c}{Molecule} & \multicolumn{1}{c}{Status\tablefootmark{a}} & \multicolumn{1}{c}{$N_{\rm det}$\tablefootmark{b}} & \multicolumn{1}{c}{Size\tablefootmark{c}} & \multicolumn{1}{c}{$T_{\mathrm{rot}}$\tablefootmark{d}} & \multicolumn{1}{c}{$N$\tablefootmark{e}} & \multicolumn{1}{c}{$F_{\rm vib}$\tablefootmark{f}} & \multicolumn{1}{c}{$\Delta V$\tablefootmark{g}} & \multicolumn{1}{c}{$V_{\mathrm{off}}$\tablefootmark{h}} \\ 
  & & & \multicolumn{1}{c}{\small ($''$)} & \multicolumn{1}{c}{\small (K)} & \multicolumn{1}{c}{\small (cm$^{-2}$)} & & \multicolumn{1}{c}{\small (km~s$^{-1}$)} & \multicolumn{1}{c}{\small (km~s$^{-1}$)} \\ 
 \hline
 NH$_2$C(O)NH$_2$, $\varv=0$ & n & 0 &  0.9 &  180 & $<$  2.3 (15) & 1.95 & 5.0 & 0.5 \\ 
\hline 
 CH$_3$NHCHO\tablefootmark{i} & t & 5 &  0.9 &  180 &  1.0 (17) & 1.26 & 5.0 & 0.5 \\ 
\hline 
 CH$_3$NCO\tablefootmark{i} & d & 64 &  1.2 &  150 &  2.2 (17) & 1.00 & 5.0 & -0.6 \\ 
\hline 
 NH$_2$CHO\tablefootmark{i}\tablefootmark{j} & d & 43 &  0.8 &  200 &  2.6 (18) & 1.17 & 5.5 & 0.2 \\ 
\hline 
 CH$_3$C(O)NH$_2$\tablefootmark{i} & d & 23 &  0.9 &  180 &  1.4 (17) & 1.23 & 5.0 & 1.5 \\ 
\hline 
 \end{tabular}
 \end{center}
 \vspace*{-2.5ex}
 \tablefoot{
 \tablefoottext{a}{d: detection, t: tentative detection, n: nondetection.}
 \tablefoottext{b}{Number of detected lines \citep[conservative estimate, see Sect.~3 of][]{Belloche16}. One line of a given species may mean a group of transitions of that species that are blended together.}
 \tablefoottext{c}{Source diameter (\textit{FWHM}).}
 \tablefoottext{d}{Rotational temperature.}
 \tablefoottext{e}{Total column density of the molecule. $x$ ($y$) means $x \times 10^y$.}
 \tablefoottext{f}{Correction factor that was applied to the column density to account for the contribution of vibrationally excited states, in the cases where this contribution was not included in the partition function of the spectroscopic predictions.}
 \tablefoottext{g}{Linewidth (\textit{FWHM}).}
 \tablefoottext{h}{Velocity offset with respect to the assumed systemic velocity of Sgr~B2(N2), $V_{\mathrm{sys}} = 74$ km~s$^{-1}$.}
 \tablefoottext{i}{The parameters were derived from the EMoCA survey by \citet{Belloche17}.}
 \tablefoottext{j}{For NH$_2$CHO, we report the parameters derived from the vibrationally excited state $\varv_{12}=1$.}
 }
 \end{table*}

\subsection{Confirmation of N-methylformamide}
\label{ss:ch3nhcho}

The ReMoCA survey provides a clear detection of N-methylformamide toward N1S.
The transitions covered by the survey are shown in 
Figs.~\ref{f:spec_ch3nhcho_ve0}, \ref{f:spec_ch3nhcho_ve1}, and 
\ref{f:spec_ch3nhcho_ve2} for $\varv=0$, $\varv_{\rm t}=1$, and $\varv_{\rm t}=2$, 
respectively. We modeled its emission assuming a temperature of 160~K. Like 
urea, many lines of CH$_3$NHCHO are contaminated by other species, but we 
count sixty clearly detected lines, marked with green stars in these 
figures. A fit to the population diagram of CH$_3$NHCHO shown in 
Fig.~\ref{f:popdiag_ch3nhcho} yields a formal rotational temperature of 
$187 \pm 17$~K (see Table~\ref{t:popfit}), but this diagram suffers from the 
same limitations as the one for urea (see Sect.~\ref{ss:urea}). This detection 
confirms the presence of N-methylformamide in the ISM, as was initially 
suggested by the tentative detection reported toward Sgr~B2(N2) on the basis 
of the EMoCA survey \citep[][]{Belloche17}.

\subsection{Other related molecules}
\label{ss:other}

In order to compare the column density derived for NH$_2$C(O)NH$_2$ to other
COMs, we derive the column densities of CH$_3$NCO, NH$_2$CHO and its 
isotopologs, and CH$_3$C(O)NH$_2$ toward N1S using the ReMoCA survey. The
parameters derived from the fits are listed in Table~\ref{t:coldens} and their
spectra are shown in Figs.~\ref{f:spec_ch3nco_ve0}--\ref{f:spec_ch3conh2_cv}.
Their population diagrams are displayed in 
Figs.~\ref{f:popdiag_ch3nco}--\ref{f:popdiag_ch3conh2}. 

We ignored the transitions of formamide with opacities higher than two to build
its population diagram because they cannot be modeled properly in the 
framework of our simple Weeds model, as can be seen in 
Fig.~\ref{f:spec_nh2cho_ve0} at 84.5~GHz, 87.9~GHz, or 102.1~GHz for instance.
The fit to the resulting population diagram yields a well-constrained 
rotational temperature of $159 \pm 5$~K (see Fig.~\ref{f:popdiag_nh2cho} and 
Table~\ref{t:popfit}) and we adopt a temperature of 160~K to produce the 
synthetic spectrum of formamide. This also motivates our choice of the 
temperature assumed for the synthetic spectra of urea (Sect.~\ref{ss:urea}) 
and N-methylformamide (Sect.~\ref{ss:ch3nhcho}). We note that the population 
diagram of the $^{13}$C isotopolog, especially the high energy range covered by 
the $\varv_{12}=1$ transitions, suggests a higher temperature, and we modeled 
this isotopolog (and the other isotopologs of formamide) assuming a temperature 
of 200~K. However, it could be that the $\varv_{12}=1$ transitions of the 
$^{13}$C isotopolog are all contaminated by emission lines of other unidentified 
species, which would bias the fit toward higher temperatures and explain the 
discrepancy with respect to the main isotopolog. 

The $^{12}$C/$^{13}$C ratio we obtain for formamide is typical for the Galactic 
center region, where ratios of around 20 to 30 have been reported 
\citep[e.g.,][]{Mueller08,Belloche16,Halfen17}. The detections of 
NH$_2$CH$^{18}$O and $^{15}$NH$_2$CHO toward Sgr~B2(N1S) are to our knowledge 
the first interstellar detections of these isotopologs. The 
$^{16}$O/$^{18}$O ratio of 140 is close to the value of 182 reported by 
\citet{Mueller16} for methanol in Sgr~B2(N2). The $^{14}$N/$^{15}$N ratio of 314 
is consistent with the value of $350 \pm 50$ derived by \citet{Thiel19b} for 
CN in the envelope of Sgr~B2, but it is somewhat lower than the value of 500 
derived by \citet{Belloche16} for ethyl cyanide in Sgr~B2(N2).

\subsection{Comparison of Sgr~B2(N1S) and Sgr~B2(N2)}

Table~\ref{t:coldens} gives the column densities obtained for Sgr~B2(N1S) on 
the basis of the ReMoCA survey. For comparison, we report in 
Table~\ref{t:coldens_n2}, in addition to the ReMoCA upper limit to the column 
density of urea, the column densities of the other species derived for 
Sgr~B2(N2) by \citet{Belloche17} on the basis of the EMoCA survey. These 
tables yield the following abundances relative to formamide: 
NH$_2$CHO\,/\,CH$_3$NCO\,/\,CH$_3$NHCHO\,/\,CH$_3$C(O)NH$_2$\,/\,NH$_2$C(O)NH$_2$ = 
1\,/\,0.086\,/\,0.090\,/\,0.14\,/\,0.0093 for Sgr~B2(N1S) and 
1\,/\,0.085\,/\,0.038\,/\,0.054\,/\,<0.00088 for Sgr~B2(N2). The two sources
have exactly the same abundance of methyl isocyanate relative to formamide,
while N-methylformamide and acetamide are $\sim$2.5 times more abundant 
in Sgr~B2(N1S) than in Sgr~B2(N2). The difference for urea is even more 
pronounced: urea is about one order of magnitude less abundant than 
CH$_3$NHCHO, CH$_3$NCO, and CH$_3$C(O)NH$_2$ in Sgr~B2(N1S), while it is at 
least 40--100 times less abundant than these three species in 
Sgr~B2(N2).

\section{Chemical modeling}
\label{s:chemistry}

To simulate the chemistry of Sgr~B2(N1S), we used the astrochemical kinetics 
code MAGICKAL \citep[Model for Astrophysical Gas and Ice Chemical Kinetics And 
Layering,][]{Garrod13} with a physical and chemical setup very similar to that
presented by \citet{Belloche17}, for which the emphasis was the production and 
destruction of CH$_3$NCO and CH$_3$NHCHO in Sgr~B2(N2). A brief description of 
the model is given here. The reader should refer to \citet{Garrod13} and 
\citet{Belloche14,Belloche17} for more detail.

\begin{table*}
\caption{Peak simulated fractional abundances and associated temperatures.}
\label{t:model}
\centering
\begin{tabular}{lcccccccc}
\hline
\hline
\noalign{\smallskip}
Molecule & \multicolumn{2}{c}{\textit{Fast} model} & & \multicolumn{2}{c}{\textit{Medium} model} & & \multicolumn{2}{c}{\textit{Slow} model} \\
\cline{2-3} \cline{5-6} \cline{8-9}
\noalign{\smallskip}
 & $n$(i)/$n$(H$_2$) & $T$ & & $n$(i)/$n$(H$_2$) & $T$ & & $n$(i)/$n$(H$_2$) & $T$ \\
 & & {\small (K)} & & & {\small (K)} & & & {\small (K)} \\
\hline
\noalign{\smallskip}

NH$_2$C(O)NH$_2$ & 3.9(-8)  & 206 & & 8.3(-8) & 200 & & 5.4(-8)  & 191  \\
CH$_3$NHCHO      & 4.4(-10) & 170 & & 2.2(-9) & 165 & & 1.7(-9)  & 158  \\
CH$_3$NCO        & 1.3(-8)  & 150 & & 9.4(-9) & 143 & & 1.1(-10) & 131  \\
NH$_2$CHO        & 1.0(-6)  & 255 & & 5.1(-7) & 157 & & 5.9(-8)  & 141  \\
CH$_3$C(O)NH$_2$ & 1.7(-8)  & 143 & & 3.2(-8) & 138 & & 2.3(-7)  & 134  \\

\hline
\end{tabular}
\tablefoot{$x$ ($y$) means $x \times 10^y$.}
\end{table*}

The model is run in two stages: (i) a cold collapse from a gas density of 
$3 \times 10^3$~cm$^{-3}$ to $2 \times 10^8$ cm$^{-3}$, during which the gas 
temperature is static, while the dust temperature falls from $\sim$18~K to 
8~K; followed by (ii) a warm up from 8~K to 400~K at fixed density, during 
which the gas and dust temperatures are assumed to be identical. The timescale 
for the initial collapse stage is approximately 10$^6$ yr. For the warm-up 
phase, here the usual three warm-up timescales (fast, medium, and slow) are 
tested, corresponding to a time of $5\times 10^4$~yr, $2 \times 10^5$~yr, or 
10$^6$~yr to reach a temperature of 200~K. The cosmic-ray ionization rate is 
held at the canonical value, $\zeta$=$1.3 \times 10^{-17}$ s$^{-1}$.

The chemical network, which is unchanged from that of \citet{Belloche17}, 
includes a full gas-phase, grain/ice-surface, and ice-mantle chemistry. The 
choice of grain-surface activation energy barriers relating to the CH$_3$NCO 
and CH$_3$NHCHO chemistry corresponds to one of the optimal model setups 
presented in that publication, labeled M4B. The only change made here to that 
model is the adjustment of several grain-surface binding energies, most 
importantly, that of methyl isocyanate. In the model of \citet{Belloche17}, 
the desorption (binding) energy, $E_{\mathrm{des}}$, used for this molecule was 
3575~K, based on a simple additive formula using values for its constituent 
atoms and groups (CH$_3$, N, C, and O). That value produced a relatively low 
desorption temperature in the hot-core models, around 75~K, which is 
substantially lower than the excitation temperature of 200~K derived from the
observations. Recently, \citet{Bertin17} determined desorption energies on 
water ice for the related molecules CH$_3$CN and CH$_3$NC of 6150~K and 
5686~K, respectively. As it is very likely that the desorption energy of 
CH$_3$NCO would be much closer to these values than to our earlier estimate, 
we instead use the value for methyl isocyanide (CH$_3$NC) as a guide, to give 
$E_{\mathrm{des}}$(CH$_3$NCO) = $E_{\mathrm{des}}$(CH$_3$NC) + $E_{\mathrm{des}}$(O) = 
6486~K. Through a similar additive method, the desorption energy for N-methyl 
formamide is also raised, to a value of 7386~K, approximately 1000~K higher 
than our previous value. As may be seen in Table~\ref{t:model}, these values 
produce peak-abundance temperatures much closer to the best-fit spectral-model 
values adopted in the present work. The desorption energies of several related 
radicals, as well as those of methyl cyanide/isocyanide themselves, are 
adjusted similarly in this model.

\subsection{Results}
\label{ss:chemistry}

\begin{figure}
\centerline{\resizebox{0.9\hsize}{!}{\includegraphics{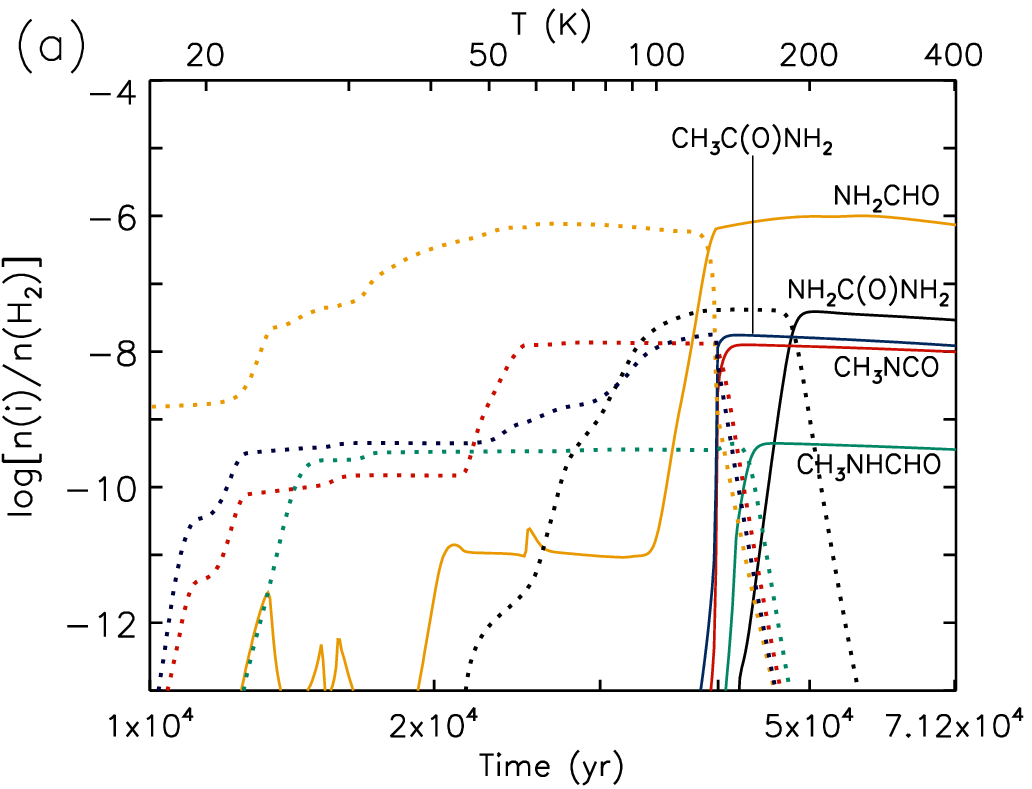}}}
\centerline{\resizebox{0.9\hsize}{!}{\includegraphics{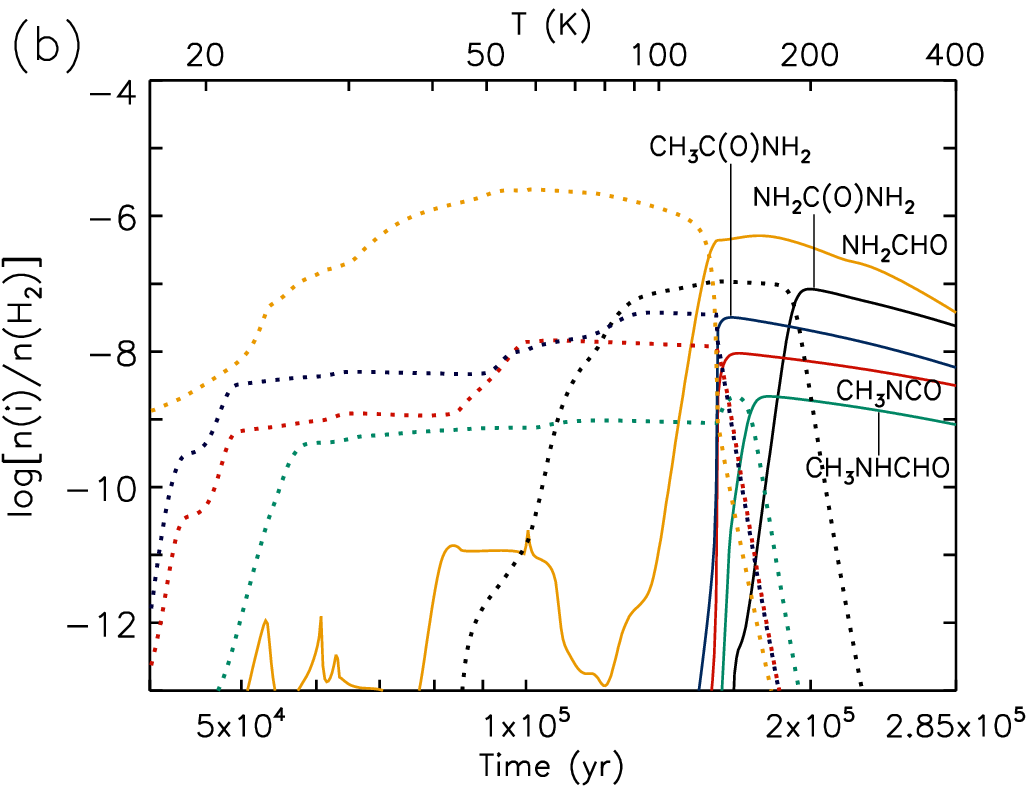}}}
\centerline{\resizebox{0.9\hsize}{!}{\includegraphics{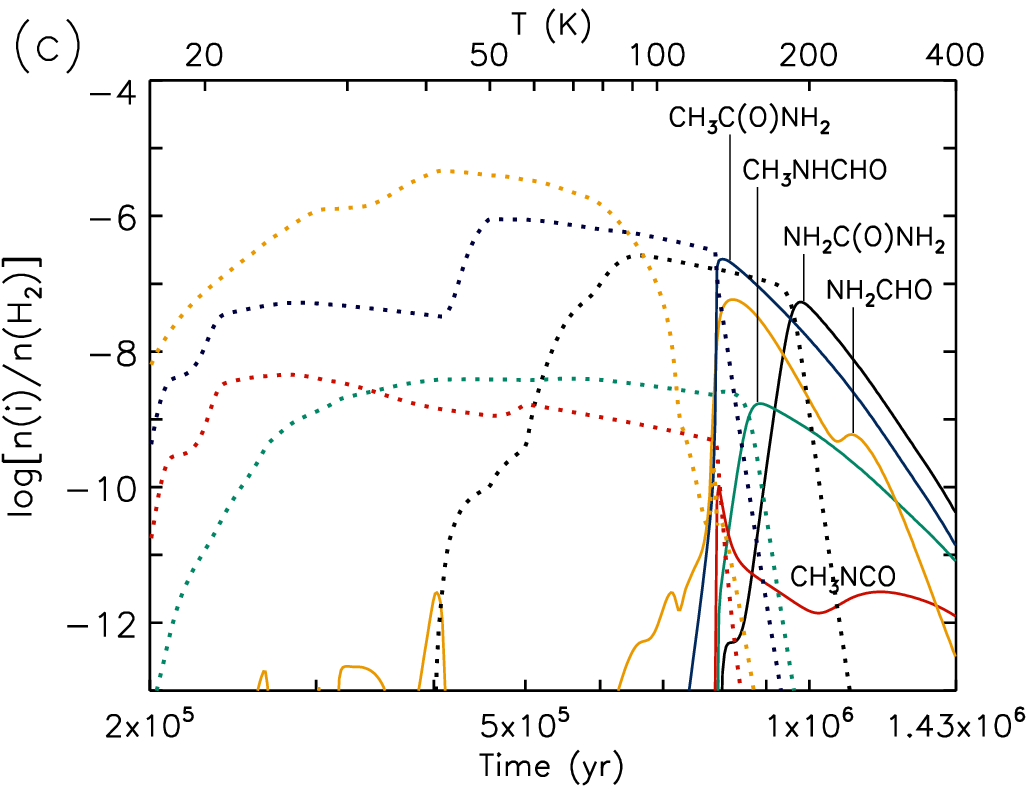}}}
\caption{Fractional abundances relative to H$_2$ with respect to time during 
the warm-up stage of the astrochemical model. Solid lines indicate gas-phase 
abundances; dashed lines of the same color indicate the same molecules in the
solid phase. Panels: (a) \textit{Fast} warm-up model; (b) \textit{Medium} 
warm-up model; (c) \textit{Slow} warm-up model.}
\label{f:model}
\end{figure}

Figure~\ref{f:model} shows the time evolution of the simulated chemical 
abundances of the five observed molecules of Table~\ref{t:coldens}, for each 
of the three warm-up scenarios (\textit{fast}, \textit{medium}, and 
\textit{slow}), while Table~\ref{t:model} shows the peak abundances and 
associated model temperatures for each of the molecules and models. The peak 
abundances are mostly produced at the desorption temperatures of the 
molecules, although in the case of NH$_2$CHO, there is some gas-phase 
production that produces a later peak in the \textit{fast} and \textit{medium} 
warm-up models, through the reaction NH$_2$ + H$_2$CO $\rightarrow$ NH$_2$CHO 
+ H proposed by \citet{Barone15}.

In the following, we compare the peak simulated abundances of the 
molecules listed in Table~\ref{t:model} with each other. The particular 
time (corresponding to a temperature in the frame of our chemical model) at 
which a molecule desorbs strongly depends on the assumption made about its 
desorption energy (see for instance the case of CH$_3$NCO discussed above). 
Therefore, the positions of the peaks of the solid curves in 
Fig.~\ref{f:model} are uncertain. However, for most species, the peak 
abundance in the gas phase stays close to the abundance of the molecule in the 
solid phase before it desorbed, so, to zeroth order, the particular time at 
which it desorbs will not affect much its peak abundance in the gas phase. 
Given that the observed rotational temperatures of the various molecules do not
differ much, we think that they in reality desorb at a similar temperature. 
This motivates our approach of comparing observed column density ratios to the 
ratios of peak abundances produced by the model in the gas phase.

In comparison to the other N-bearing molecules listed in Table \ref{t:model}, 
the abundance of urea is higher in the model than in the observational data. 
However, between each of the models, its peak abundance is fairly stable, while 
the peak abundances of the other chemical species are more variable. In the 
\textit{medium} model, urea is around an order of magnitude less abundant than 
formamide (NH$_2$CHO), while the observations toward Sgr~B2(N1S) suggest a 
ratio of around two orders of magnitude. The \textit{fast} model better 
reproduces the 
observational ratio, while the \textit{slow} model is worse. On the other 
hand, urea is more abundant than acetamide (CH$_3$C(O)NH$_2$) in the 
\textit{fast} and \textit{medium} models, contrary to the observations, while 
the ratio of these two molecules in the \textit{slow} model is in modest 
agreement with observations. 

In all of the models, urea is mainly produced directly from NH$_2$CHO in the 
grain-surface ices; hydrogen abstraction by radicals, including NH$_2$ itself, 
produces the radical NH$_2$CO, with which NH$_2$ reacts to form 
NH$_2$C(O)NH$_2$. Acetamide is mainly formed by a similar mechanism on the 
grains, ending in the addition of a methyl group. However, in the 
\textit{slow} model, the addition of NH$_2$ to the CH$_3$CO radical, which is 
derived from acetaldehyde, gives a major boost to the abundance of acetamide 
on the grains at around 50~K and is responsible for its greater abundance in 
the gas phase once the solid-phase material has desorbed.

The observations suggest that methyl isocyanate and N-methylformamide should 
be similar in abundance, but the models produce a range of ratios. The 
\textit{medium} model comes closest to the observed ratio between the two, as 
well as to the ratios with other molecules such as formamide and acetamide. 
The adjustment to the binding energy of CH$_3$NCO provides something of an 
improvement to the model results, allowing a somewhat higher abundance to be 
achieved than in the models of \citet{Belloche17}, while also providing a more 
plausible temperature of peak abundance in comparison to the observational 
excitation temperature.

\section{Discussion}
\label{s:discussion}

\subsection{Astronomical detection}

We presented in Sect.~\ref{ss:urea} an identification of urea that is 
based on the assignment of nine clearly detected lines (see 
Fig.~\ref{f:urea_det}). While there is no strict criterion to decide how many 
lines are necessary to have a robust detection of a molecule in emission in a
nearly confusion-limited spectrum, we consider that our detailed modeling of 
the full spectrum of Sgr~B2(N1S) that includes contributions from all 
molecules identified so far
makes the detection of these nine lines, which are well reproduced by our LTE
model of the urea emission, a sufficient number to secure the identification
of the molecule. For comparison, we claimed only a tentative detection of
N-methylformamide toward Sgr~B2(N2) in our previous work using the EMoCA survey 
because there were only five clearly detected lines assigned to this molecule 
\citep[][]{Belloche17}. With now sixty lines of N-methylformamide identified
in the ReMoCA spectrum of Sgr~B2(N1S), the identification of N-methylformamide
in the ISM is robustly confirmed.

\citet{Xue19} presented a quantitative method to evaluate the degree
to which a line can be claimed as clearly assigned to a molecule. While we 
appreciate the
efforts made to propose a quantitative estimator (their P and D factors), we 
did not attempt to apply this method here for two reasons. First, this method 
does not take into account the systematic uncertainties that may affect the 
observed spectra, such as the uncertainty related to the level of the baseline 
in a spectrum that is as close to the confusion limit as the spectrum we 
analyzed here, or the uncertainty related to the attenuation by the dust, 
which we account for in our radiative transfer model as well as we can. A 
second limitation of the method of \citet{Xue19} is that the frequency range 
over which the P and D factors are computed for each line is, as stated by the 
authors themselves, crucial for calculating these factors. There is some 
degree of arbitrariness in the selection of these frequency ranges, especially 
in the case of confusion-limited spectra where (partial) blends frequently 
occur. Rather than relying blindly on these quantitative estimators, we 
consider that the visualization of the nine lines clearly assigned to urea in
Fig.~\ref{f:urea_det} along with the LTE synthetic spectrum of urea and the
synthetic spectrum that includes contributions from all molecules are 
sufficient to assess the robustness of the identification.

\citet{Remijan14} attempted a fit to the data set of their multitelescope 
search for urea mentioned in Sect.~\ref{s:introduction}. They obtained 
a column density of $\sim$$8 \times 10^{14}$~cm$^{-2}$ for urea toward 
Sgr~B2(N1), assuming a rotational temperature of $\sim$80~K and a source size 
between 2$\arcsec$ and 3.25$\arcsec$ depending on the telescope. This column 
density is more than one order of magnitude lower than the column density we 
derive from the ReMoCA survey for a source size of $2\arcsec$. A fit of 
the ReMoCA data with a temperature of 80~K does not significantly change the 
column density that we derive for urea. An explanation for the discrepancy 
between \citeauthor{Remijan14}'s result and ours could be that the urea 
emission region is more compact than the size that they and we assumed. 
Because of the small beam of the ReMoCA survey, reducing the size of the urea 
emission region by a factor of two would increase its column density by 40$\%$ 
only, while it would imply a column density a factor of four times higher in 
\citeauthor{Remijan14}'s analysis. This would reduce the discrepancy by a 
factor of three, which is not sufficient to resolve it.

Another possibility is that there is an issue with the calculation of the 
column density reported by \citet{Remijan14} or with the spectroscopic 
predictions they used. To test this hypothesis, we used the spectroscopic 
predictions obtained in Sect.~\ref{s:spectro} to compute a synthetic spectrum 
for the IRAM 30\,m telescope as \citeauthor{Remijan14} did in their Table 2. 
With their parameters (a source size of 3.25$\arcsec$, a linewidth of 
7~km~s$^{-1}$, a column density of $8 \times 10^{14}$~cm$^{-2}$, a temperature of 
80~K, and a beam size of 20.2$\arcsec$\footnote{This angular size 
underestimates the actual beam size (HPBW) of the IRAM 30\,m telescope, which 
is about $23.9\arcsec$ at 102.9~GHz.}), we obtain a peak brightness 
temperature of 69~mK at 102.864~GHz, about 15\% lower than predicted by 
\citet{Remijan14} at this frequency. The origin of this difference is 
unclear, but it is small and does not explain the discrepancy mentioned in the 
previous paragraph.

A third reason is that we account for the contribution of vibrationally excited
states to the partition function of urea, while \citet{Remijan14} did not. 
This increases the column density that we derive for urea by a factor of 1.9 
at a temperature of 160~K (see Table~\ref{t:coldens}). Ignoring this 
vibrational correction would reduce the discrepancy by a factor of 1.9, which 
is insufficient to resolve it completely.

Finally, another reason why our model delivers a higher column density is 
that we take into account the strong background continuum emission in the
radiative transfer equation. For instance, the continuum level toward 
Sgr~B2(N1S) at 102.864~GHz is 50~K in the ReMoCA survey, while the observed 
line has a peak of $\sim$13~K only. A fit of the ReMoCA line at 102.864~GHz 
without accounting for the background continuum emission yields a column 
density a factor of 1.5 lower. Because the continuum emission is even more 
opaque toward the peak of Sgr~B2(N1), observations targeting this peak with a 
larger beam such as the ones performed by \citet{Remijan14} will collect more 
dust-attenuated urea emission and thereby underestimate the true strength of 
the urea lines even further. 

Combined together, these four reasons may explain the discrepancy between
the column density of urea derived from the ReMoCA survey and the 
column density claimed by \citet{Remijan14} for this molecule.

\begin{figure}
\centerline{\resizebox{1.0\hsize}{!}{\includegraphics[angle=0]{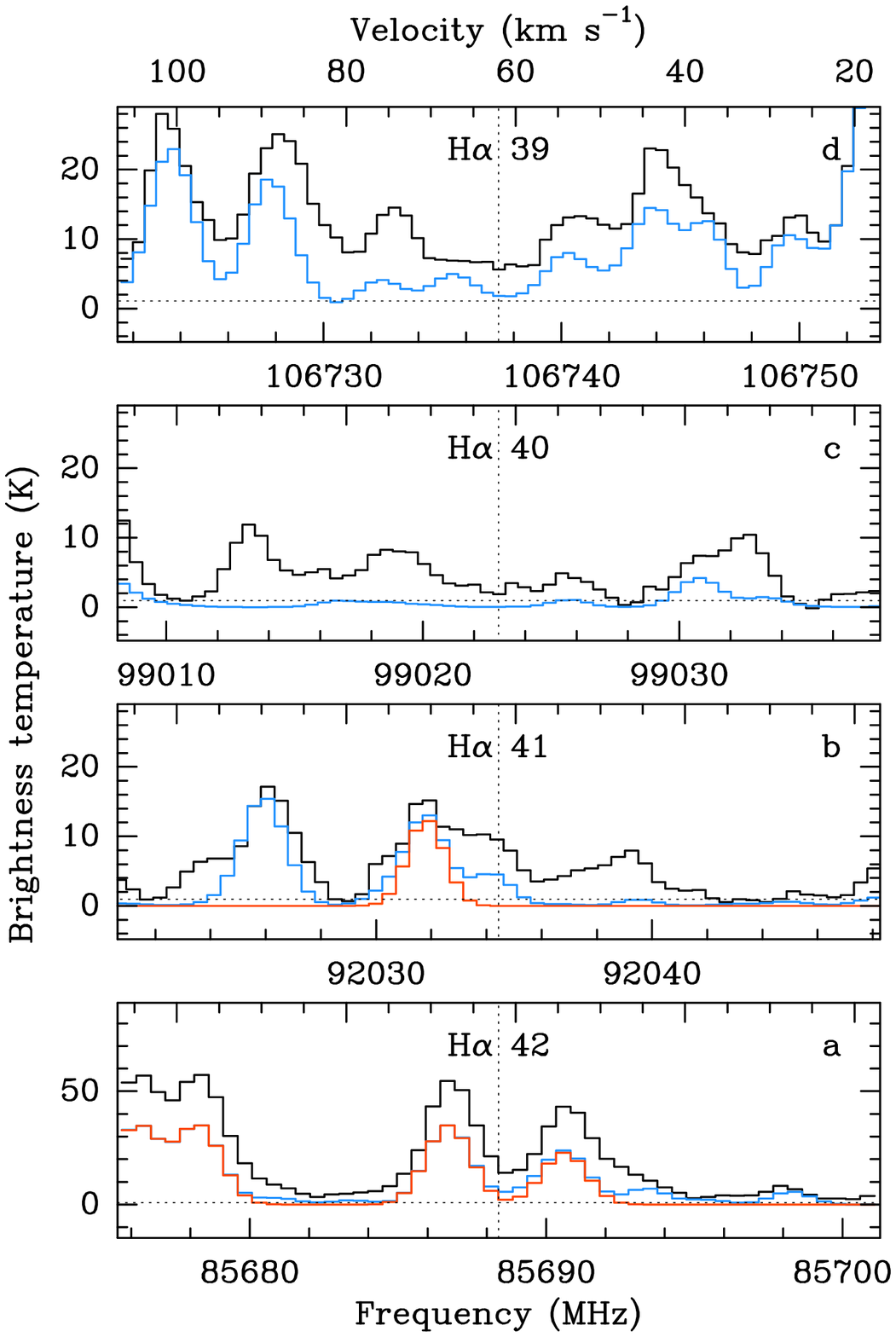}}}
\caption{Portions of the continuum-removed ReMoCA spectrum of 
Sgr~B2(N1S) that cover the frequencies of H$\alpha$ recombination lines. The
frequency of each H$\alpha$ line is marked with a dotted line in each panel
and the name of the transition is indicated. The lower and upper axes 
of each panel are labeled in frequency and velocity, respectively, assuming a 
systemic velocity of 62~km~s$^{-1}$. The velocity range is the same for all 
panels. The blue spectrum represents our current (preliminary) LTE model that 
includes contributions from all identified molecules. It is overlaid 
on the ALMA spectrum displayed in black. The red spectrum in panel \textbf{a} 
is our current LTE synthetic spectrum of C$_2$H$_3$CN, $\varv_{11}=1$ while the 
red spectrum in panel \textbf{b} is our best-fit LTE synthetic spectrum of 
urea.}
\label{f:halpha}
\end{figure}

A point of concern could be the potential contamination of the line
we assigned to urea at 92031.8 MHz by the H$\alpha$~41 recombination line at 
92034.4 MHz, as briefly mentioned by \citet{Remijan14}. This recombination 
line is prominent in the IRAM 30\,m spectrum of Sgr B2(N1) at a poor angular 
resolution of $\sim$27$\arcsec$ \citep[see Fig. 2 of][]{Belloche13}. 
Figure~\ref{f:halpha} shows the portions of the ReMoCA spectrum of Sgr~B2(N1S) 
that cover the frequencies of all H$\alpha$ recombination lines that fall 
in the ReMoCA band. The synthetic spectrum of C$_2$H$_3$CN $\varv_{11}=1$,
which dominates the emission detected around the frequency of the H$\alpha$ 42 
recombination line, is displayed in red in Fig.~\ref{f:halpha}a. Our 
preliminary model obviously underestimates the strength of the four 
C$_2$H$_3$CN lines in this portion of the ReMoCA survey. If we rescale it to 
fit the observed line strengths then there is no room left for H$\alpha$ 42 
emission. Likewise, we do not detect any obvious H$\alpha$ 40 emission in 
Fig.~\ref{f:halpha}c, with no signal detected above 2~K at the rest frequency 
of this transition. The four spectra of Fig.~\ref{f:halpha} are displayed with 
the same velocity axis. Therefore, any H$\alpha$ emission present toward 
Sgr~B2(N1S) should manifest itself with the same lineshape in each panel. No 
such pattern is obvious in this figure. We can thus safely conclude that at 
the high angular resolution of the ReMoCA survey and toward the specific 
position Sgr~B2(N1S), the ReMoCA spectrum is not significantly contaminated by 
H$\alpha$ emission. We conclude that the assignment of the transition at 
92031.8~MHz to urea is robust and certainly not affected by any contamination 
from H$\alpha$ emission. The prominent H$\alpha$ emission seen in the 
single-dish spectrum around this frequency is either filtered out by the 
interferometer, or arises from regions that do not enclose Sgr~B2(N1S).

\subsection{Chemical modeling}

In general, the model with the medium warm-up timescale appears best to 
reproduce the observational relative abundances of the five molecules of 
interest, considering that the peak absolute abundance of urea varies little 
between the models. The large relative abundance of urea in all cases suggests 
that the production mechanism for this molecule may be too efficient on the 
grains. Solid-phase 
urea is seen to be produced most strongly after around 55~K, which is the 
temperature at which production and diffusion of NH$_2$ becomes rapid. 
Production of NH$_2$ is caused by the abstraction of H from solid-phase 
ammonia by OH radicals \citep[see][]{Garrod13}, although the activation 
barriers and rates for such processes are poorly constrained. However, a 
somewhat 
higher barrier to the abstraction of H from NH$_3$ may be a possible solution 
to lower the total urea production. Although this might also reduce the 
production of acetamide to some degree, the latter molecule has other 
formation pathways, especially at lower temperatures, and thus would likely be 
less affected, whereas all significant urea production in the present models 
comes from the NH$_2$ + NH$_2$CO route. The interstellar detection of urea may 
therefore give further constraint to the chemistry of ammonia-related species.

The production of urea through UV-induced solid-phase chemistry was recently 
studied as part of an investigation into the formation of peptide-like 
molecules on interstellar grains \citep[][]{Ligterink18}. These authors found 
that UV irradiation of CH$_4$:HNCO ice mixtures could produce several of the 
species detected observationally in the present work, with the NH$_2$CO 
radical being a key intermediary, in agreement with our model findings.
\citeauthor{Ligterink18} suggested that, in their experiments, this radical 
should be produced 
by the reaction NH$_2$ + CO $\rightarrow$ NH$_2$CO. This reaction is included 
in our chemical network, with an activation energy barrier arbitrarily set to 
2500~K, in line with typical values for the reaction of H with CO 
\citep[][]{Garrod08}. In the models, the reaction is found to be negligible 
compared with both the abstraction of hydrogen from NH$_2$CHO and (at lower 
temperature) the addition of H to HNCO, although the adoption of a lower 
barrier could change this result. 

\citeauthor{Ligterink18} also state that the hydrogenation of CH$_3$NCO to 
N-methylformate is not consistent with their nondetection of the latter 
molecule (in spite of detecting the former). The models of \citet{Belloche17} 
suggested that either this process or the direct addition of radicals to form 
CH$_3$NHCHO are plausible to explain observations, although in the present 
models hydrogenation dominates. The toxicity of methyl isocyanate makes
experimentation with its pure ice a  challenging prospect, so 
unambiguous evidence of the efficacy or otherwise of this route is not yet 
forthcoming.

Although our models do not consider mechanisms for native gas-phase production 
of urea, a theoretical study by \citet{Jeanvoine19} recently considered the 
possible reaction between protonated hydroxylamine (NH$_2$OH$_2$$^+$) and 
formamide (NH$_2$CHO) to form protonated urea, which, if formed, could be 
converted to neutral urea through electronic recombination. Those authors 
found that the structure of the ionic product (with the correct mass) 
resulting from the suggested reaction would be unlikely ultimately to produce 
urea without significant structural re-arrangement; however, they suggest that 
the process may be more efficient if occurring in or upon an ice mantle. No 
astrochemical models currently exist that explicitly treat ion-molecule 
chemistry in the solid phase. 

As mentioned in Sect.~\ref{ss:chemistry}, in comparison to related chemical 
species, the peak abundance of urea obtained with the models presented here is 
somewhat higher than the observations would seem to support. However, one 
further destruction mechanism in the gas phase that is not currently included 
in the chemical network might ameliorate this disagreement. The efficiency of 
electronic dissociative recombination of protonated molecules in producing the 
neutral molecule plus a hydrogen atom is generally found only to be on the 
order of a few percent for the larger highly-saturated complex organic 
molecules for which experiments have been conducted 
\citep[e.g.,][]{Geppert06,Hamberg10}. This means that the gas-phase 
protonation of complex organics typically leads to the destruction of such 
molecules, thus limiting their lifetimes and fractional abundances. 
\citet{Taquet15} suggested that protonated complex organics may instead react 
efficiently with neutral ammonia, passing on the proton and leaving the 
structure of the larger molecule intact. This mechanism is not included in the 
network used here. However, the thermodynamic viability of the process is 
dependent on the neutral complex molecule having a lower proton affinity than 
ammonia. The most recent determinations of this quantity for urea 
\citep[207.8~kcal/mol;][]{Zheng02} and ammonia 
\citep[204.0~kcal/mol;][]{Hunter98} indicate that this mechanism would have a 
substantial endothermicity ($\sim$1900~K) for this pairing. Indeed, the 
reverse process would be energetically favorable, suggesting that neutral urea 
should instead be protonated by an encounter with NH$_4$$^+$, leading to an 
enhanced destruction rate over other complex organics. The proton affinity of 
acetamide \citep[206.4~kcal/mol;][]{Hunter98} suggests that it too 
should behave in this way. 
The rates of neither the forward nor the reverse process are well studied in 
general, and the relatively small difference in proton affinity between urea 
and ammonia (compared with other possible pairings) means that the efficiency 
of proton transfer from the latter molecule to the former should perhaps be 
determined using more detailed methods than the crude thermodynamic argument 
above. However, even the simple absence of a proton-transfer mechanism between 
protonated urea and ammonia (due to its endothermicity), without the inclusion 
of the reverse mechanism, may be sufficient to decrease the relative abundance 
of urea found in the models versus other complex organics, if the protonated 
forms of those species suffer no such difficulty in reacting with ammonia.
This would also make such abundances strongly dependent on the gas-phase 
abundance of ammonia in individual sources. A greater abundance of gas-phase 
ammonia (and thus of protonated ammonia) toward Sgr~B2(N2), versus 
Sgr~B2(N1S), could be one explanation for the lower abundance of urea toward
N2.

\citet{Raunier04} claimed a tentative identification of urea and formamide in
interstellar ices toward the protostellar source NGC~7538 IRS9 (see 
Sect.~\ref{s:introduction}). Unfortunately, they do not make any quantitative 
statement about the relative abundance of these two molecules in that source. 
Our best-fit chemical model for Sgr~B2(N1S) (Fig.~\ref{f:model}b) suggests 
that the abundance of formamide with respect to urea in the ices should be 
even higher than in the gas phase so, given the abundance ratio we measure in 
the gas phase toward Sgr~B2(N1S), we expect urea to be at least two orders of 
magnitude less abundant than  formamide in the ice mantles. It would be 
interesting to test whether the tentative identifications claimed by 
\citet{Raunier04} in the ices are consistent with such a low abundance of 
urea relative to formamide.

\section{Conclusions}
\label{s:conclusions}

The high angular resolution of the ReMoCA spectral line survey recently 
conducted with ALMA toward Sgr~B2(N) allows us to beat the spectral confusion 
limit that our previous survey, EMoCA, reached toward the main hot core 
embedded in Sgr~B2(N). One difficulty that arises, though, is that, even at 
3~mm, the dust emission starts to be optically thick on these small scales
($\sim 4000$~au), thus obscuring the molecular line emission and preventing us 
from deriving the molecular composition of the very inner region. However, a 
careful inspection of the datacube has revealed that the molecular lines 
detected toward the position Sgr~B2(N1S), which is offset from the continuum 
peak by $\sim 8000$~au and therefore suffers less from the dust obscuration, 
have full widths at half maximum of only $\sim$5~km~s$^{-1}$. This is similar 
to the linewidths that characterize the spectrum of the secondary hot core 
Sgr~B2(N2). In the same way as EMoCA allowed us to make significant progress 
in the determination of the molecular composition of Sgr~B2(N2), ReMoCA now 
allows us to detect less abundant species toward the main hot core of 
Sgr~B2(N). We took this opportunity to search for urea toward this source. The 
main results of this article are the following:

\begin{enumerate}
\item The rotational spectrum of urea was revisited and the ground as well as 
three low-lying excited vibrational states were analyzed. The results of 
this spectroscopic study, which will be reported separately, were instrumental 
for the present search for urea in space.
\item We report the identification of urea toward Sgr~B2(N1S) with nine 
clearly detected lines in its vibrational ground state and first vibrationally
excited state. This is the first secure detection of urea in the ISM.
\item We confirm the interstellar detection of N-methylformamide, which was
recently tentatively identified on the basis of the EMoCA survey. Sixty 
lines of this molecule are clearly detected toward Sgr~B2(N1S) in the ReMoCA
data.
\item We report the first interstellar detections of NH$_2$CH$^{18}$O and 
$^{15}$NH$_2$CHO.
\item We find that urea is about two orders of magnitude less 
abundant than formamide toward Sgr~B2(N1S), and about one order of
magnitude less abundant 
than N-methylformamide, methyl isocyanate, and acetamide, which have similar 
column densities.
\item Urea is not detected toward the secondary hot core Sgr~B2(N2). The 
nondetection implies that urea is at least one order of magnitude less 
abundant in Sgr~B2(N2) than in Sgr~B2(N1S) with respect to the other four 
species.
\item Our chemical modeling roughly reproduces the relative abundances of 
formamide, N-methylformamide, methyl isocyanate, and acetamide measured in 
Sgr~B2(N1S) and Sgr~B2(N2). However, it overproduces urea by at least one 
order of magnitude. An insufficiently high barrier to hydrogen abstraction 
from solid-phase ammonia by OH radicals may be the reason for this 
overproduction. The likely absence of an efficient gas-phase proton-transfer 
mechanism from protonated urea to ammonia could also result in more rapid 
destruction of urea than other complex organics.
\end{enumerate}
The differences in the chemical composition of Sgr~B2(N1S) and Sgr~B2(N2) with
urea being at least one order of magnitude less abundant with respect to the
other four species in the latter source is puzzling and not yet understood. A
careful comparison of the entire COM composition of both sources may help us to
find out the reasons for this astonishing difference.

\begin{acknowledgements}
We thank the referee, Anthony Remijan, for his critical review of our 
article that helped to clarify several aspects of our analysis.
This paper makes use of the following ALMA data: 
ADS/JAO.ALMA\#2016.1.00074.S. 
ALMA is a partnership of ESO (representing its member states), NSF (USA), and 
NINS (Japan), together with NRC (Canada), NSC and ASIAA (Taiwan), and KASI 
(Republic of Korea), in cooperation with the Republic of Chile. The Joint ALMA 
Observatory is operated by ESO, AUI/NRAO, and NAOJ. The interferometric data 
are available in the ALMA archive at https://almascience.eso.org/aq/.
Part of this work has been carried out within the Collaborative
Research Centre 956, sub-project B3, funded by the Deutsche
Forschungsgemeinschaft (DFG) -- project ID 184018867.
\end{acknowledgements}

\begin{appendix}
\label{appendix}
\section{Complementary figures: Spectra}
\label{a:spectra}

Figures~\ref{f:spec_nh2conh2_ve0}--\ref{f:spec_ch3conh2_cv} show the
transitions of NH$_2$C(O)NH$_2$, CH$_3$NHCHO, CH$_3$NCO, NH$_2$CHO, 
CH$_3$C(O)NH$_2$, and some of their isotopologs or vibrationally excited
states that are covered by the ReMoCA survey and contribute significantly to 
the signal detected toward Sgr~B2(N1S). 

\begin{figure*}
\centerline{\resizebox{0.82\hsize}{!}{\includegraphics[angle=0]{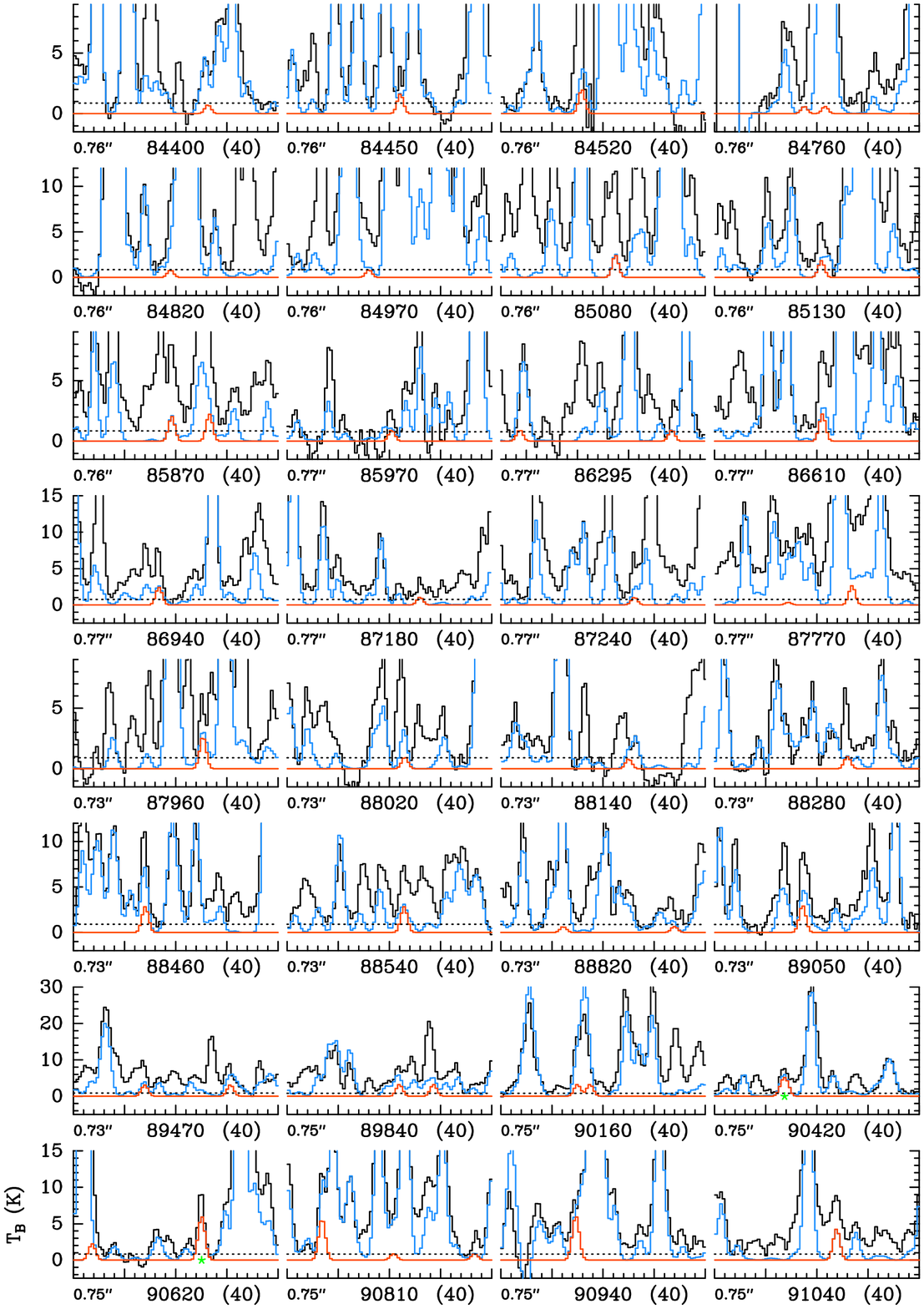}}}
\caption{Transitions of NH$_2$C(O)NH$_2$, $\varv = 0$ covered by our ALMA 
survey. The best-fit LTE synthetic spectrum of NH$_2$C(O)NH$_2$, $\varv = 0$ 
is displayed in red and overlaid on the observed spectrum of Sgr~B2(N1S) shown 
in black. The blue synthetic spectrum contains the contributions of all 
molecules identified in our survey so far, including the species shown in red. 
The central frequency and width are indicated in MHz below each panel. The 
angular resolution (HPBW) is also indicated. The y-axis is labeled in 
brightness temperature units (K). The dotted line indicates the $3\sigma$ 
noise level. The green stars mark the transitions that we consider as clearly 
detected.
}
\label{f:spec_nh2conh2_ve0}
\end{figure*}

\clearpage
\begin{figure*}
\addtocounter{figure}{-1}
\centerline{\resizebox{0.82\hsize}{!}{\includegraphics[angle=0]{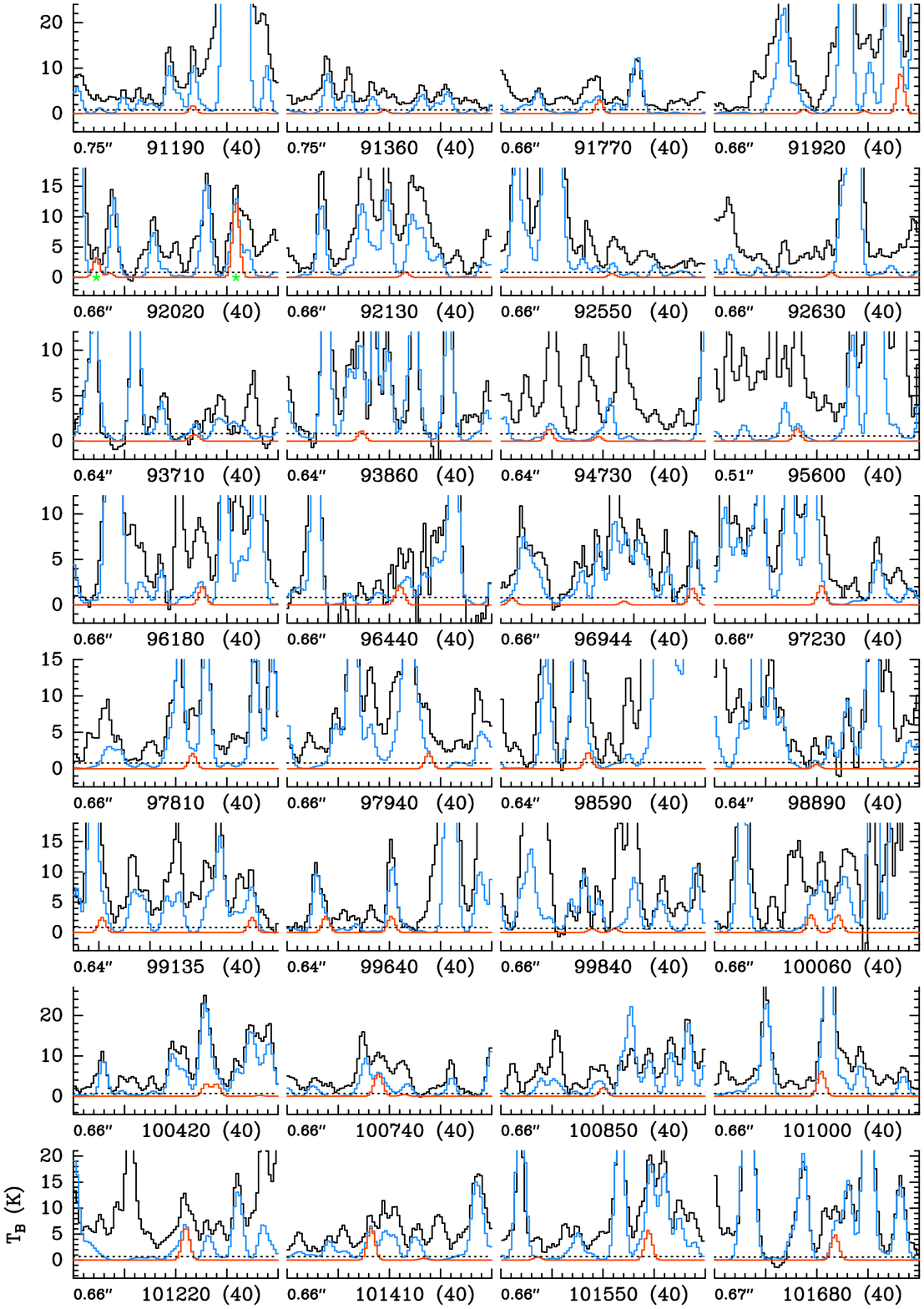}}}
\caption{continued.
}
\end{figure*}

\clearpage
\begin{figure*}
\addtocounter{figure}{-1}
\centerline{\resizebox{0.82\hsize}{!}{\includegraphics[angle=0]{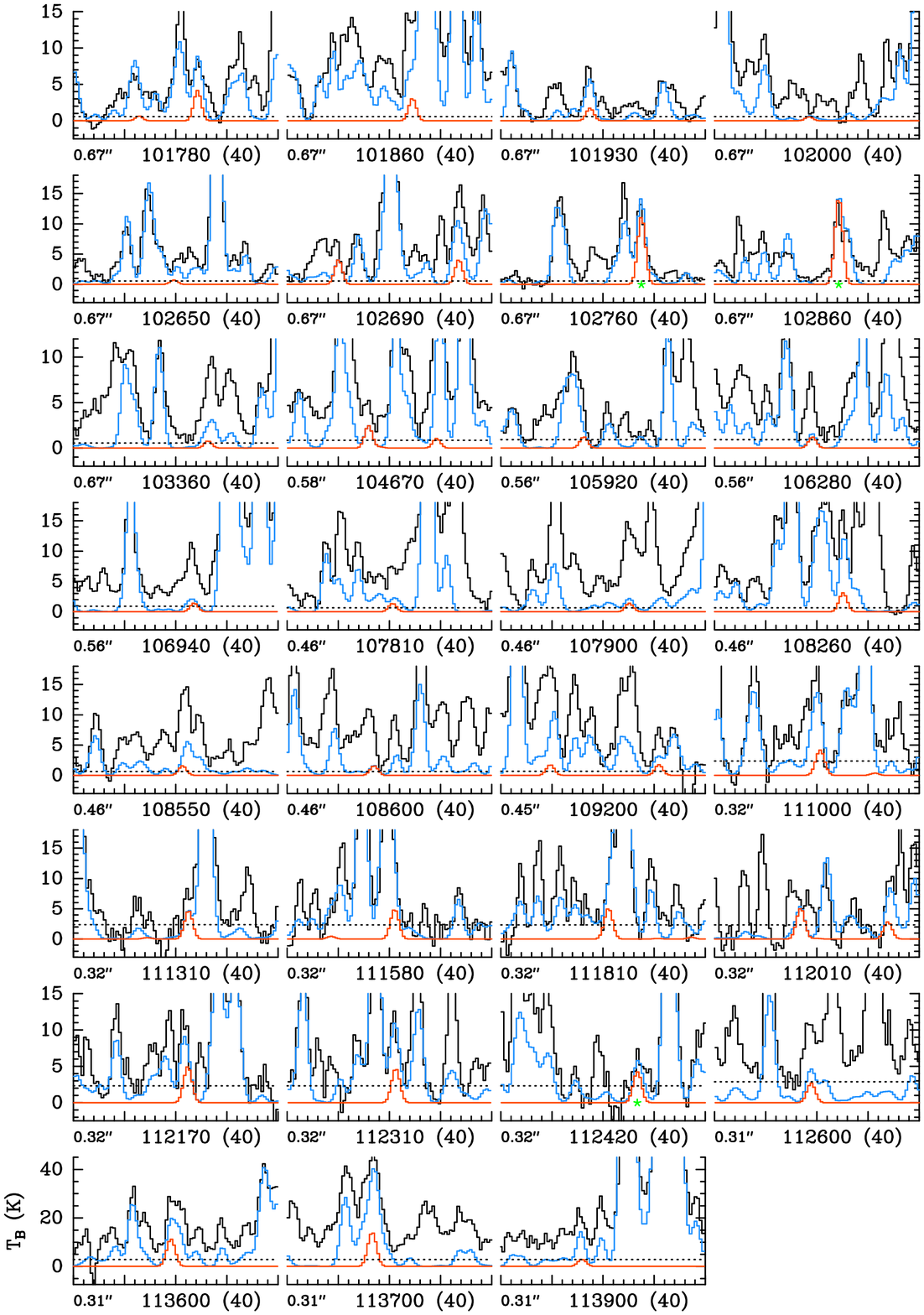}}}
\caption{continued.
}
\end{figure*}

\clearpage
\begin{figure*}
\centerline{\resizebox{0.82\hsize}{!}{\includegraphics[angle=0]{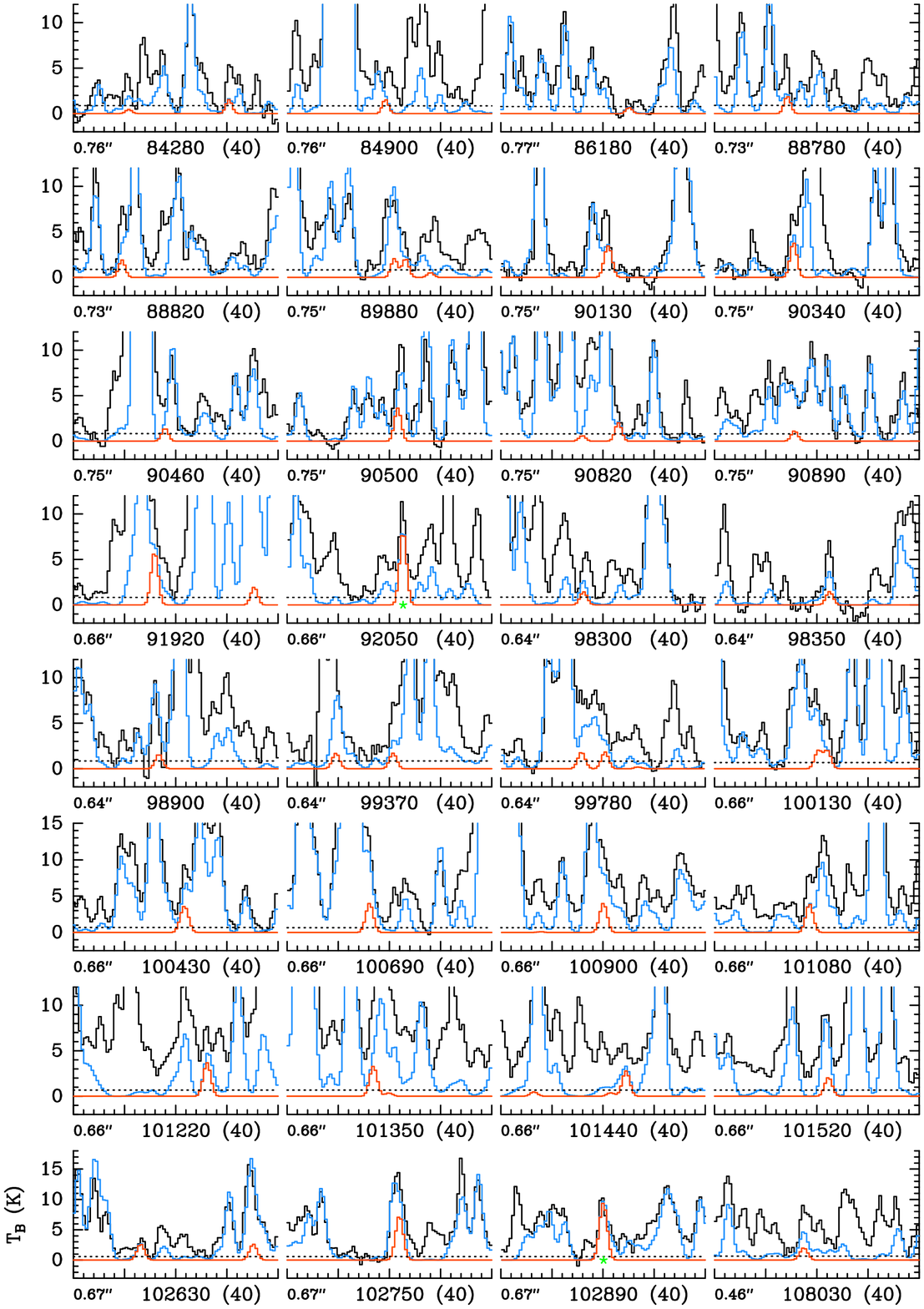}}}
\caption{Same as Fig.~\ref{f:spec_nh2conh2_ve0} for NH$_2$C(O)NH$_2$, 
$\varv = 1$.
}
\label{f:spec_nh2conh2_ve1}
\end{figure*}

\clearpage
\begin{figure*}
\addtocounter{figure}{-1}
\centerline{\resizebox{0.82\hsize}{!}{\includegraphics[angle=0]{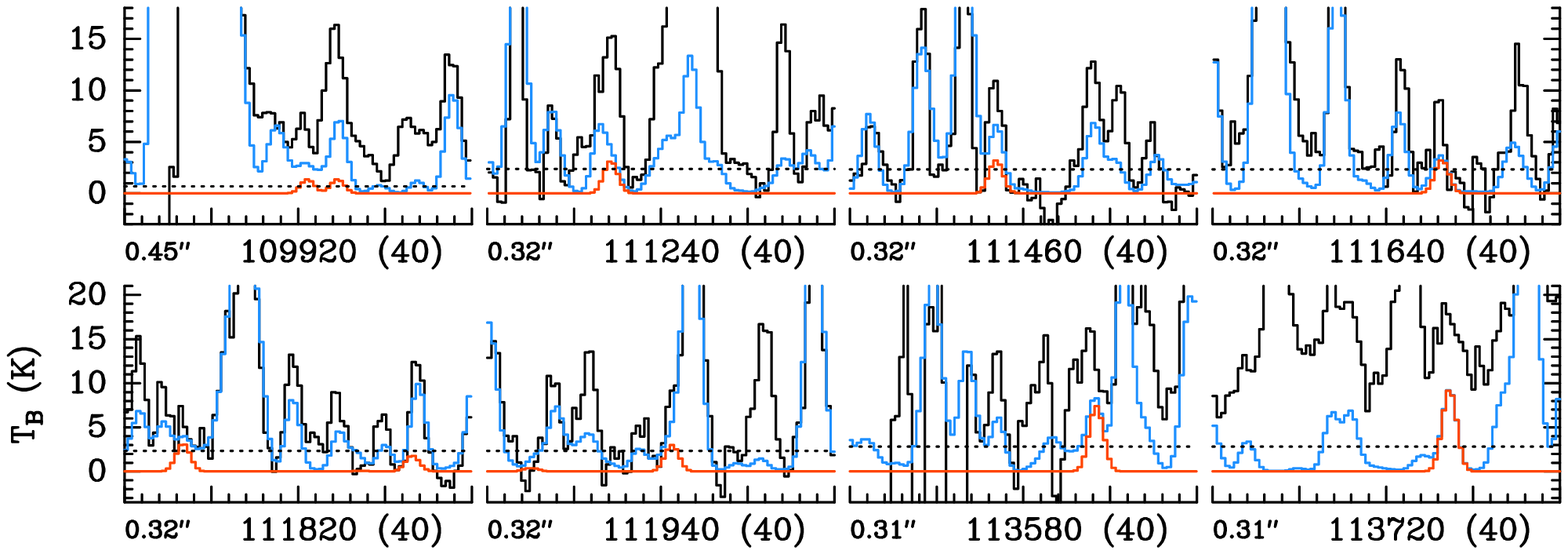}}}
\caption{continued.
}
\end{figure*}

\clearpage
\begin{figure*}
\centerline{\resizebox{0.82\hsize}{!}{\includegraphics[angle=0]{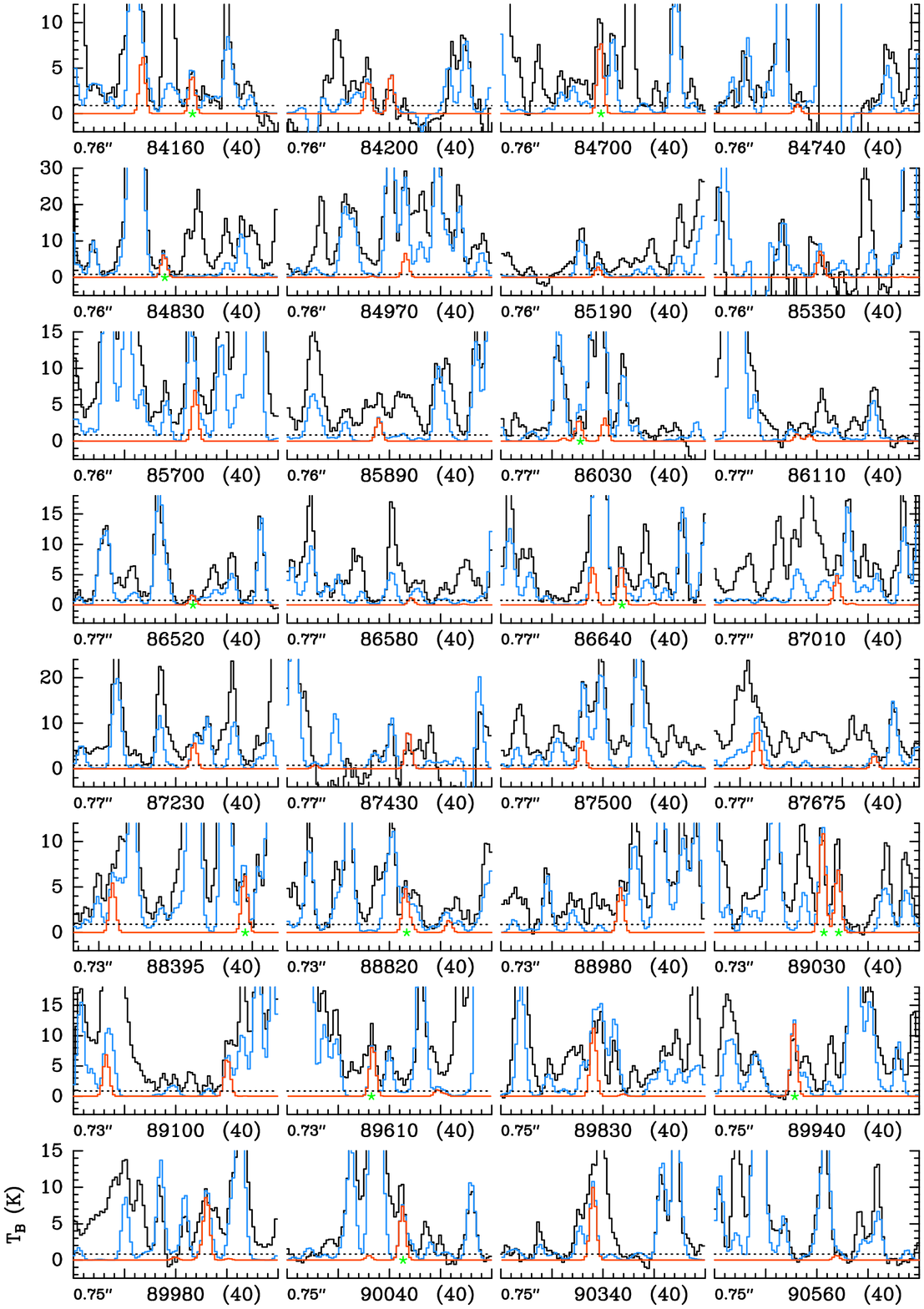}}}
\caption{Same as Fig.~\ref{f:spec_nh2conh2_ve0} for CH$_3$NHCHO, $\varv = 0$.
}
\label{f:spec_ch3nhcho_ve0}
\end{figure*}

\clearpage
\begin{figure*}
\addtocounter{figure}{-1}
\centerline{\resizebox{0.82\hsize}{!}{\includegraphics[angle=0]{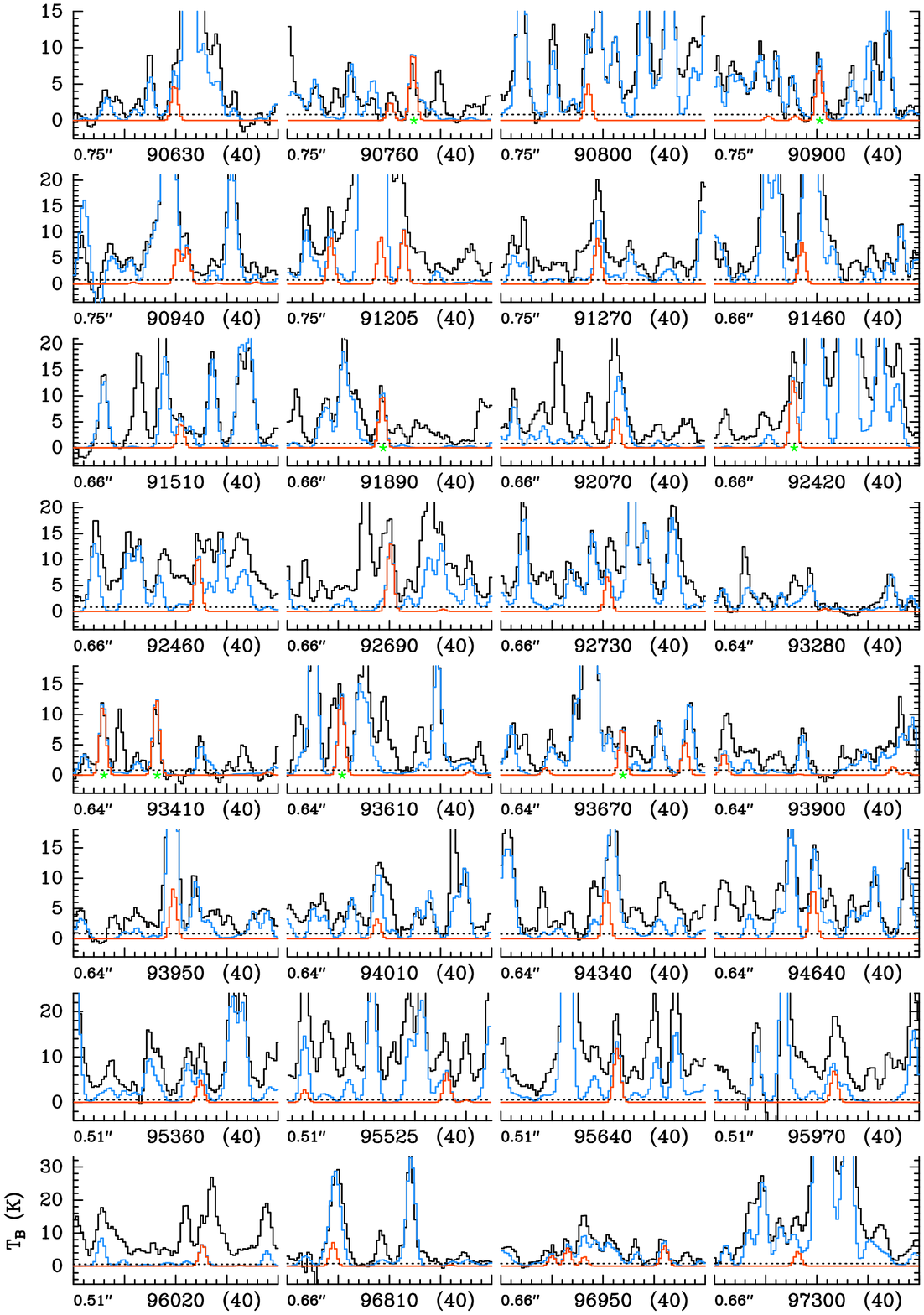}}}
\caption{continued.
}
\end{figure*}

\clearpage
\begin{figure*}
\addtocounter{figure}{-1}
\centerline{\resizebox{0.82\hsize}{!}{\includegraphics[angle=0]{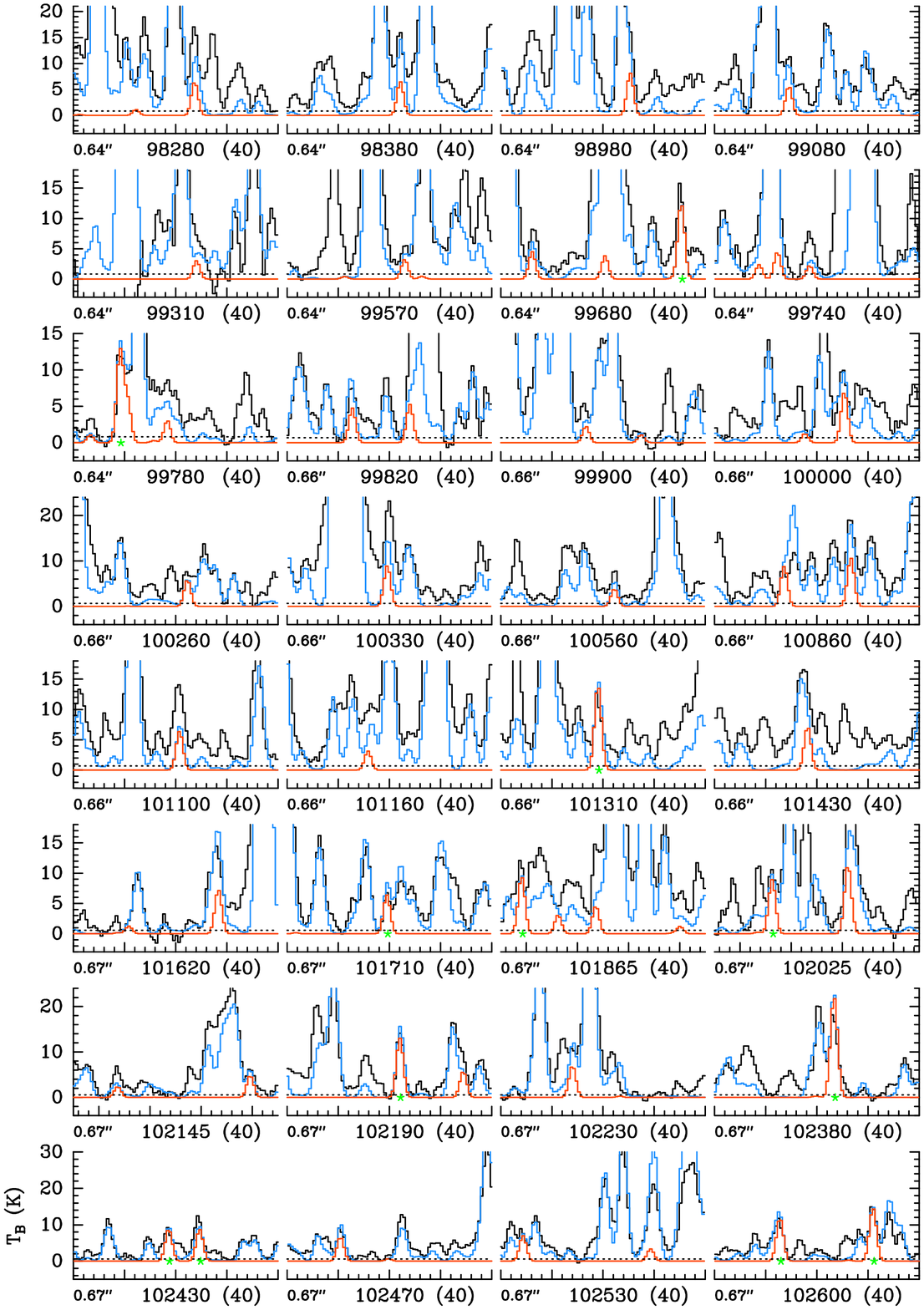}}}
\caption{continued.
}
\end{figure*}

\clearpage
\begin{figure*}
\addtocounter{figure}{-1}
\centerline{\resizebox{0.82\hsize}{!}{\includegraphics[angle=0]{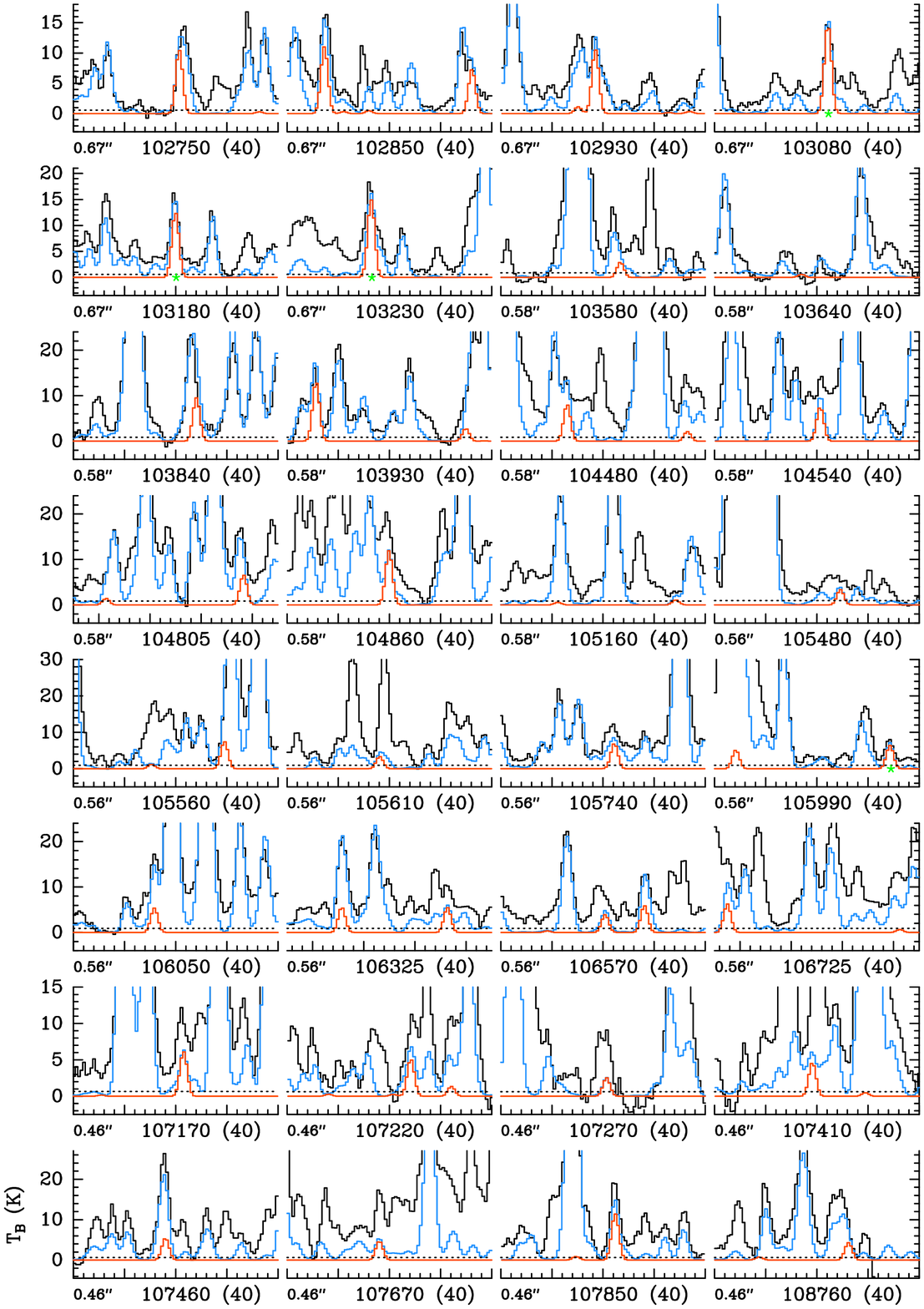}}}
\caption{continued.
}
\end{figure*}

\clearpage
\begin{figure*}
\addtocounter{figure}{-1}
\centerline{\resizebox{0.82\hsize}{!}{\includegraphics[angle=0]{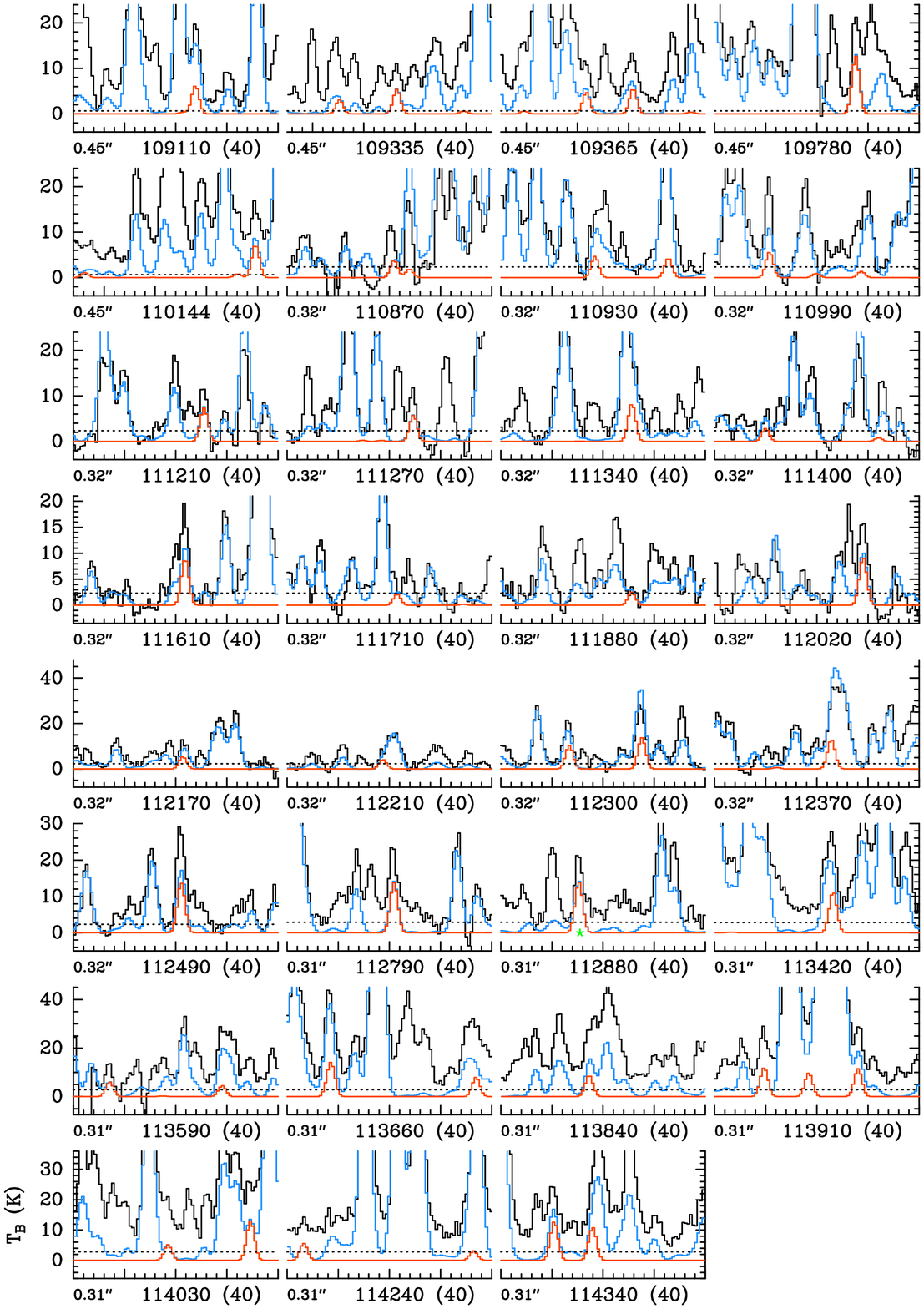}}}
\caption{continued.
}
\end{figure*}

\clearpage
\begin{figure*}
\centerline{\resizebox{0.82\hsize}{!}{\includegraphics[angle=0]{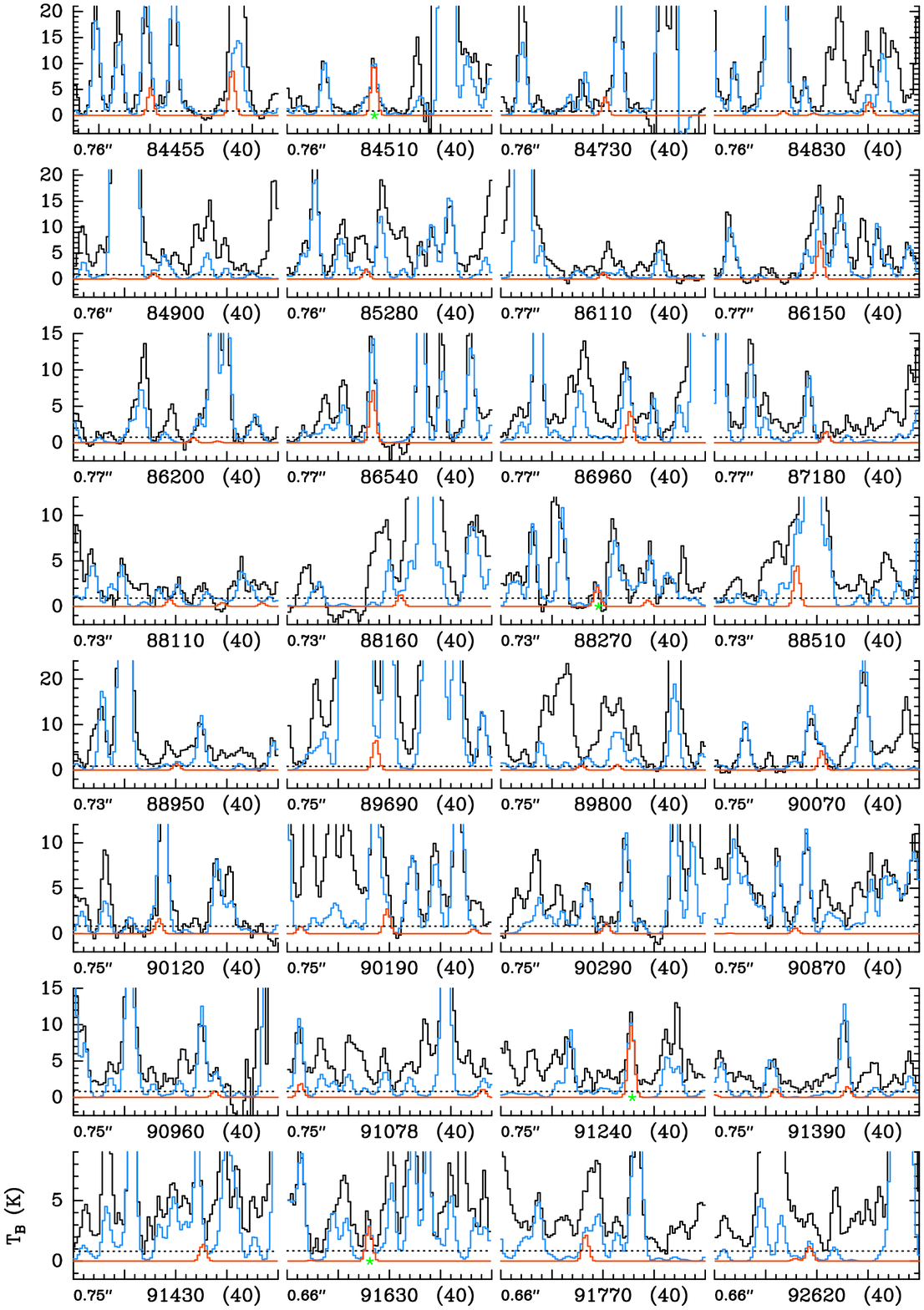}}}
\caption{Same as Fig.~\ref{f:spec_nh2conh2_ve0} for CH$_3$NHCHO, 
$\varv_{\rm t} = 1$.
}
\label{f:spec_ch3nhcho_ve1}
\end{figure*}

\clearpage
\begin{figure*}
\addtocounter{figure}{-1}
\centerline{\resizebox{0.82\hsize}{!}{\includegraphics[angle=0]{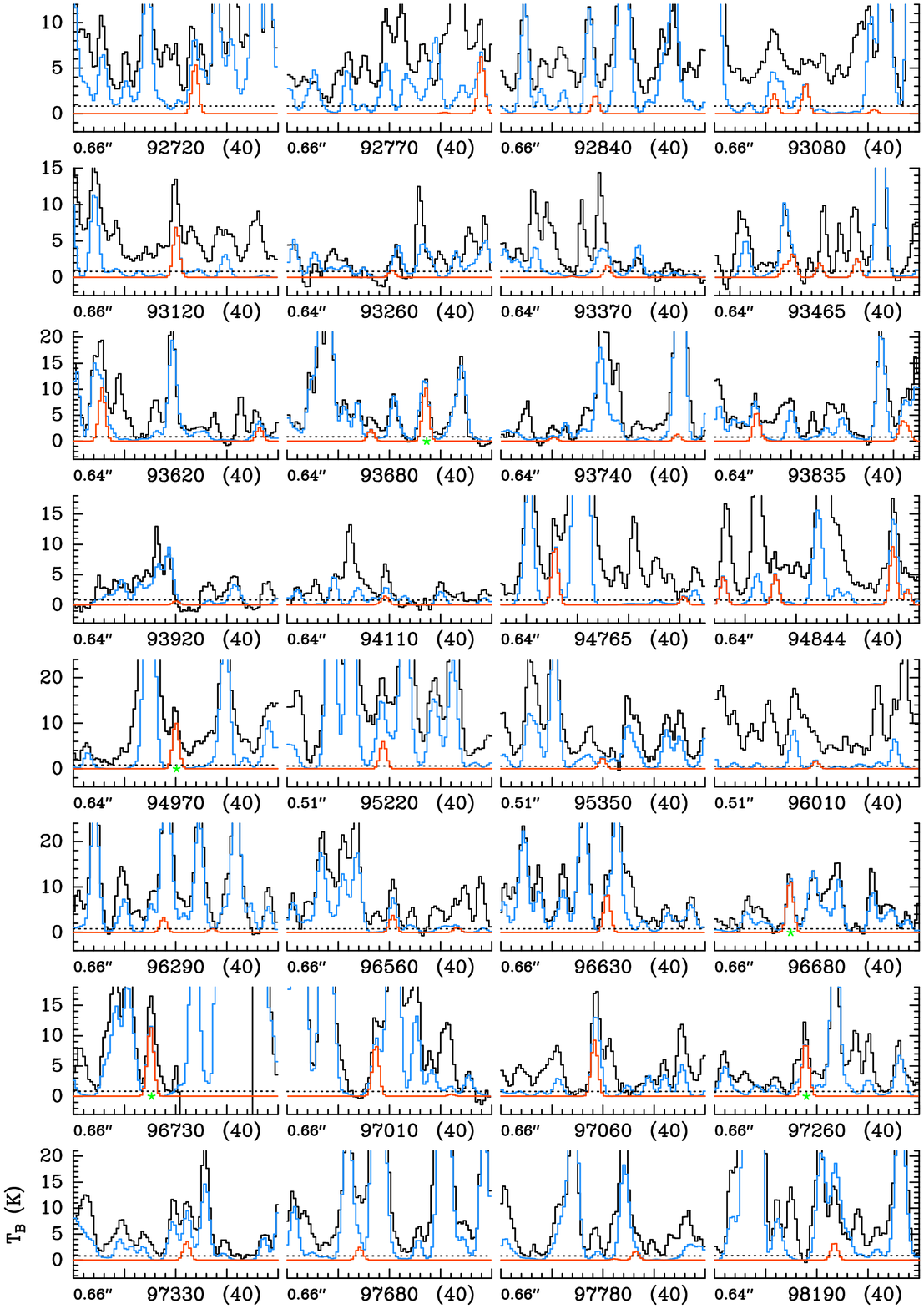}}}
\caption{continued.
}
\end{figure*}

\clearpage
\begin{figure*}
\addtocounter{figure}{-1}
\centerline{\resizebox{0.82\hsize}{!}{\includegraphics[angle=0]{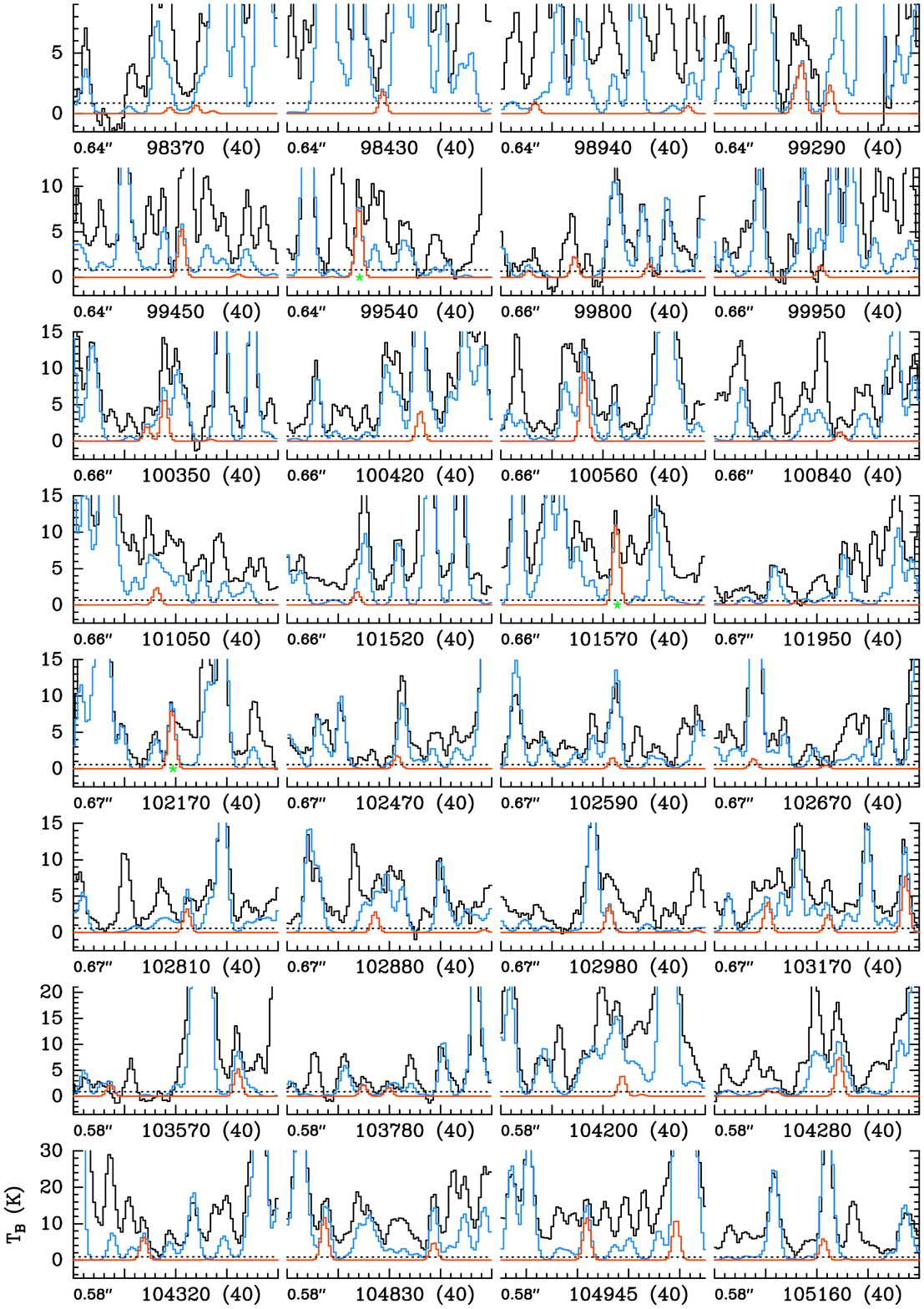}}}
\caption{continued.
}
\end{figure*}

\clearpage
\begin{figure*}
\addtocounter{figure}{-1}
\centerline{\resizebox{0.82\hsize}{!}{\includegraphics[angle=0]{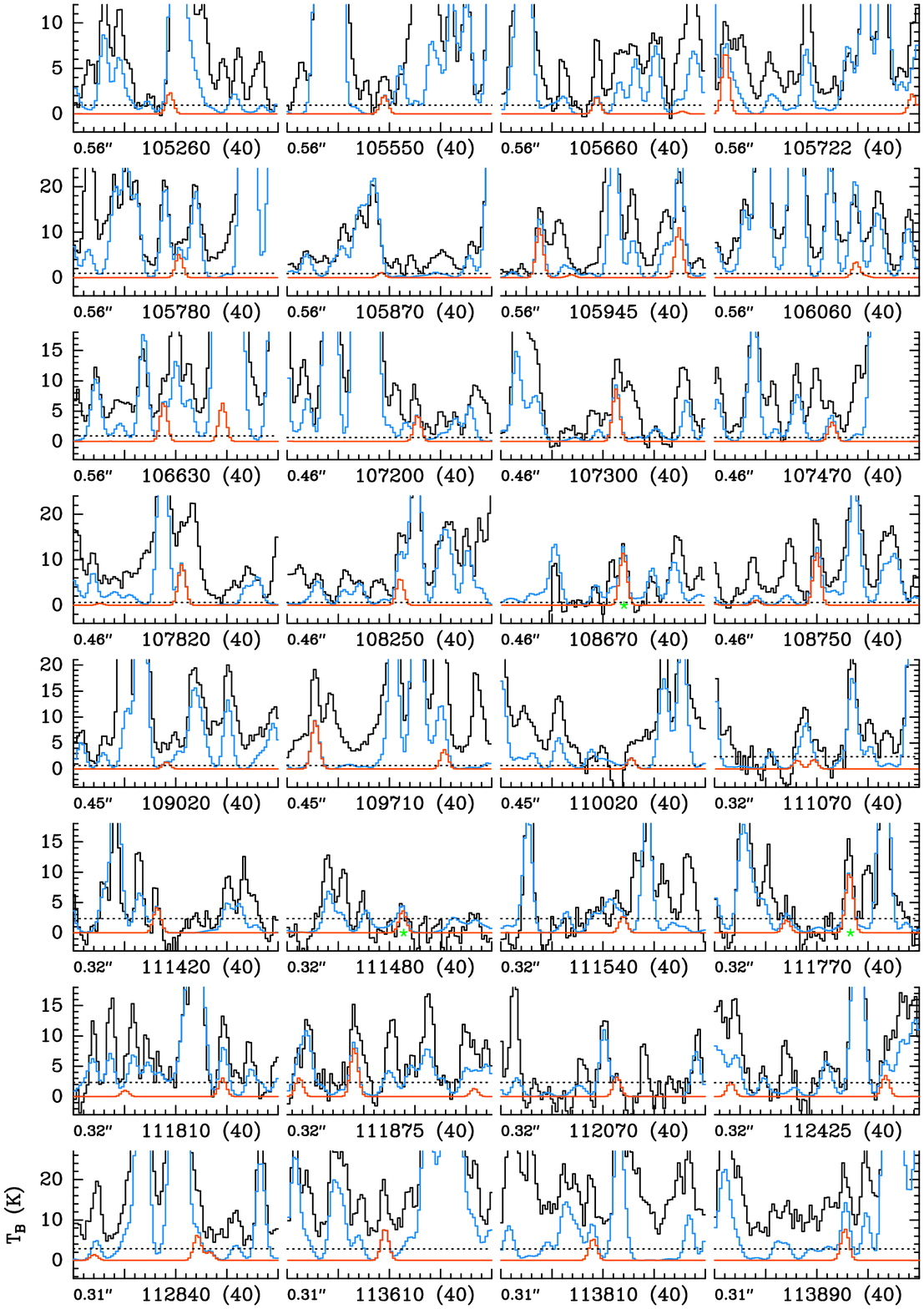}}}
\caption{continued.
}
\end{figure*}

\clearpage
\begin{figure*}
\centerline{\resizebox{0.82\hsize}{!}{\includegraphics[angle=0]{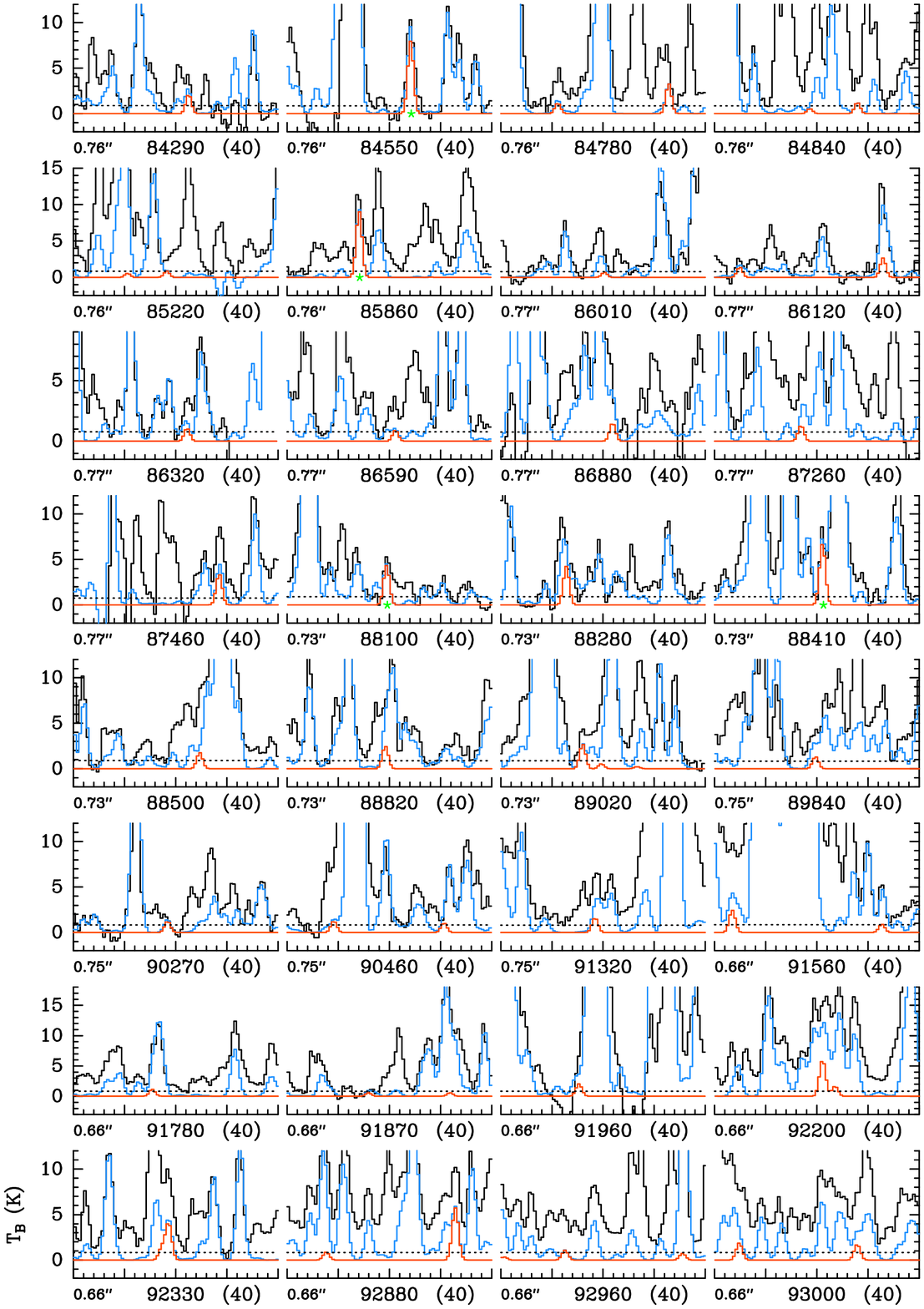}}}
\caption{Same as Fig.~\ref{f:spec_nh2conh2_ve0} for CH$_3$NHCHO, 
$\varv_{\rm t} = 2$.
}
\label{f:spec_ch3nhcho_ve2}
\end{figure*}

\clearpage
\begin{figure*}
\addtocounter{figure}{-1}
\centerline{\resizebox{0.82\hsize}{!}{\includegraphics[angle=0]{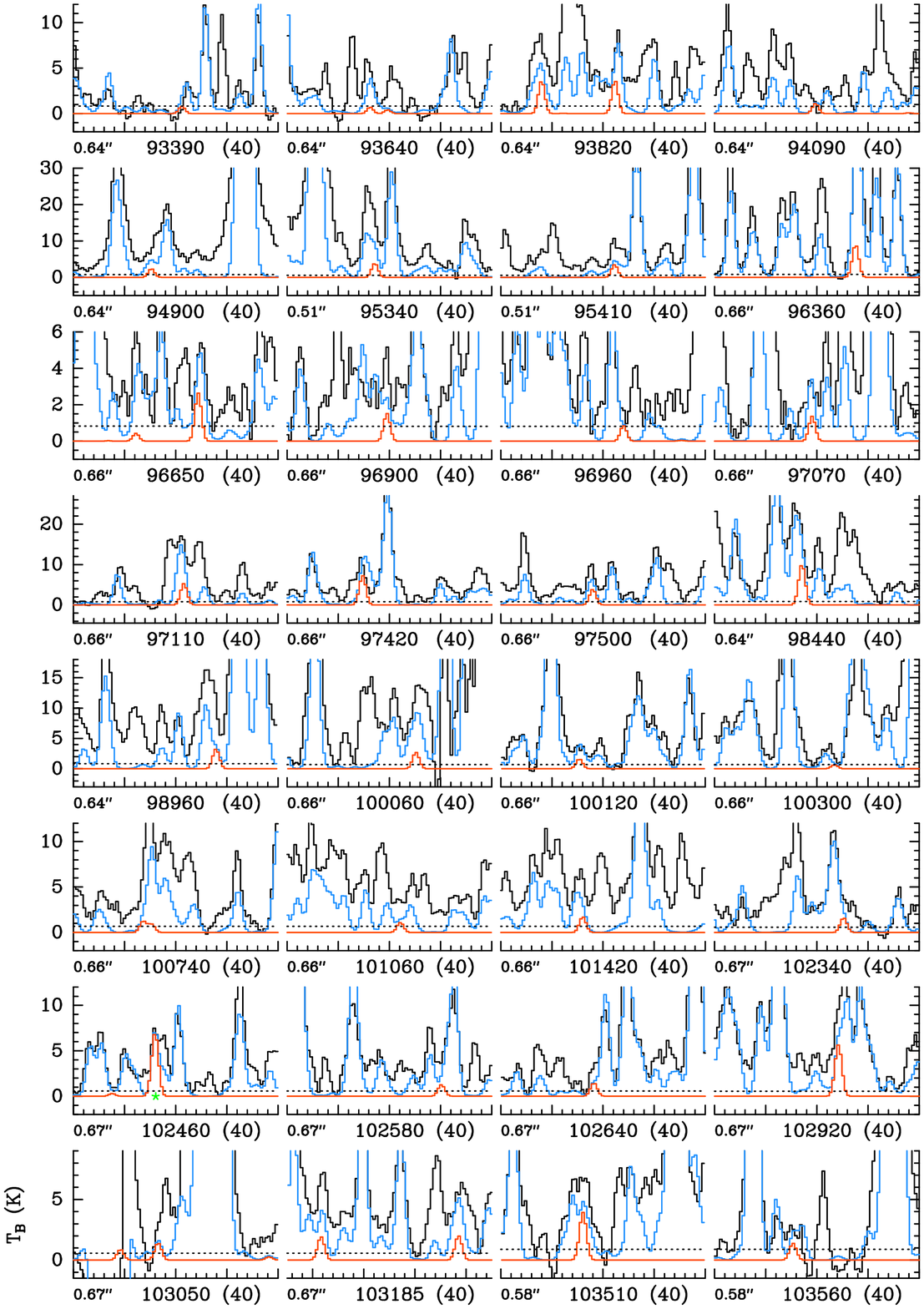}}}
\caption{continued.
}
\end{figure*}

\clearpage
\begin{figure*}
\addtocounter{figure}{-1}
\centerline{\resizebox{0.82\hsize}{!}{\includegraphics[angle=0]{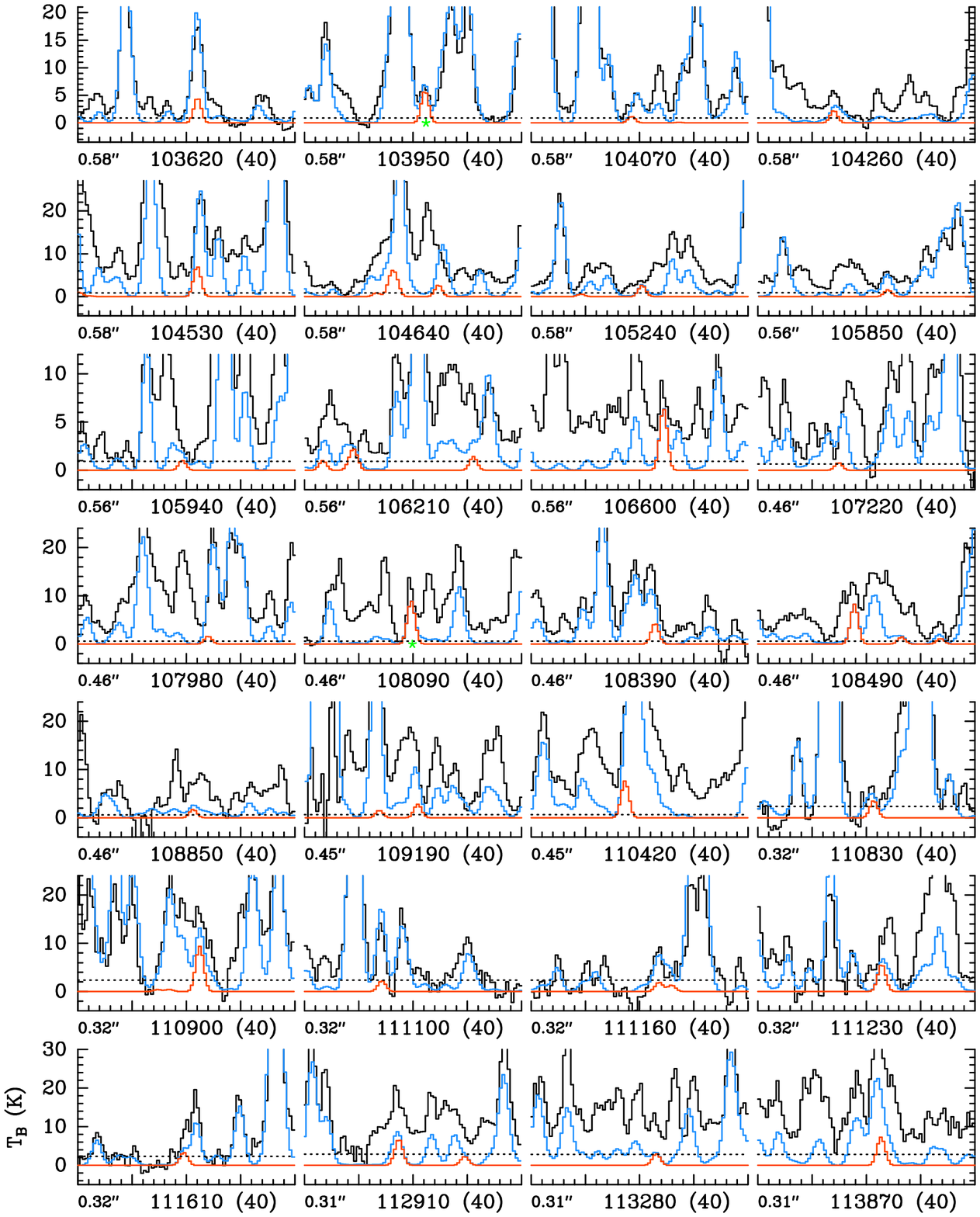}}}
\caption{continued.
}
\end{figure*}

\clearpage
\begin{figure*}
\centerline{\resizebox{0.82\hsize}{!}{\includegraphics[angle=0]{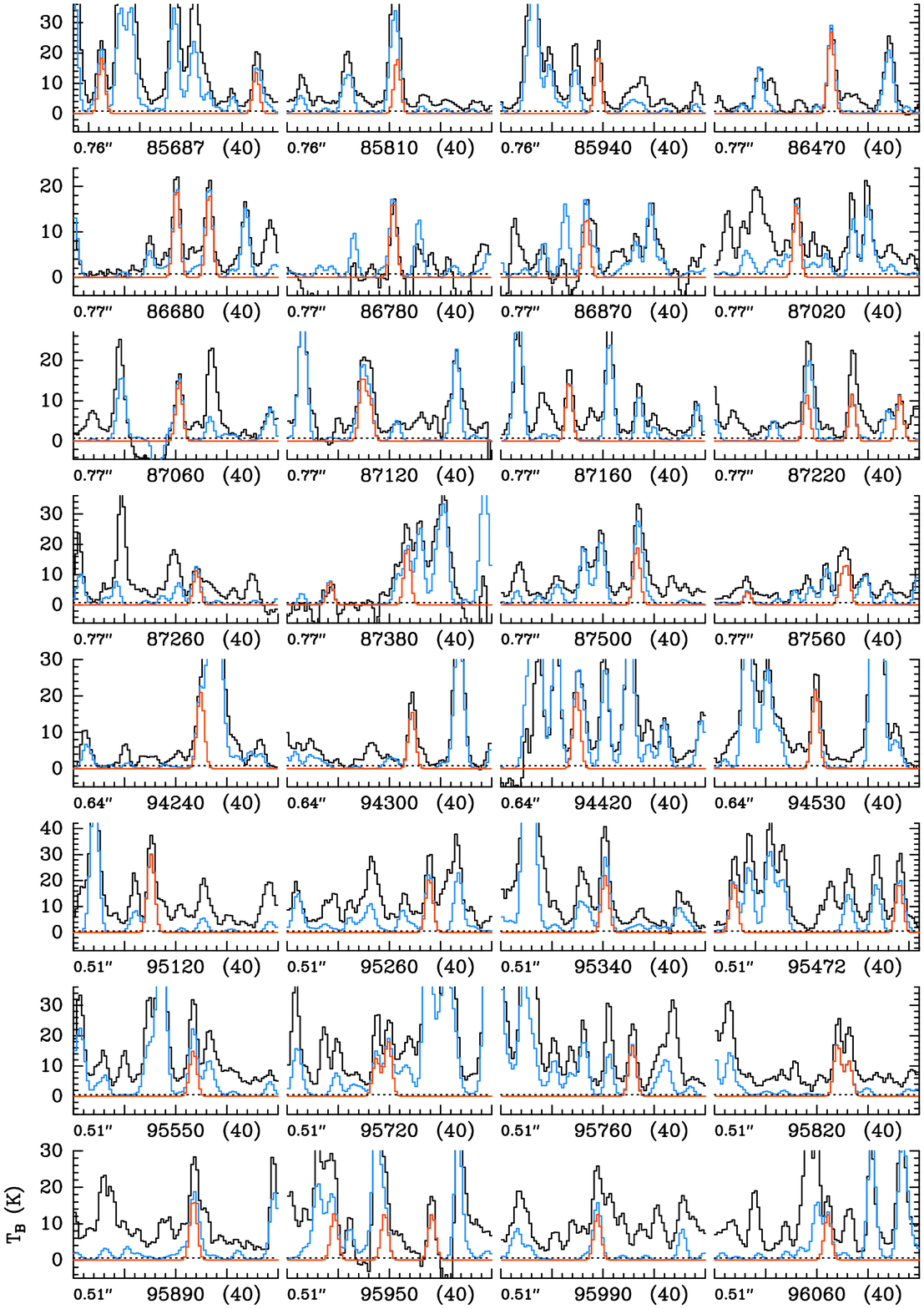}}}
\caption{Same as Fig.~\ref{f:spec_nh2conh2_ve0} for CH$_3$NCO, $\varv = 0$.
}
\label{f:spec_ch3nco_ve0}
\end{figure*}

\clearpage
\begin{figure*}
\addtocounter{figure}{-1}
\centerline{\resizebox{0.82\hsize}{!}{\includegraphics[angle=0]{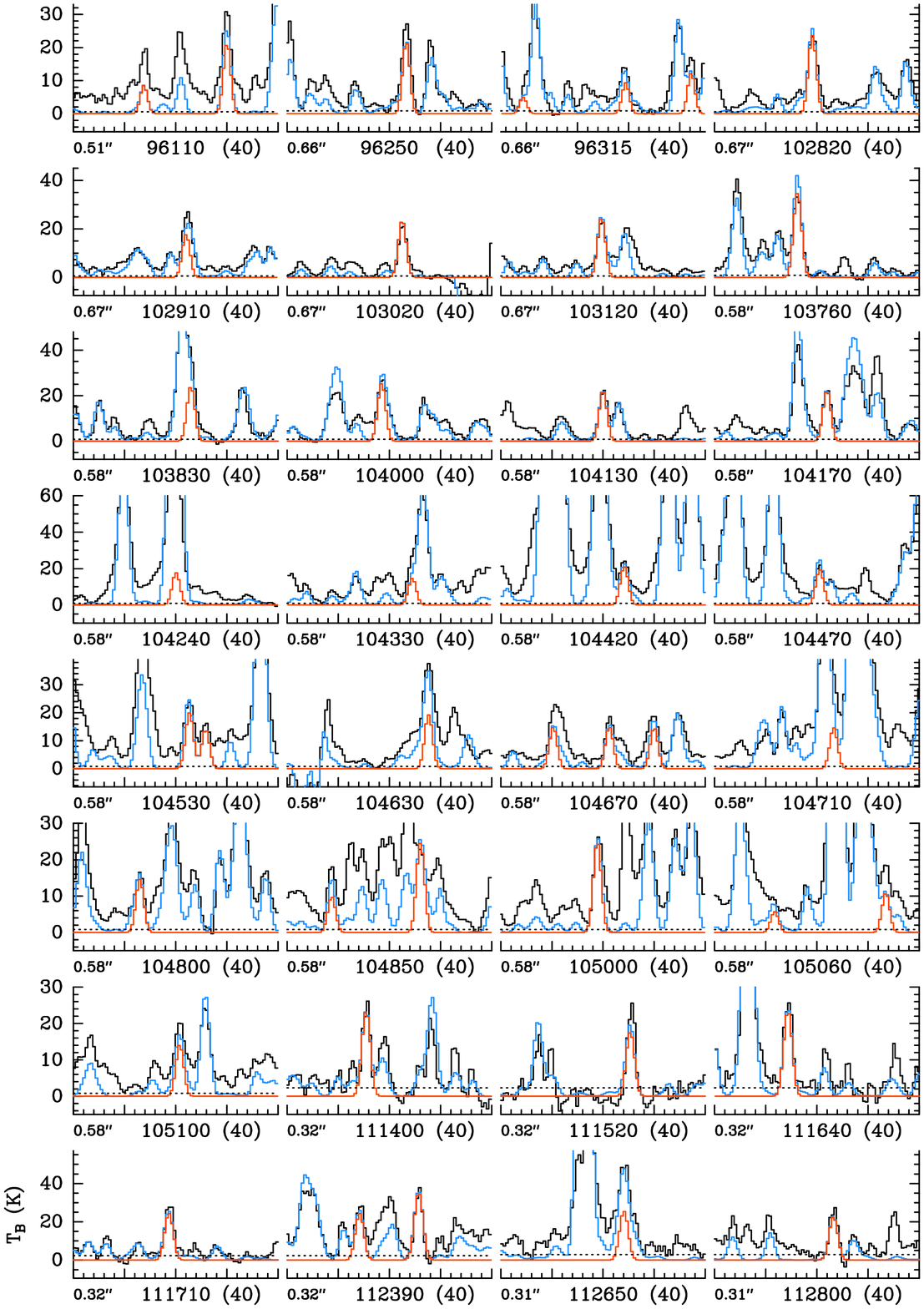}}}
\caption{continued.
}
\end{figure*}

\clearpage
\begin{figure*}
\addtocounter{figure}{-1}
\centerline{\resizebox{0.82\hsize}{!}{\includegraphics[angle=0]{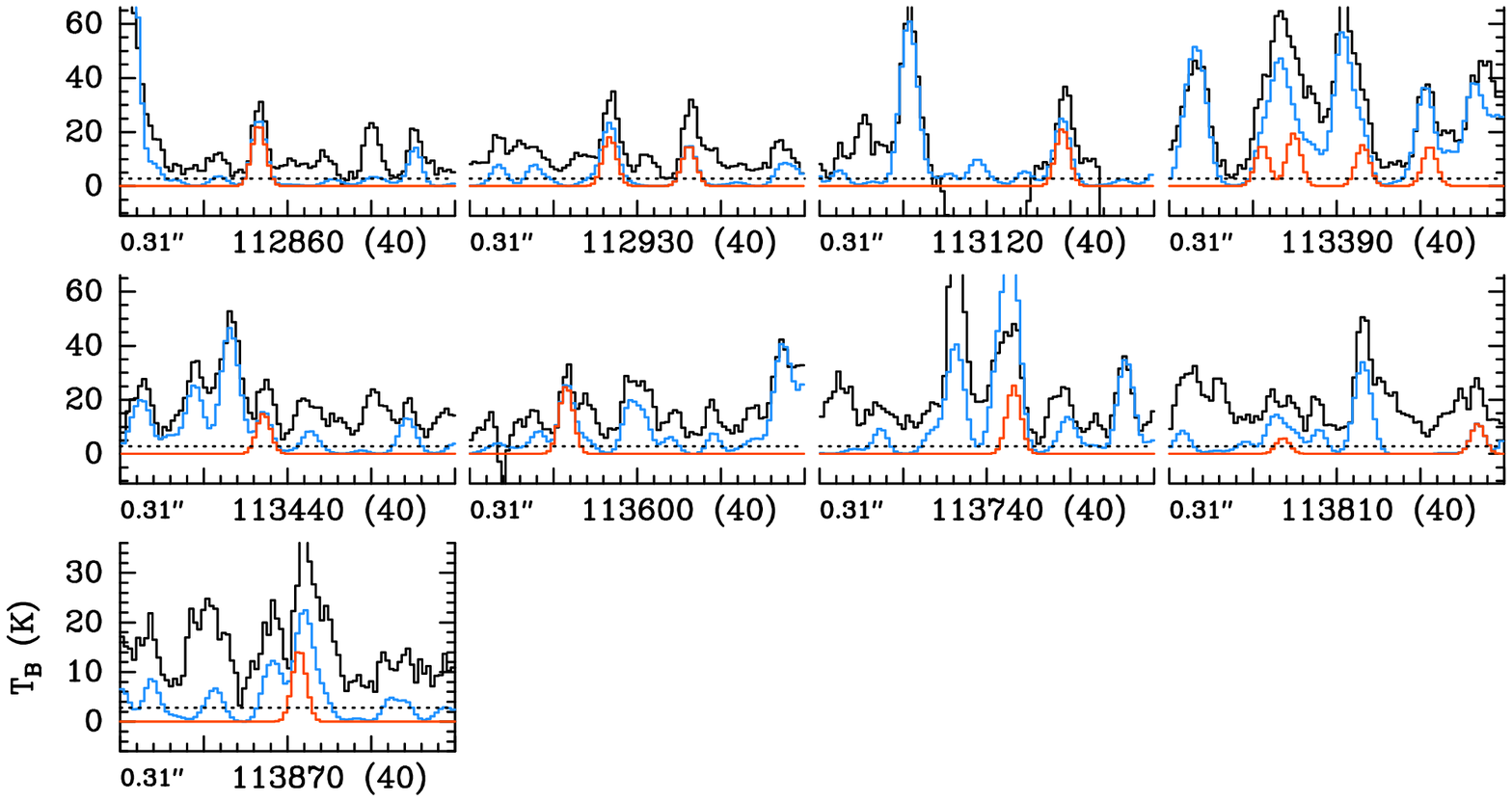}}}
\caption{continued.
}
\end{figure*}

\clearpage
\begin{figure*}
\centerline{\resizebox{0.82\hsize}{!}{\includegraphics[angle=0]{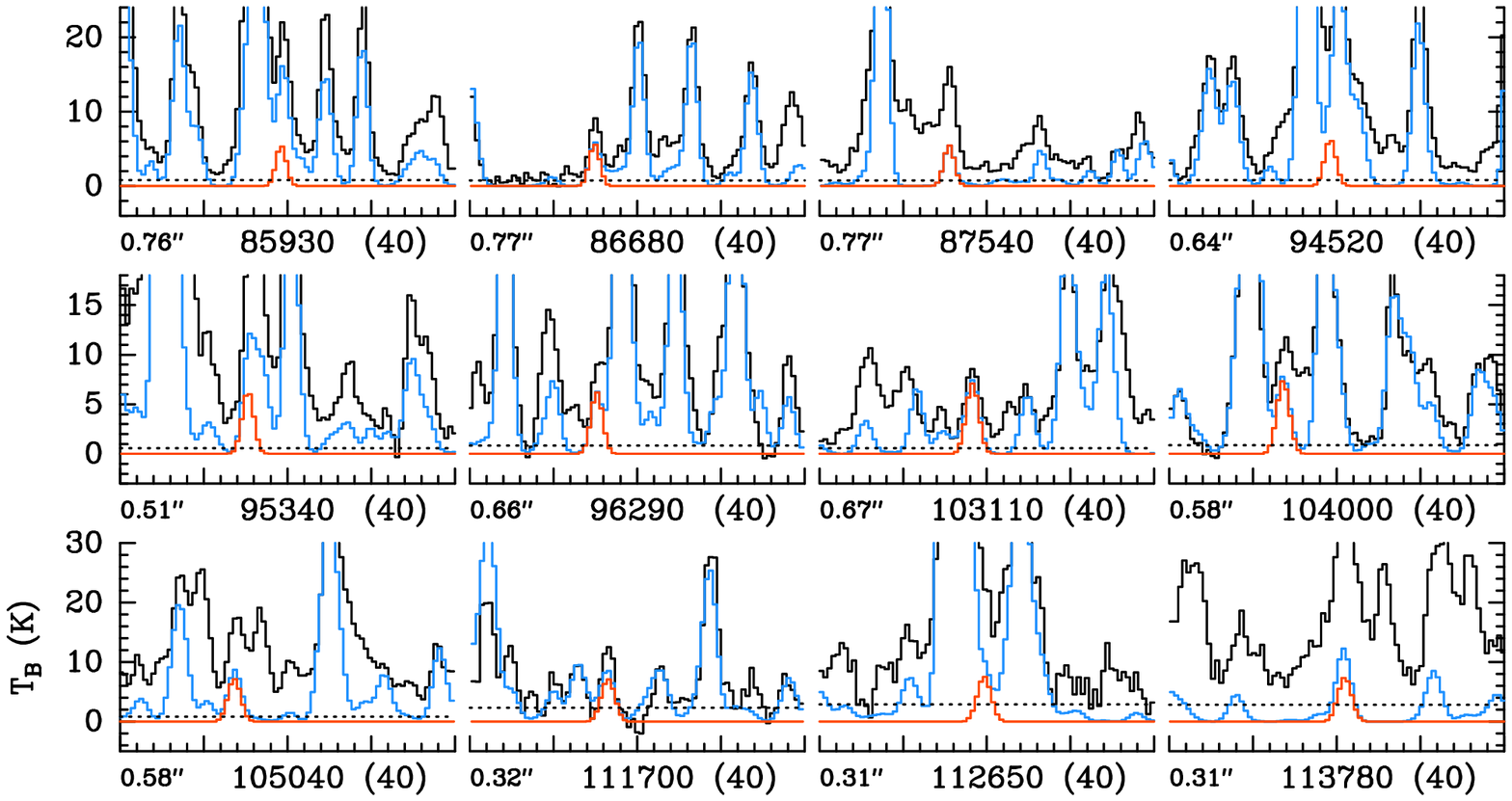}}}
\caption{Same as Fig.~\ref{f:spec_nh2conh2_ve0} for CH$_3$NCO, 
$\varv_{\rm b} = 1$.
}
\label{f:spec_ch3nco_ve1}
\end{figure*}

\clearpage
\begin{figure*}
\centerline{\resizebox{0.82\hsize}{!}{\includegraphics[angle=0]{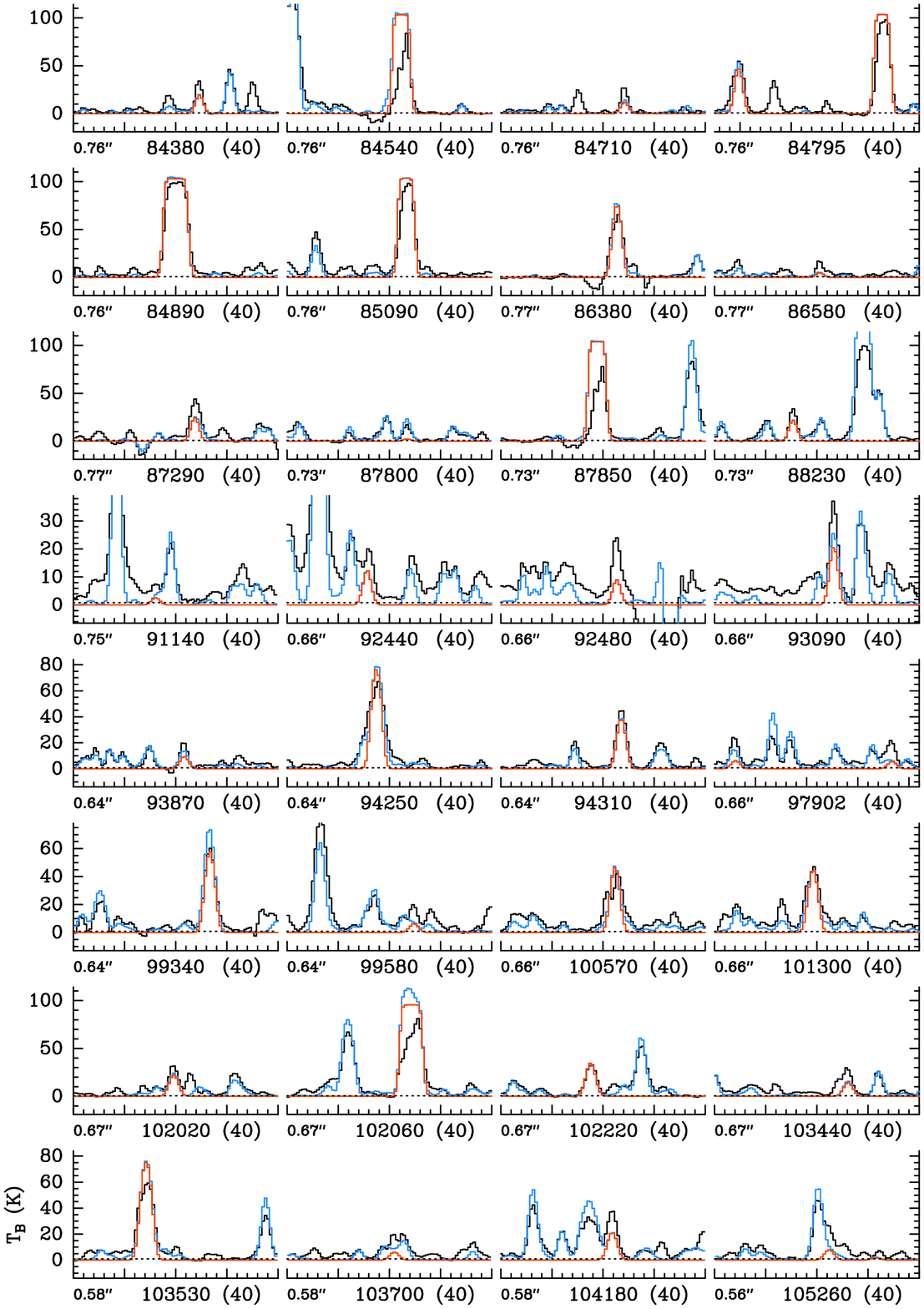}}}
\caption{Same as Fig.~\ref{f:spec_nh2conh2_ve0} for NH$_2$CHO, $\varv = 0$.
}
\label{f:spec_nh2cho_ve0}
\end{figure*}

\clearpage
\begin{figure*}
\addtocounter{figure}{-1}
\centerline{\resizebox{0.82\hsize}{!}{\includegraphics[angle=0]{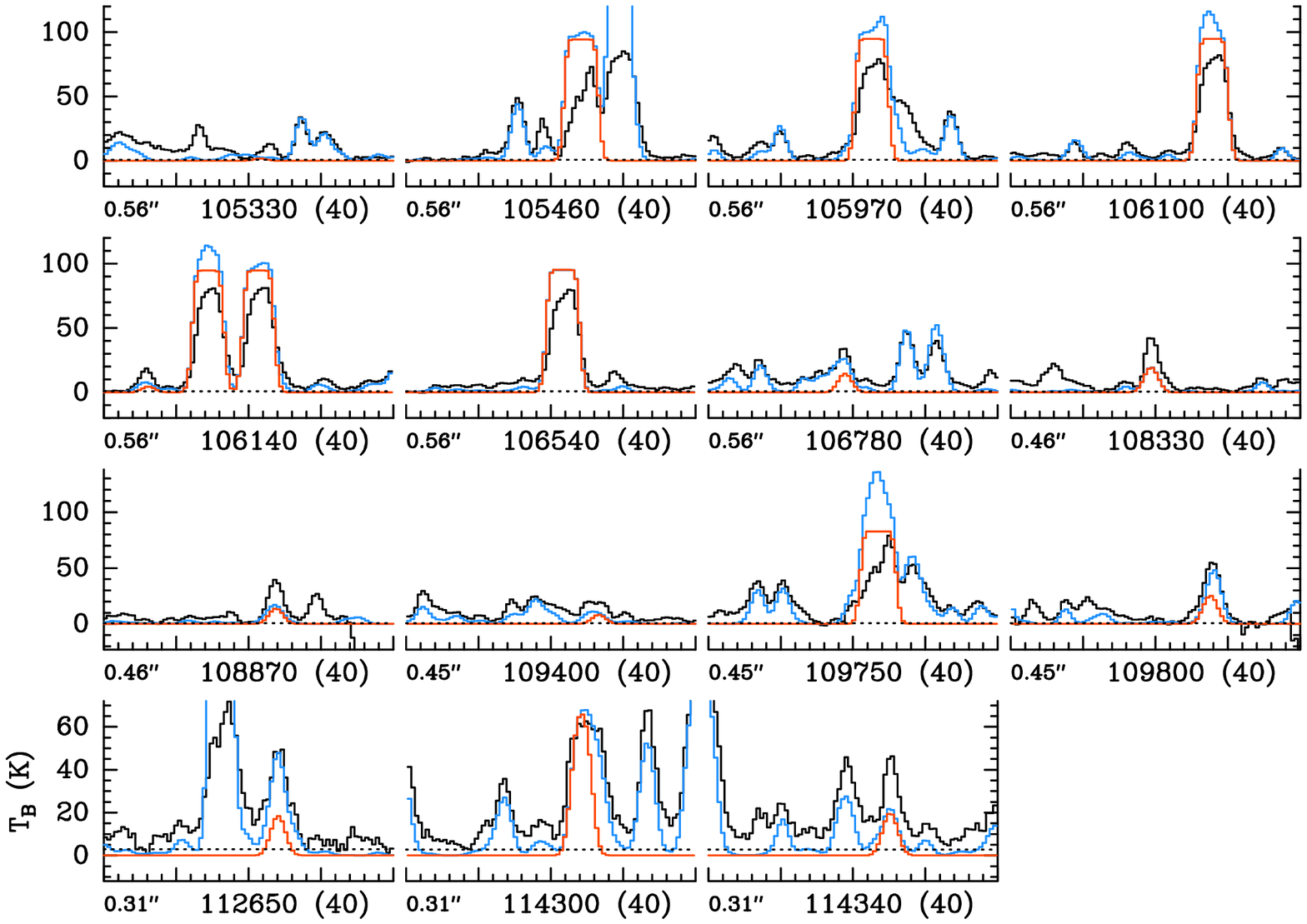}}}
\caption{continued.
}
\end{figure*}

\clearpage
\begin{figure*}
\centerline{\resizebox{0.82\hsize}{!}{\includegraphics[angle=0]{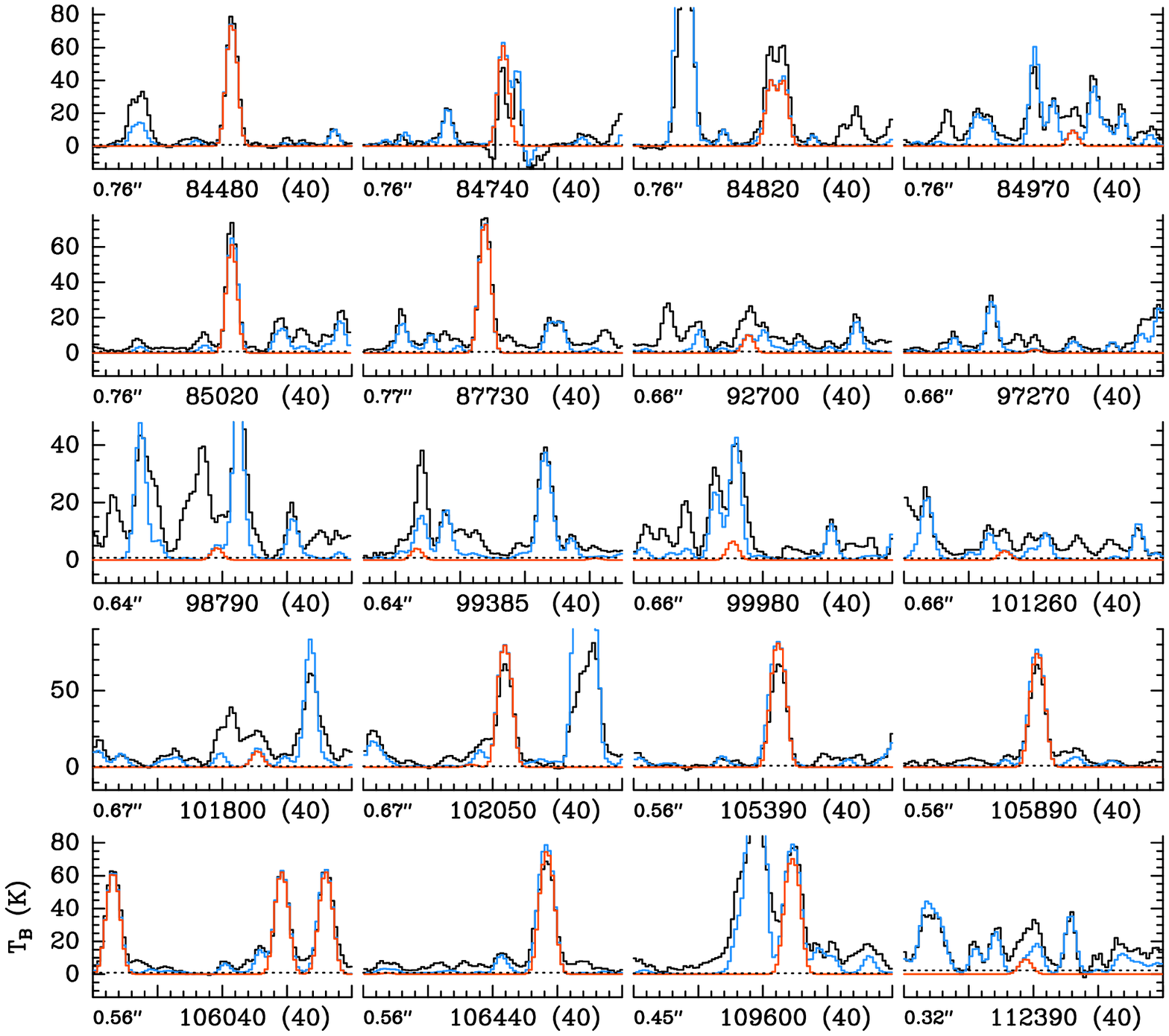}}}
\caption{Same as Fig.~\ref{f:spec_nh2conh2_ve0} for NH$_2$CHO, $\varv_{12} = 1$.
}
\label{f:spec_nh2cho_v12e1}
\end{figure*}

\clearpage
\begin{figure*}
\centerline{\resizebox{0.82\hsize}{!}{\includegraphics[angle=0]{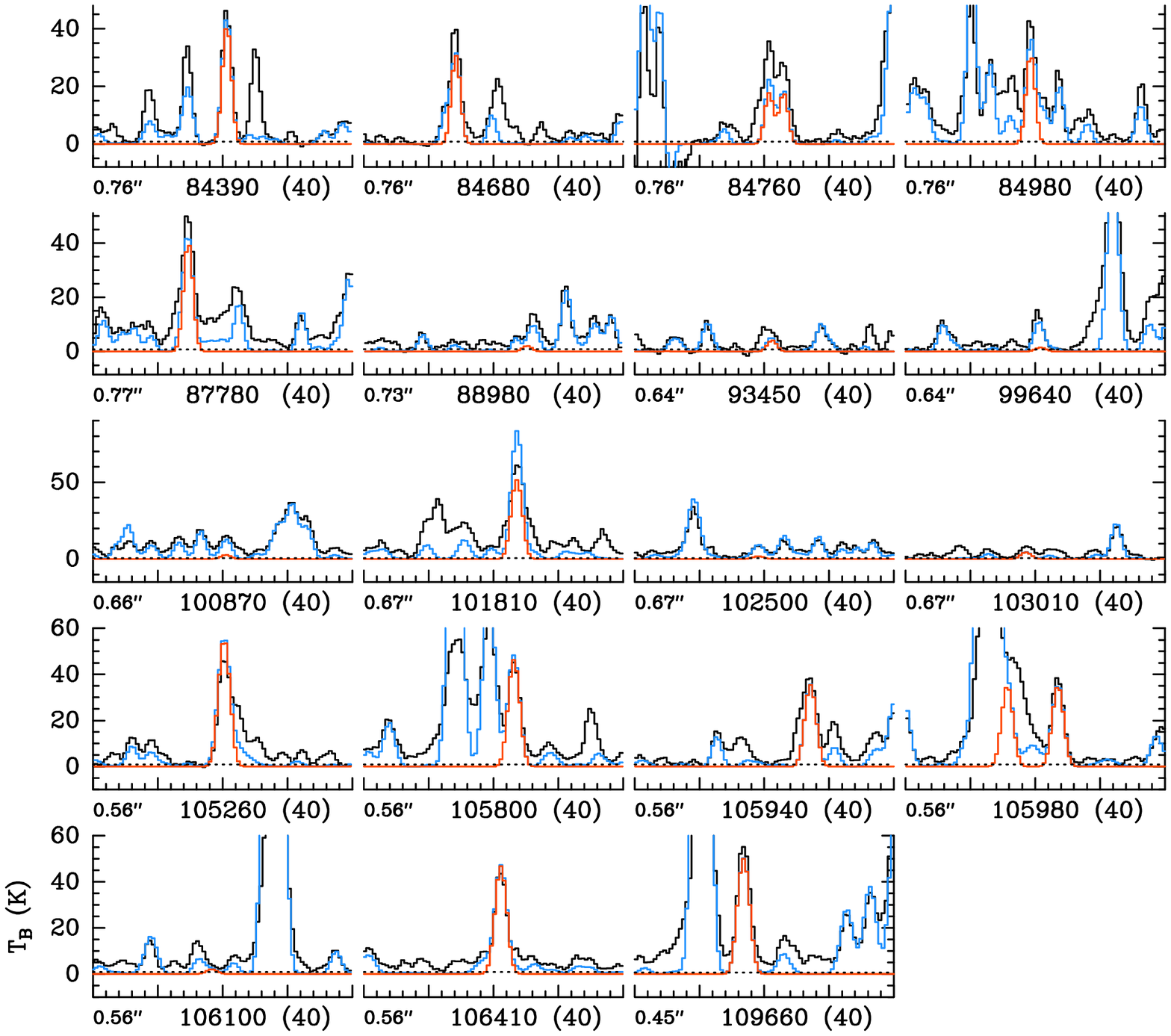}}}
\caption{Same as Fig.~\ref{f:spec_nh2conh2_ve0} for NH$_2$$^{13}$CHO, 
$\varv = 0$.
}
\label{f:spec_nh2cho_13c_ve0}
\end{figure*}

\clearpage
\begin{figure*}
\centerline{\resizebox{0.82\hsize}{!}{\includegraphics[angle=0]{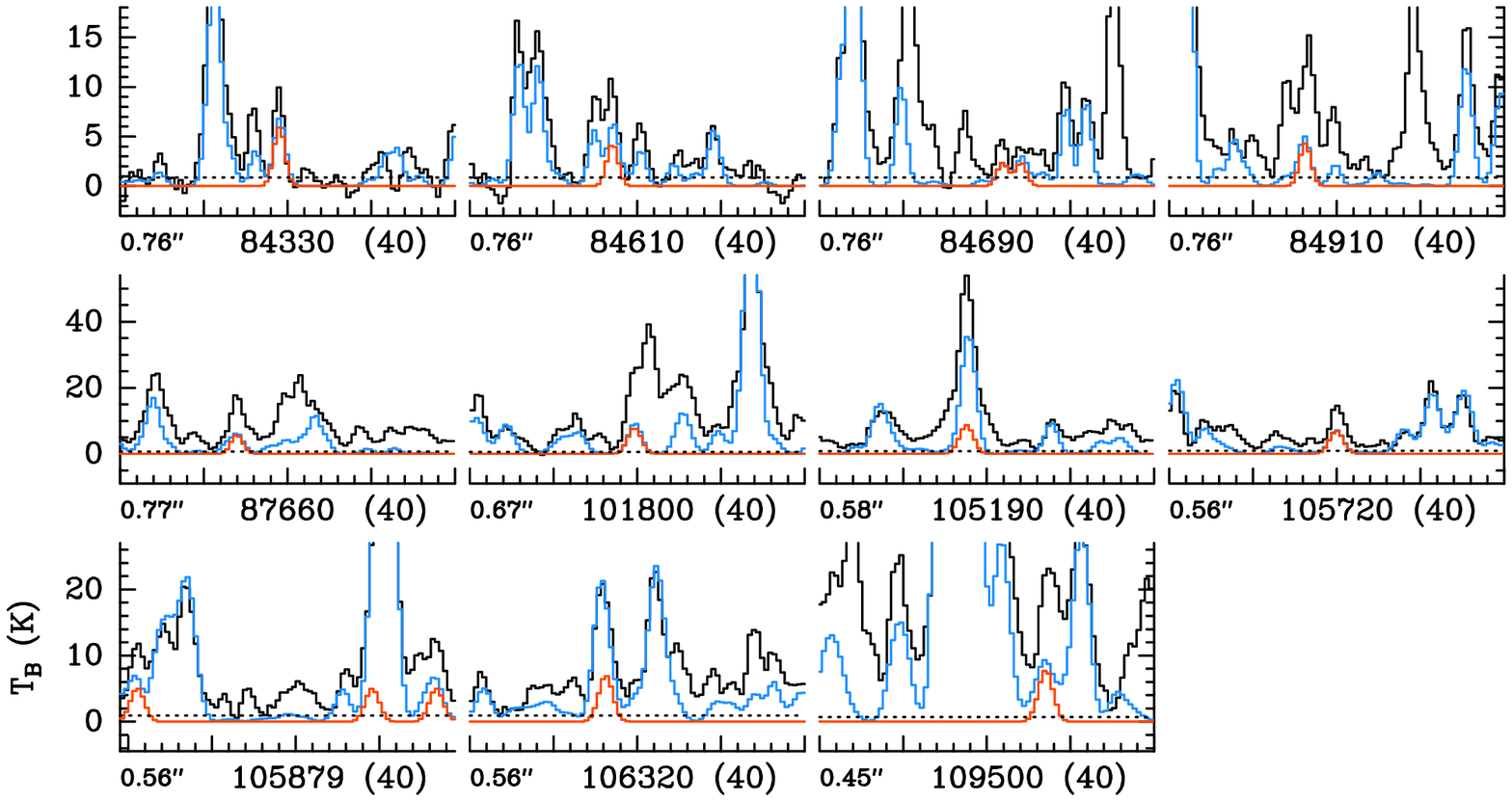}}}
\caption{Same as Fig.~\ref{f:spec_nh2conh2_ve0} for NH$_2$$^{13}$CHO, 
$\varv_{12} = 1$.
}
\label{f:spec_nh2cho_13c_v12e1}
\end{figure*}

\clearpage
\begin{figure*}
\centerline{\resizebox{0.82\hsize}{!}{\includegraphics[angle=0]{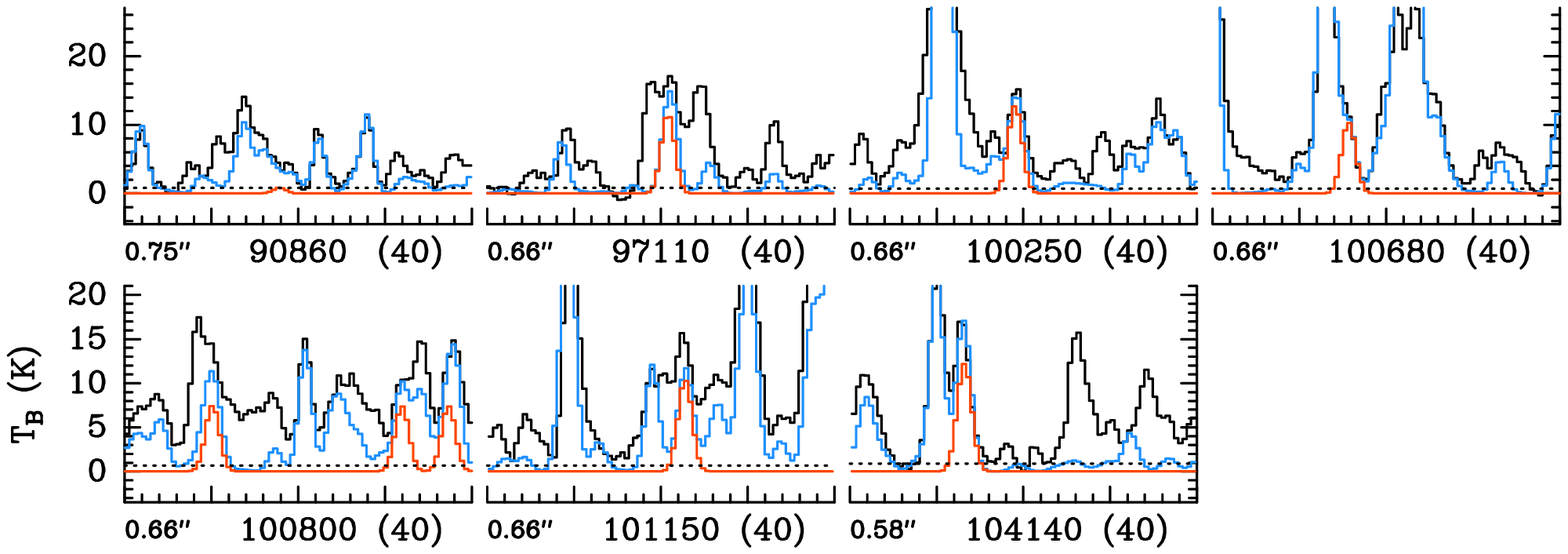}}}
\caption{Same as Fig.~\ref{f:spec_nh2conh2_ve0} for NH$_2$CH$^{18}$O, 
$\varv = 0$.
}
\label{f:spec_nh2cho_18o_ve0}
\end{figure*}

\clearpage
\begin{figure*}
\centerline{\resizebox{0.82\hsize}{!}{\includegraphics[angle=0]{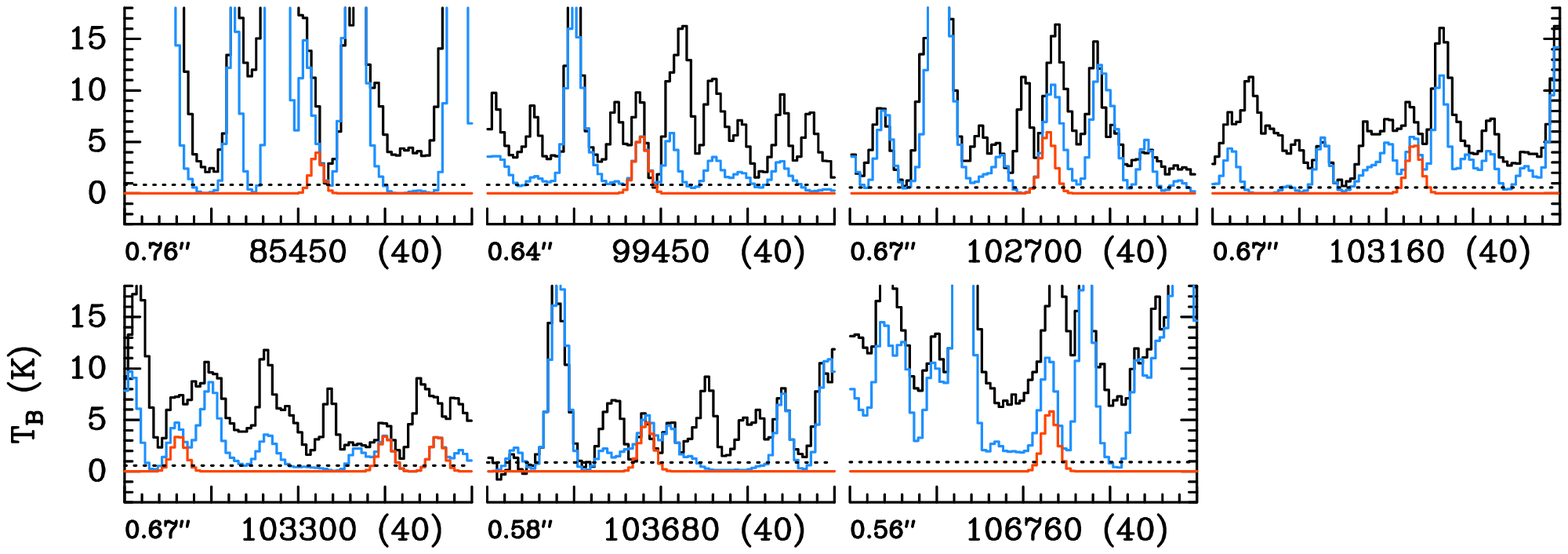}}}
\caption{Same as Fig.~\ref{f:spec_nh2conh2_ve0} for $^{15}$NH$_2$CHO, 
$\varv = 0$.
}
\label{f:spec_nh2cho_15n_ve0}
\end{figure*}

\clearpage
\begin{figure*}
\centerline{\resizebox{0.82\hsize}{!}{\includegraphics[angle=0]{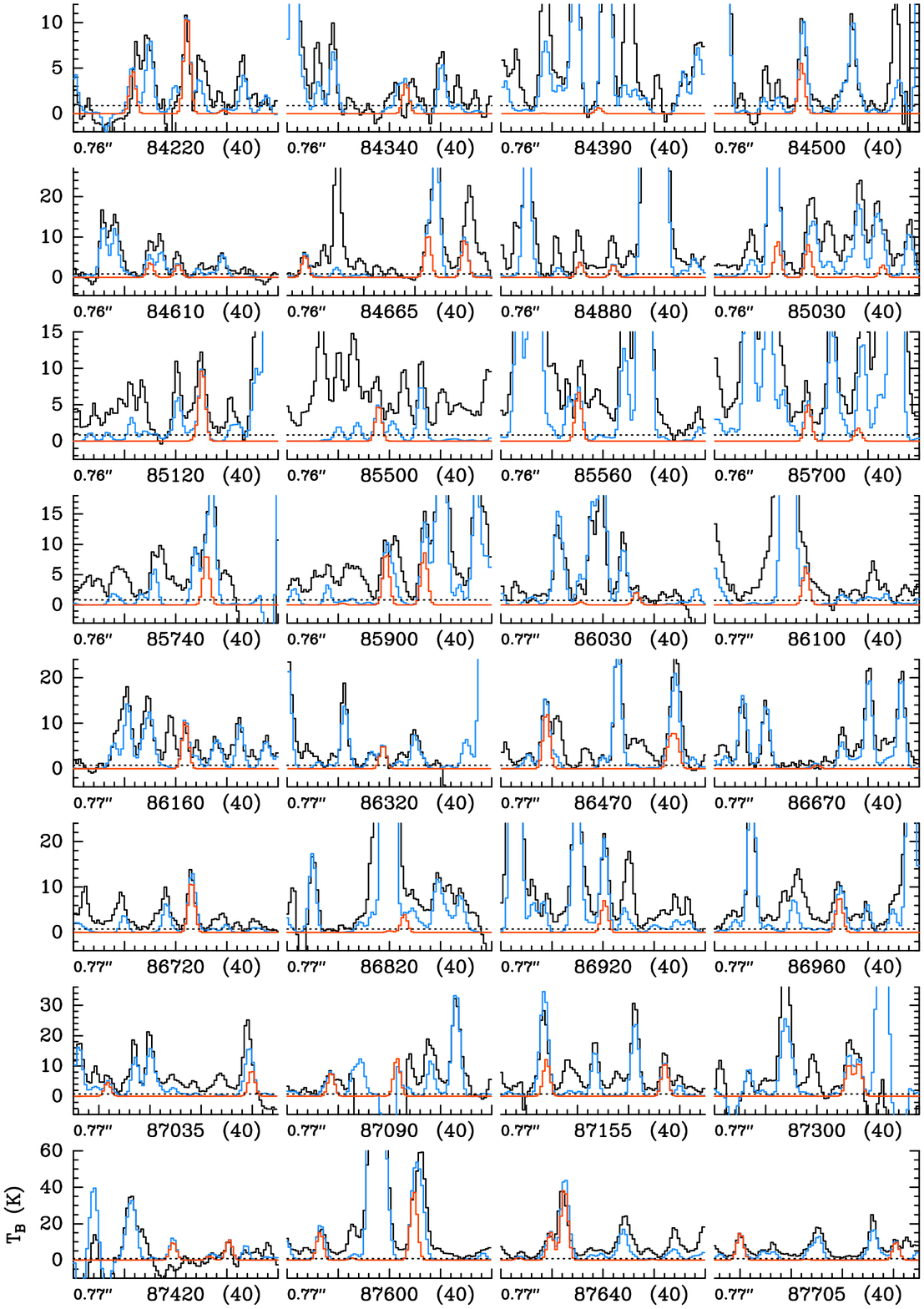}}}
\caption{Same as Fig.~\ref{f:spec_nh2conh2_ve0} for CH$_3$C(O)NH$_2$, 
$\varv = 0$.
}
\label{f:spec_ch3conh2_ve0}
\end{figure*}

\clearpage
\begin{figure*}
\addtocounter{figure}{-1}
\centerline{\resizebox{0.82\hsize}{!}{\includegraphics[angle=0]{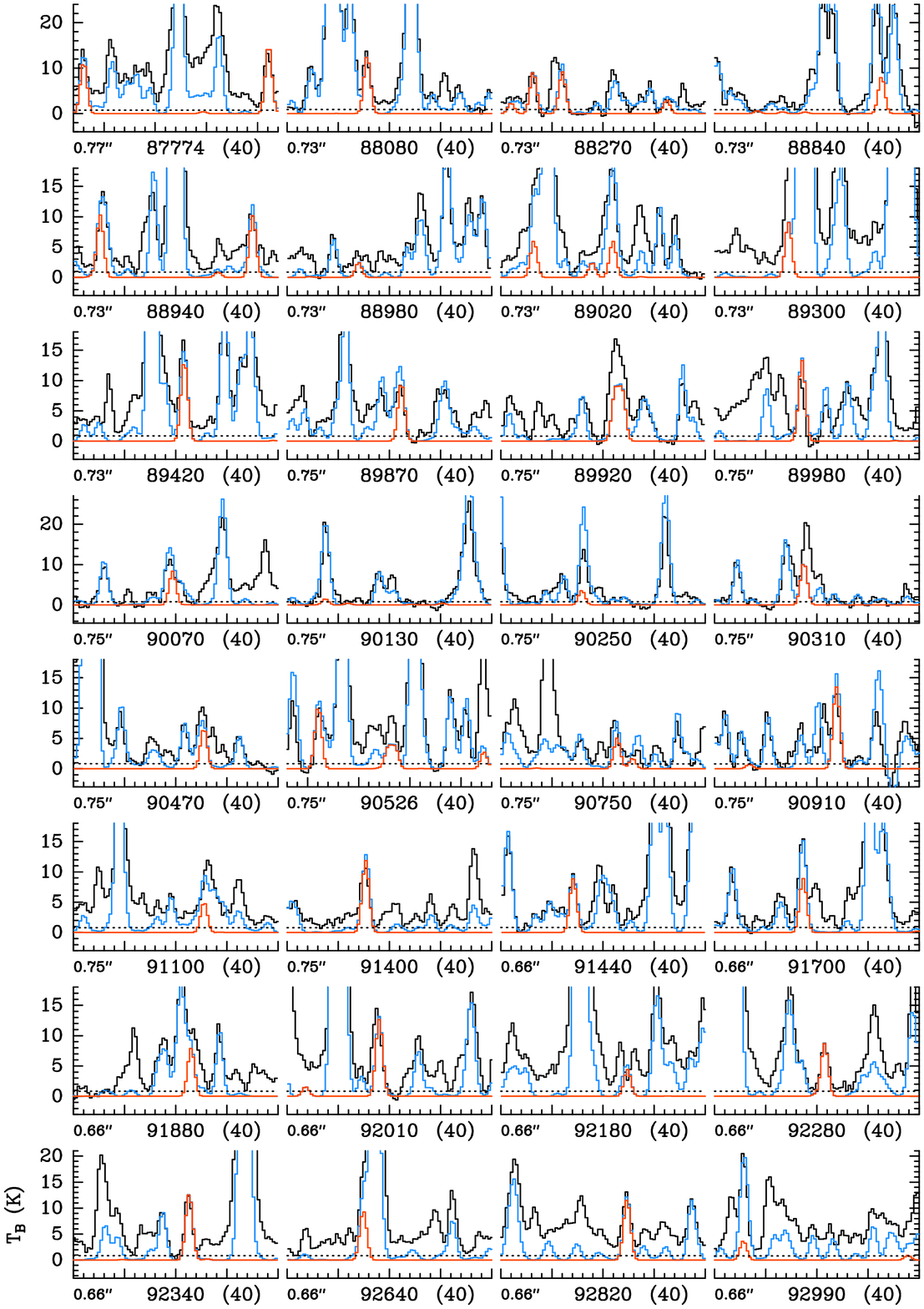}}}
\caption{continued.
}
\end{figure*}

\clearpage
\begin{figure*}
\addtocounter{figure}{-1}
\centerline{\resizebox{0.82\hsize}{!}{\includegraphics[angle=0]{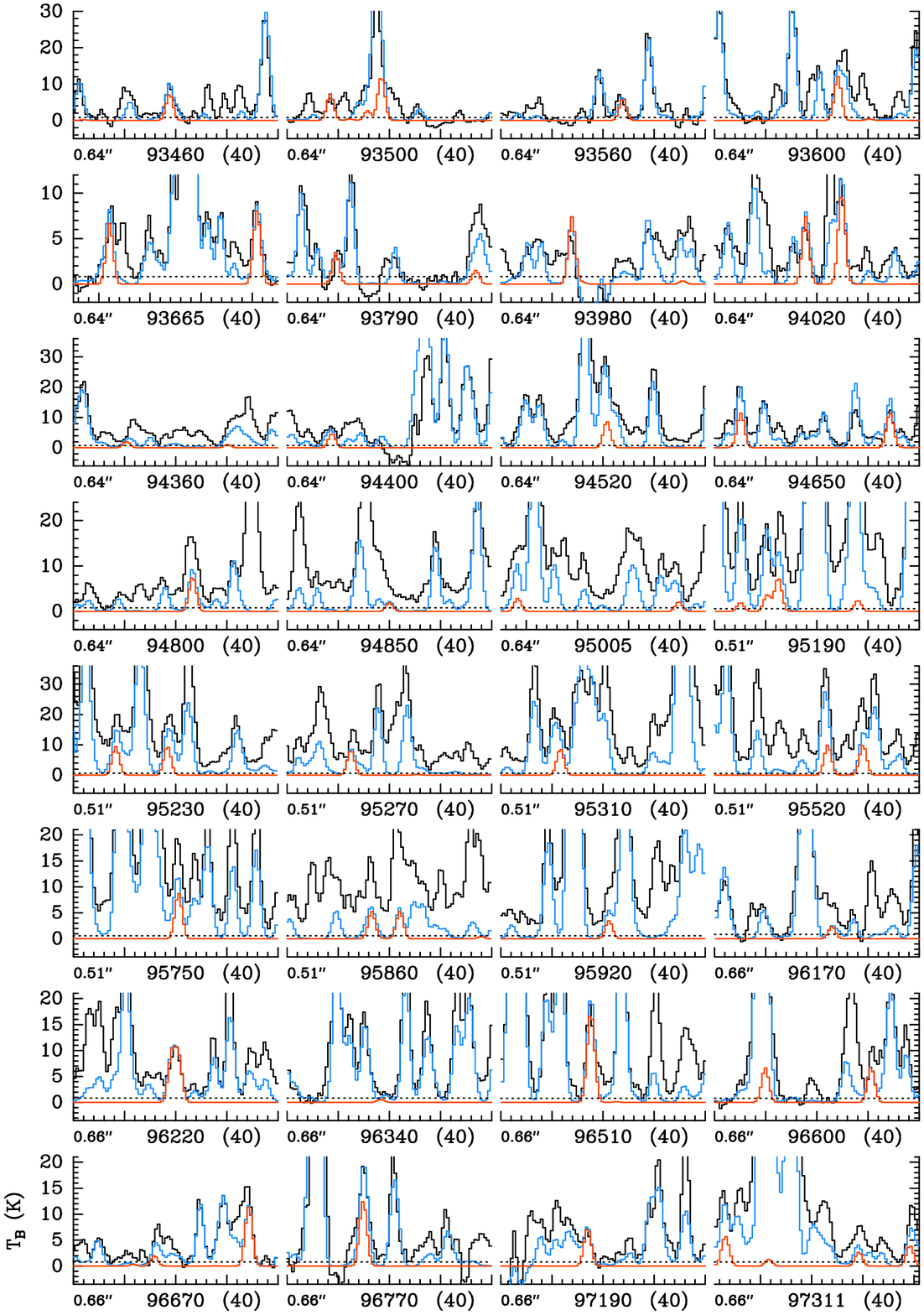}}}
\caption{continued.
}
\end{figure*}

\clearpage
\begin{figure*}
\addtocounter{figure}{-1}
\centerline{\resizebox{0.82\hsize}{!}{\includegraphics[angle=0]{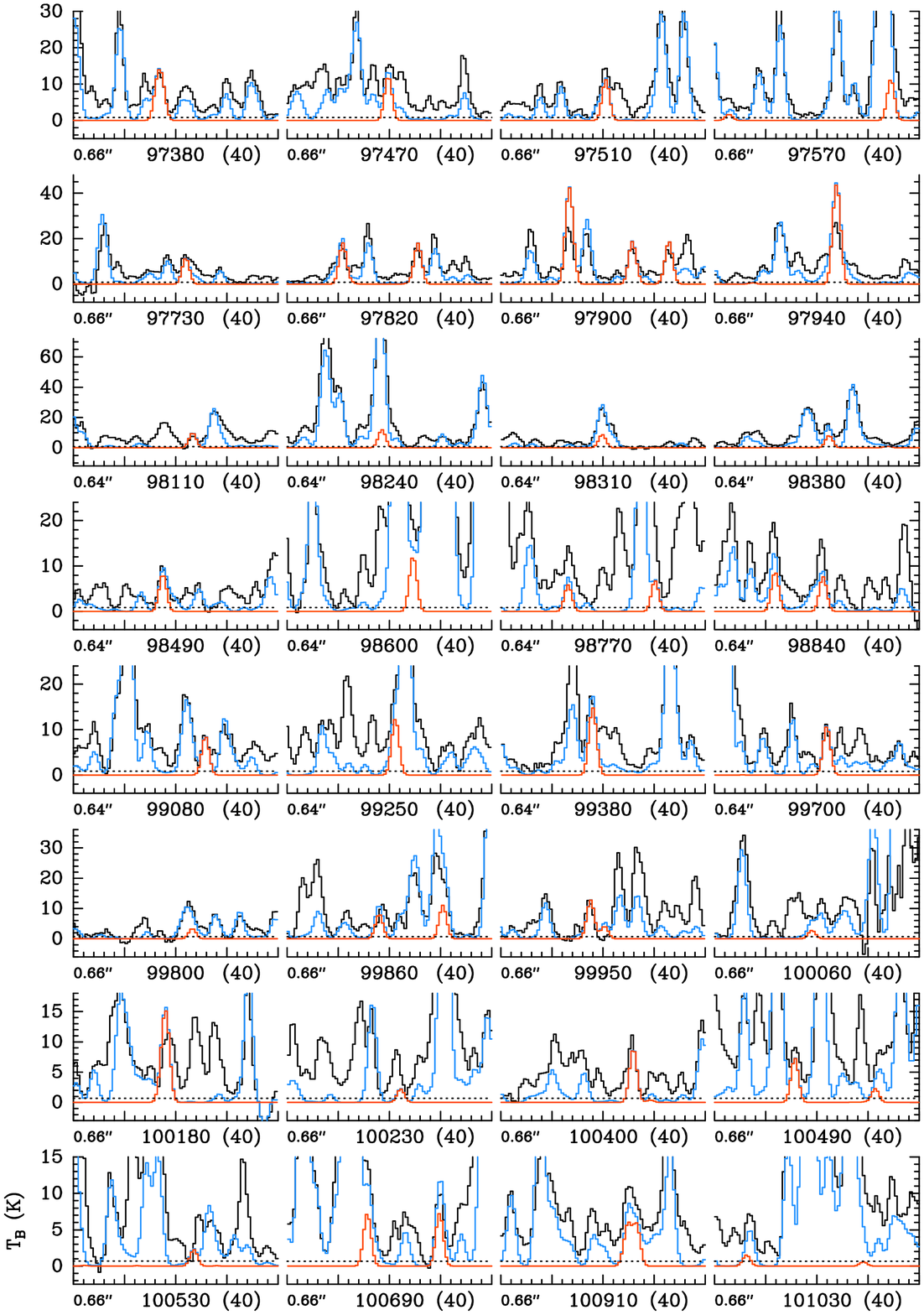}}}
\caption{continued.
}
\end{figure*}

\clearpage
\begin{figure*}
\addtocounter{figure}{-1}
\centerline{\resizebox{0.82\hsize}{!}{\includegraphics[angle=0]{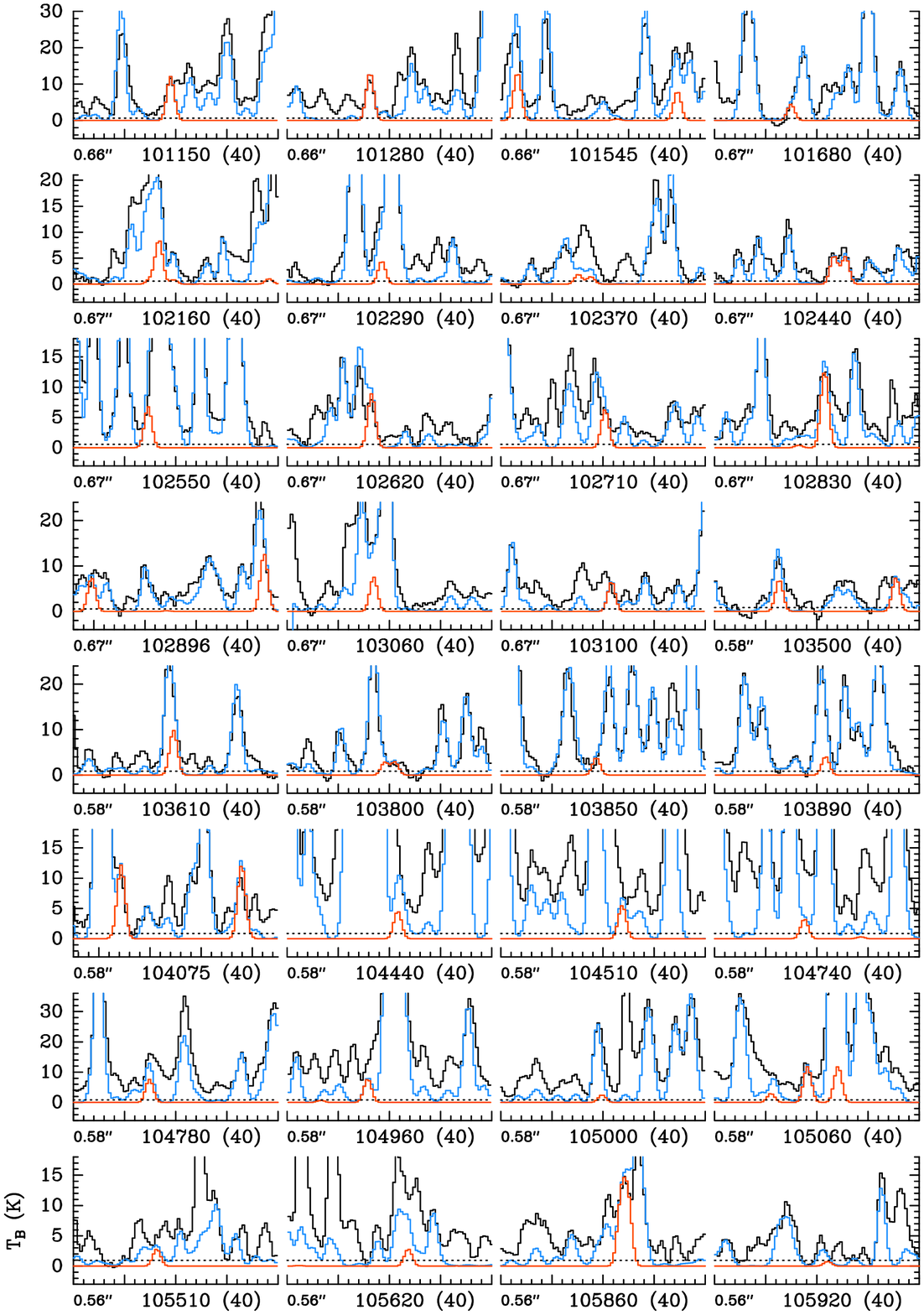}}}
\caption{continued.
}
\end{figure*}

\clearpage
\begin{figure*}
\addtocounter{figure}{-1}
\centerline{\resizebox{0.82\hsize}{!}{\includegraphics[angle=0]{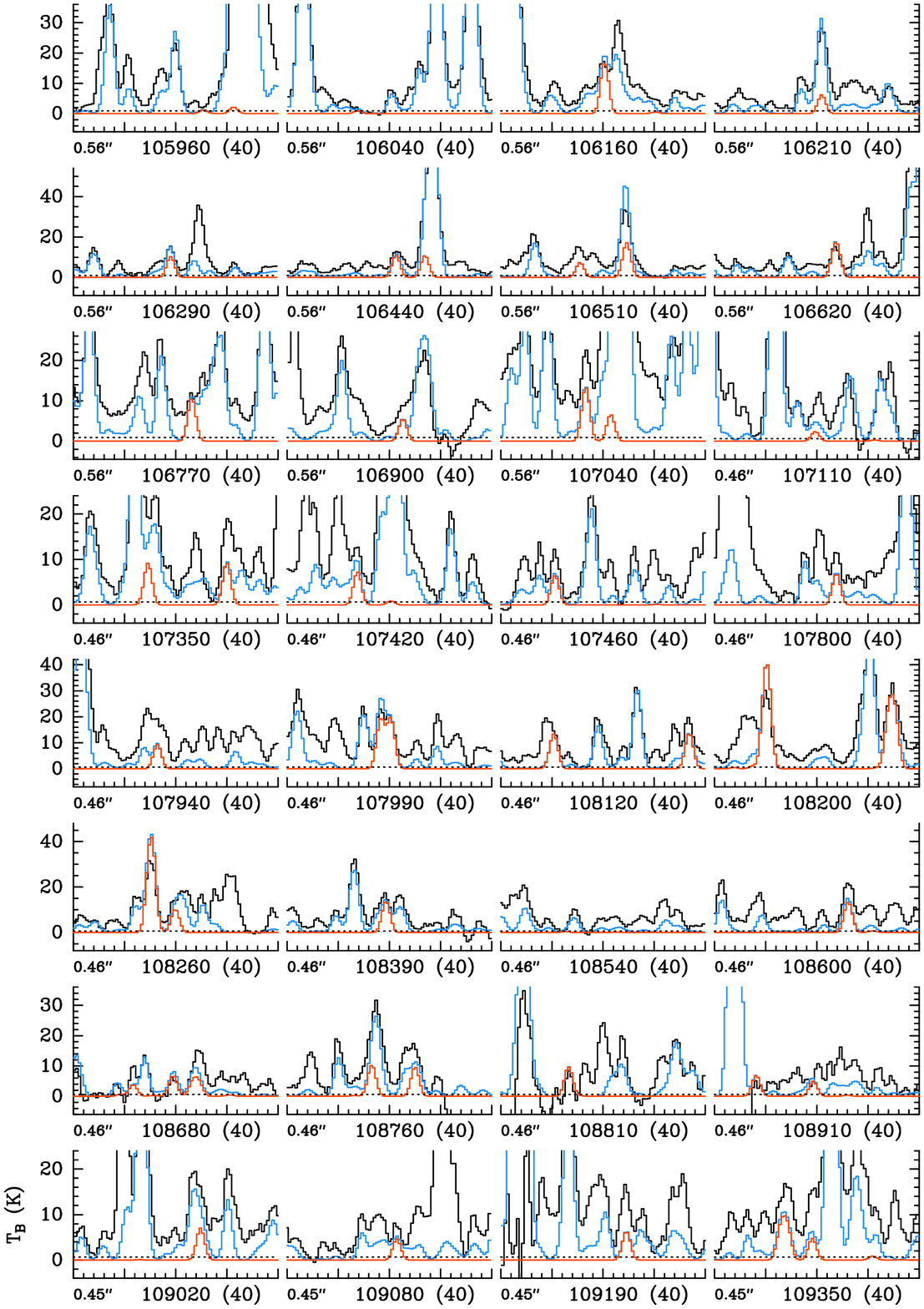}}}
\caption{continued.
}
\end{figure*}

\clearpage
\begin{figure*}
\addtocounter{figure}{-1}
\centerline{\resizebox{0.82\hsize}{!}{\includegraphics[angle=0]{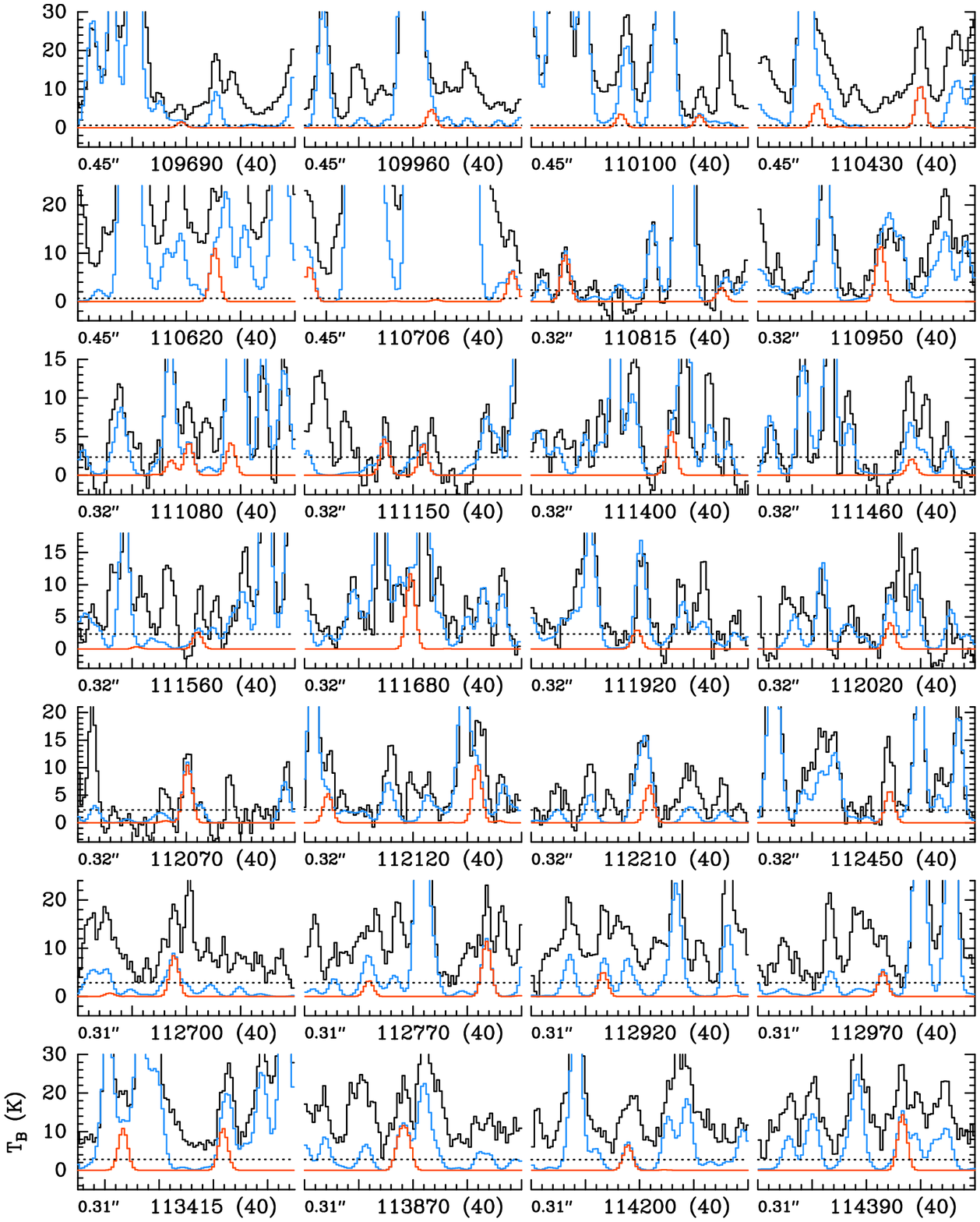}}}
\caption{continued.
}
\end{figure*}

\clearpage
\begin{figure*}
\centerline{\resizebox{0.82\hsize}{!}{\includegraphics[angle=0]{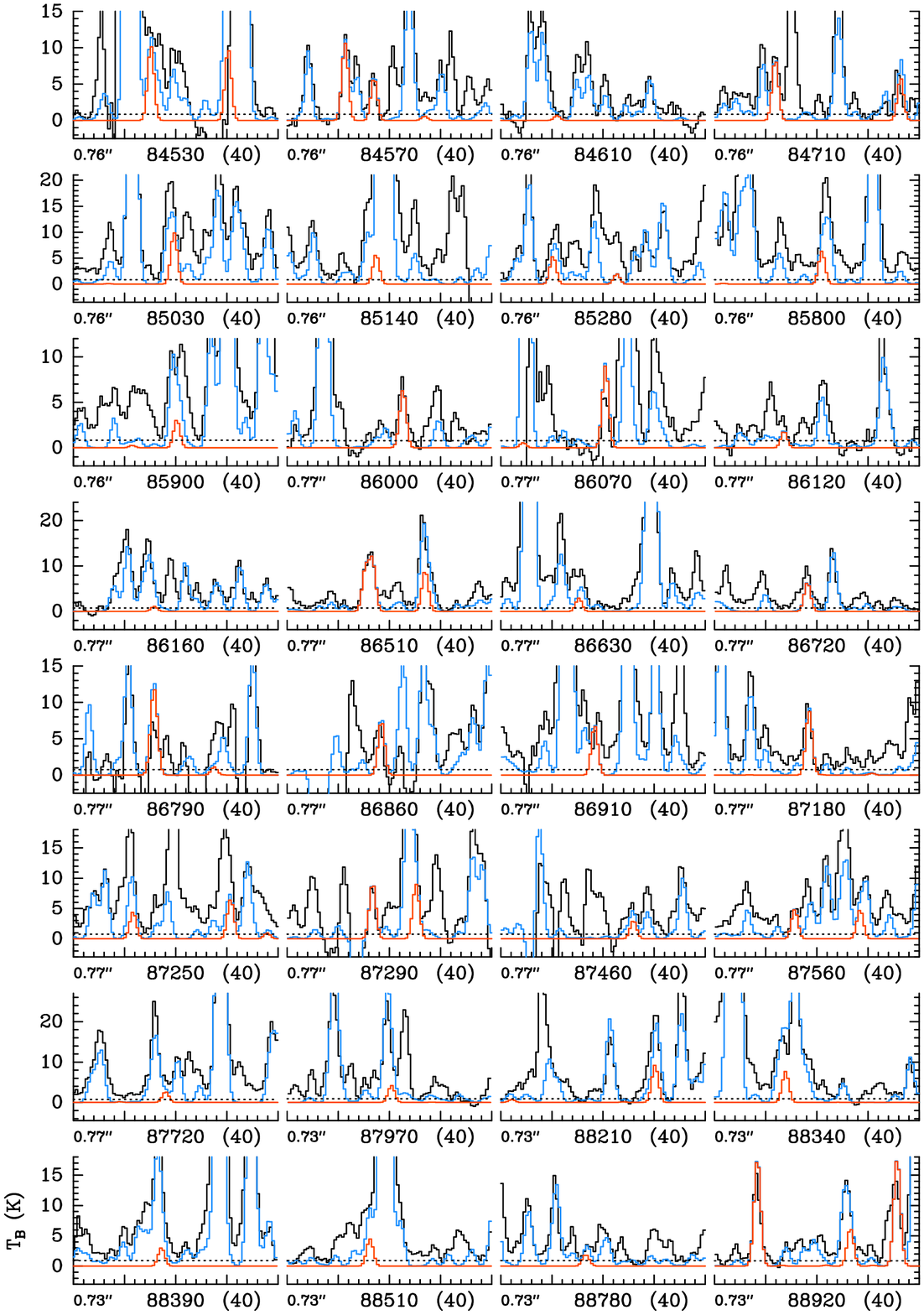}}}
\caption{Same as Fig.~\ref{f:spec_nh2conh2_ve0} for CH$_3$C(O)NH$_2$, 
$\varv_{\rm t} = 1$.
}
\label{f:spec_ch3conh2_ve1}
\end{figure*}

\clearpage
\begin{figure*}
\addtocounter{figure}{-1}
\centerline{\resizebox{0.82\hsize}{!}{\includegraphics[angle=0]{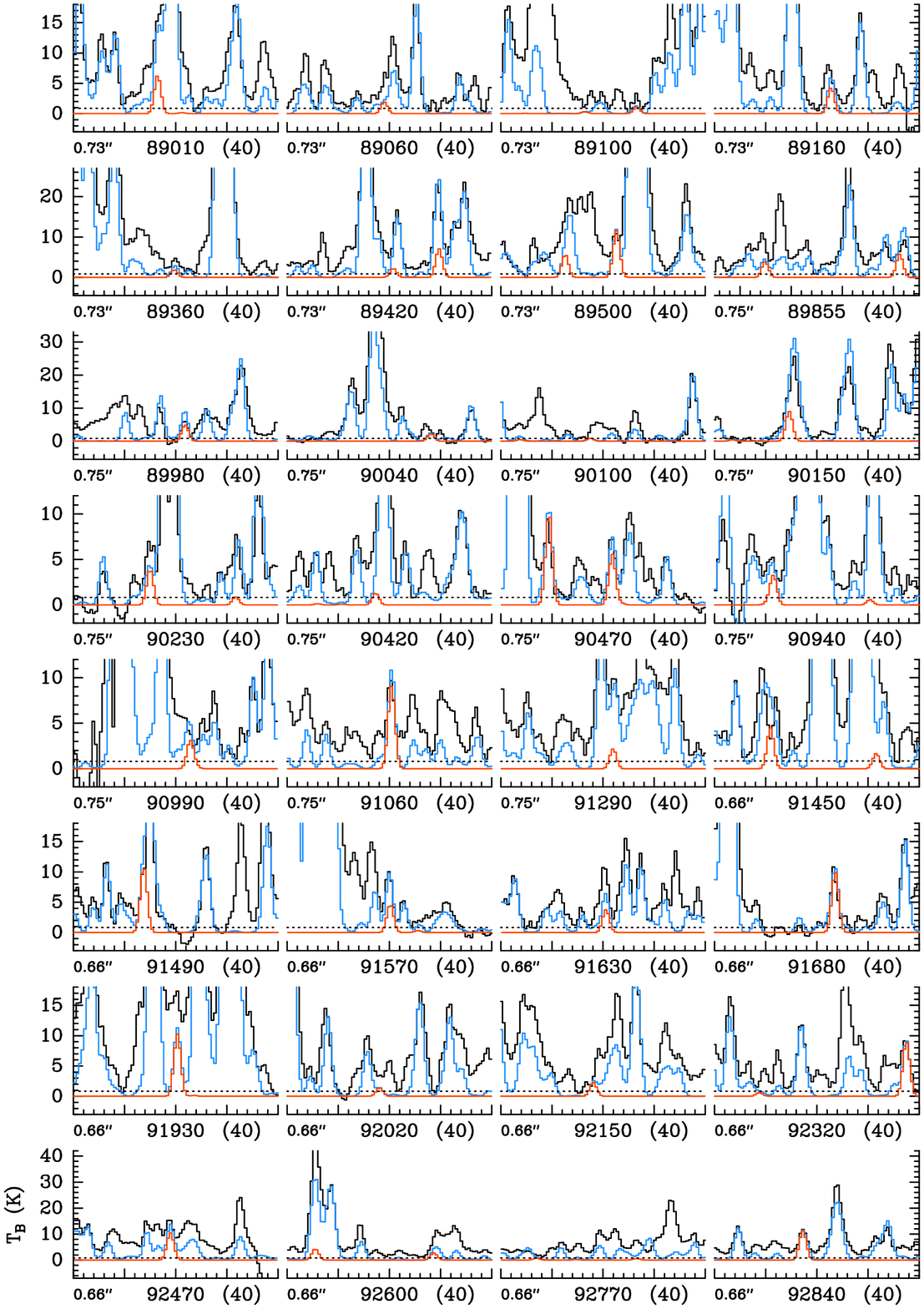}}}
\caption{continued.
}
\end{figure*}

\clearpage
\begin{figure*}
\addtocounter{figure}{-1}
\centerline{\resizebox{0.82\hsize}{!}{\includegraphics[angle=0]{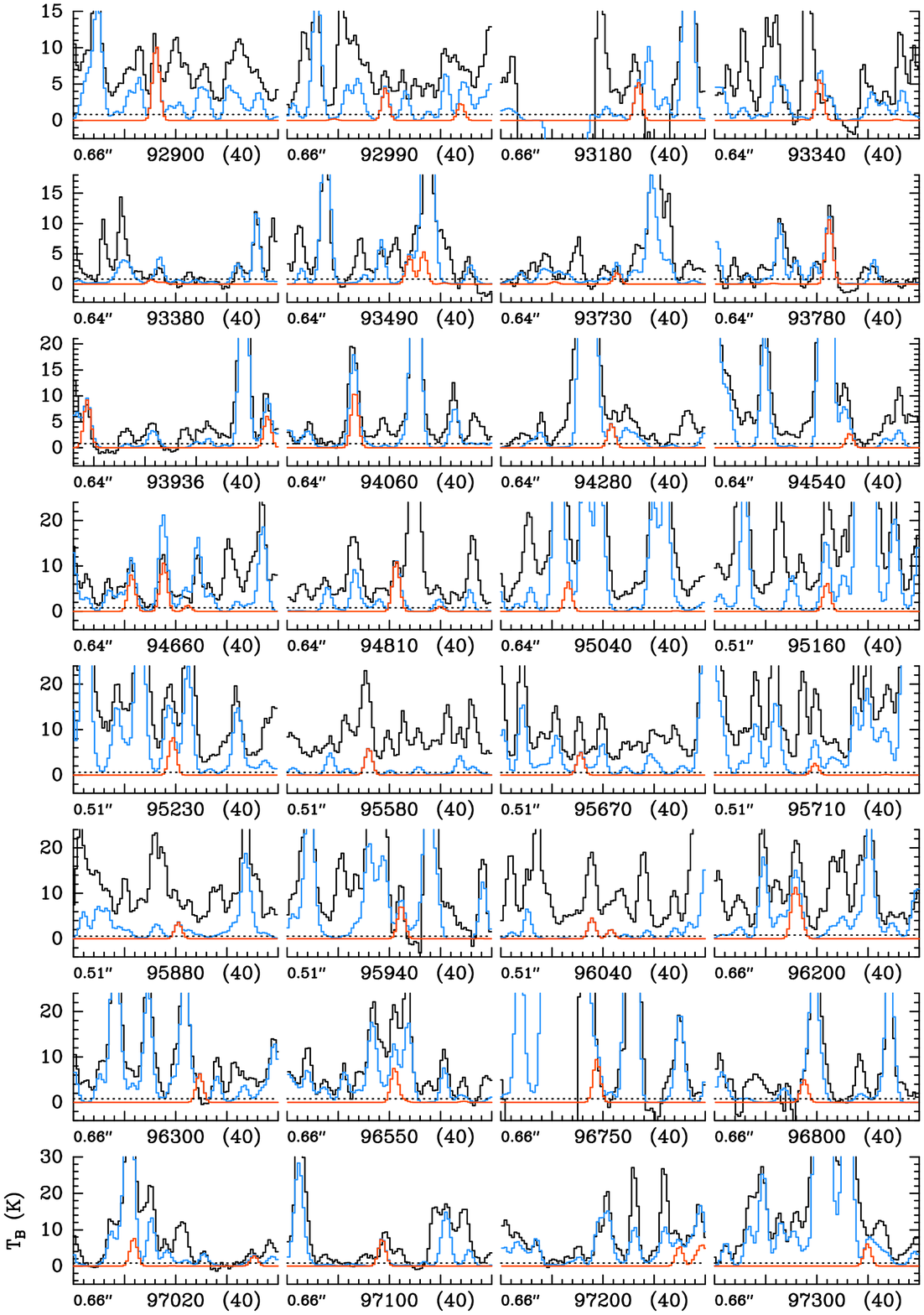}}}
\caption{continued.
}
\end{figure*}

\clearpage
\begin{figure*}
\addtocounter{figure}{-1}
\centerline{\resizebox{0.82\hsize}{!}{\includegraphics[angle=0]{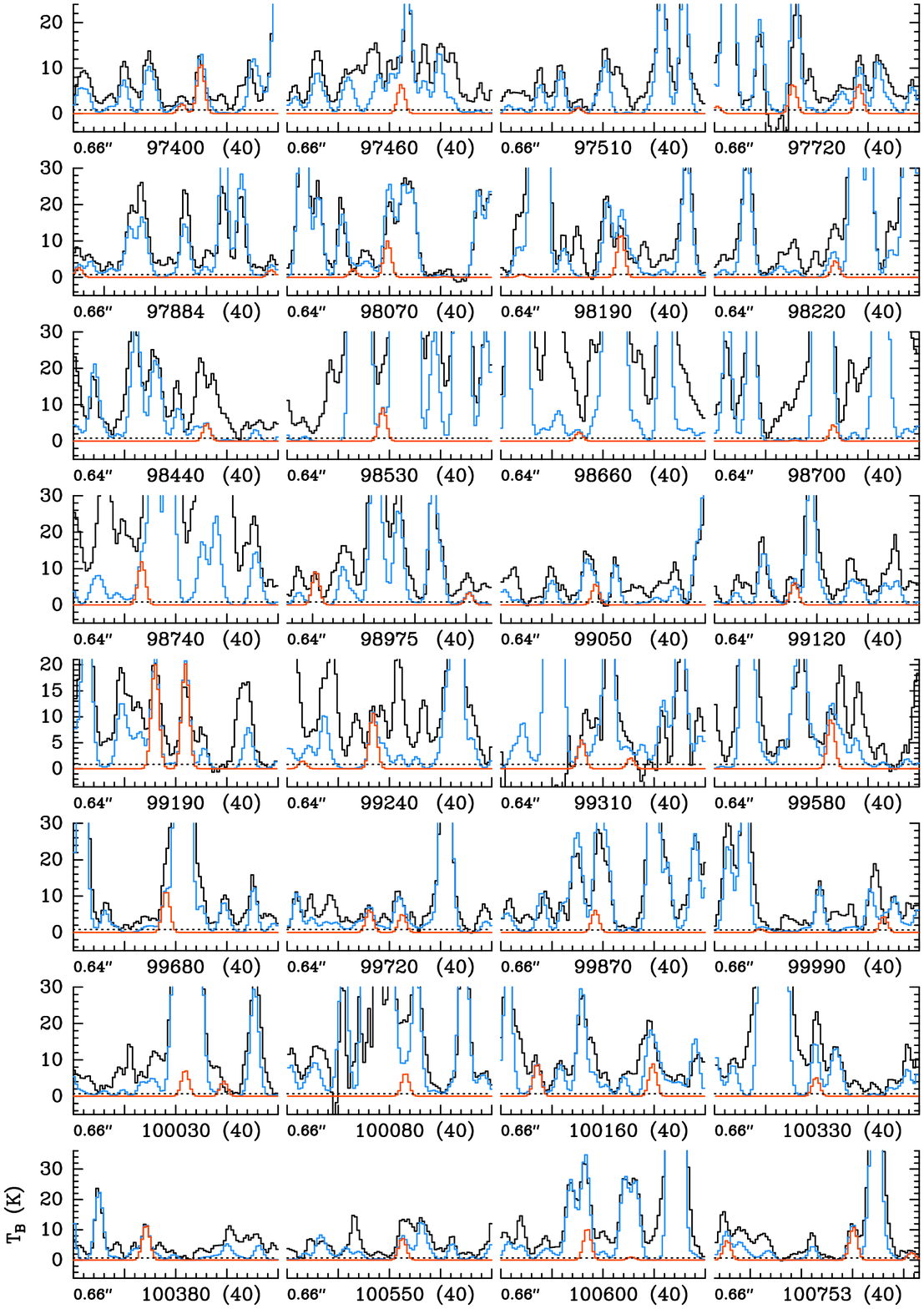}}}
\caption{continued.
}
\end{figure*}

\clearpage
\begin{figure*}
\addtocounter{figure}{-1}
\centerline{\resizebox{0.82\hsize}{!}{\includegraphics[angle=0]{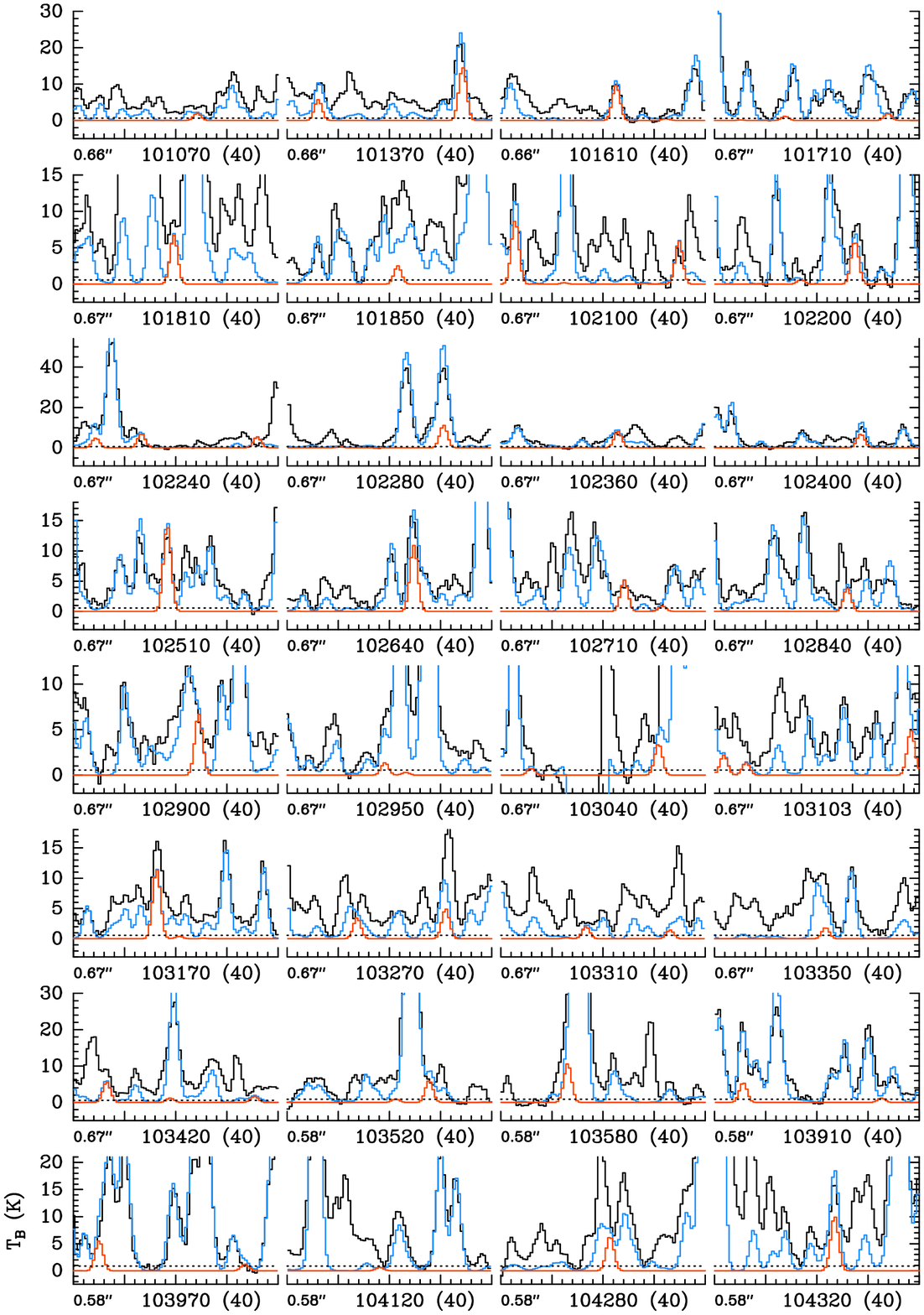}}}
\caption{continued.
}
\end{figure*}

\clearpage
\begin{figure*}
\addtocounter{figure}{-1}
\centerline{\resizebox{0.82\hsize}{!}{\includegraphics[angle=0]{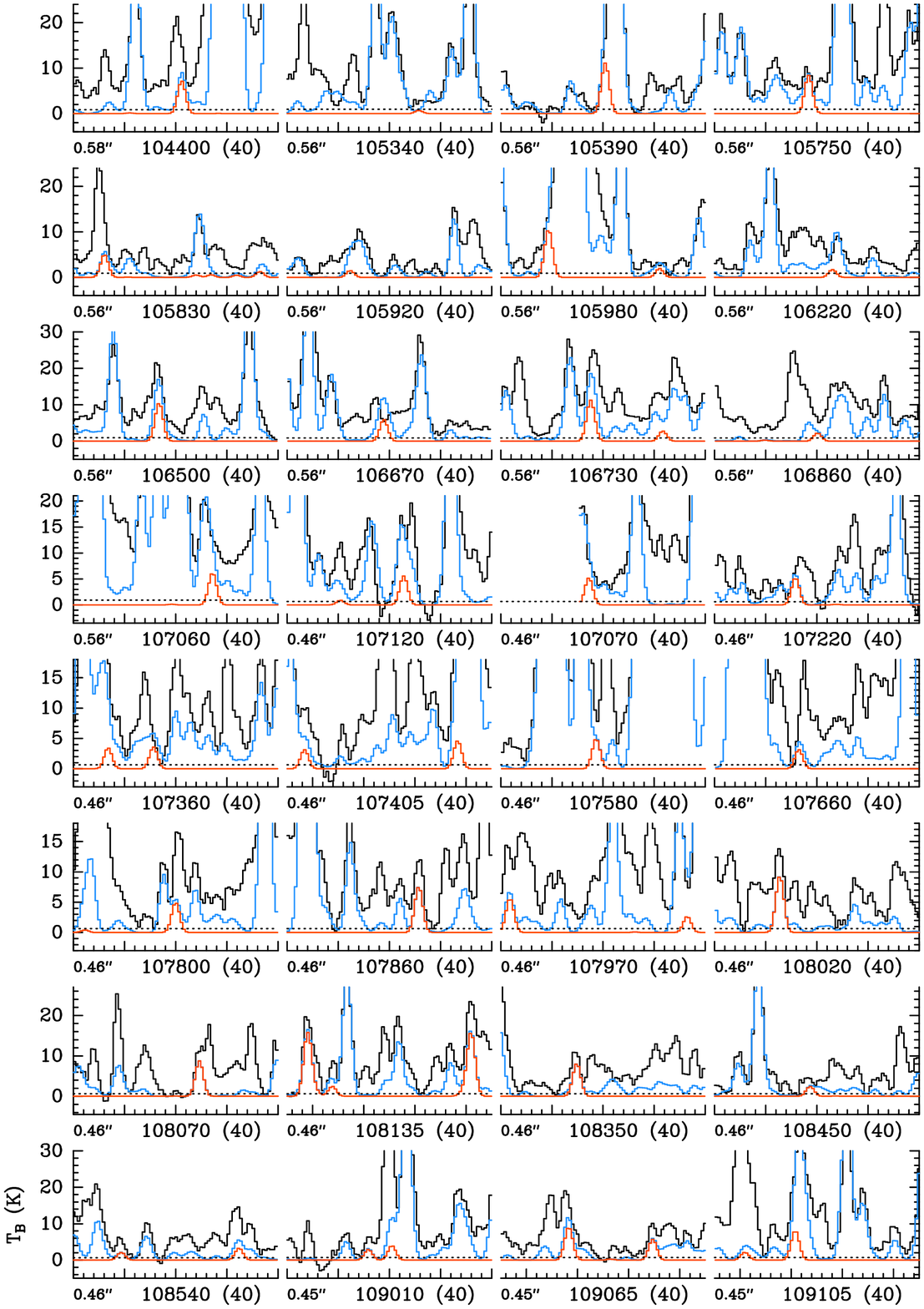}}}
\caption{continued.
}
\end{figure*}

\clearpage
\begin{figure*}
\centerline{\resizebox{0.82\hsize}{!}{\includegraphics[angle=0]{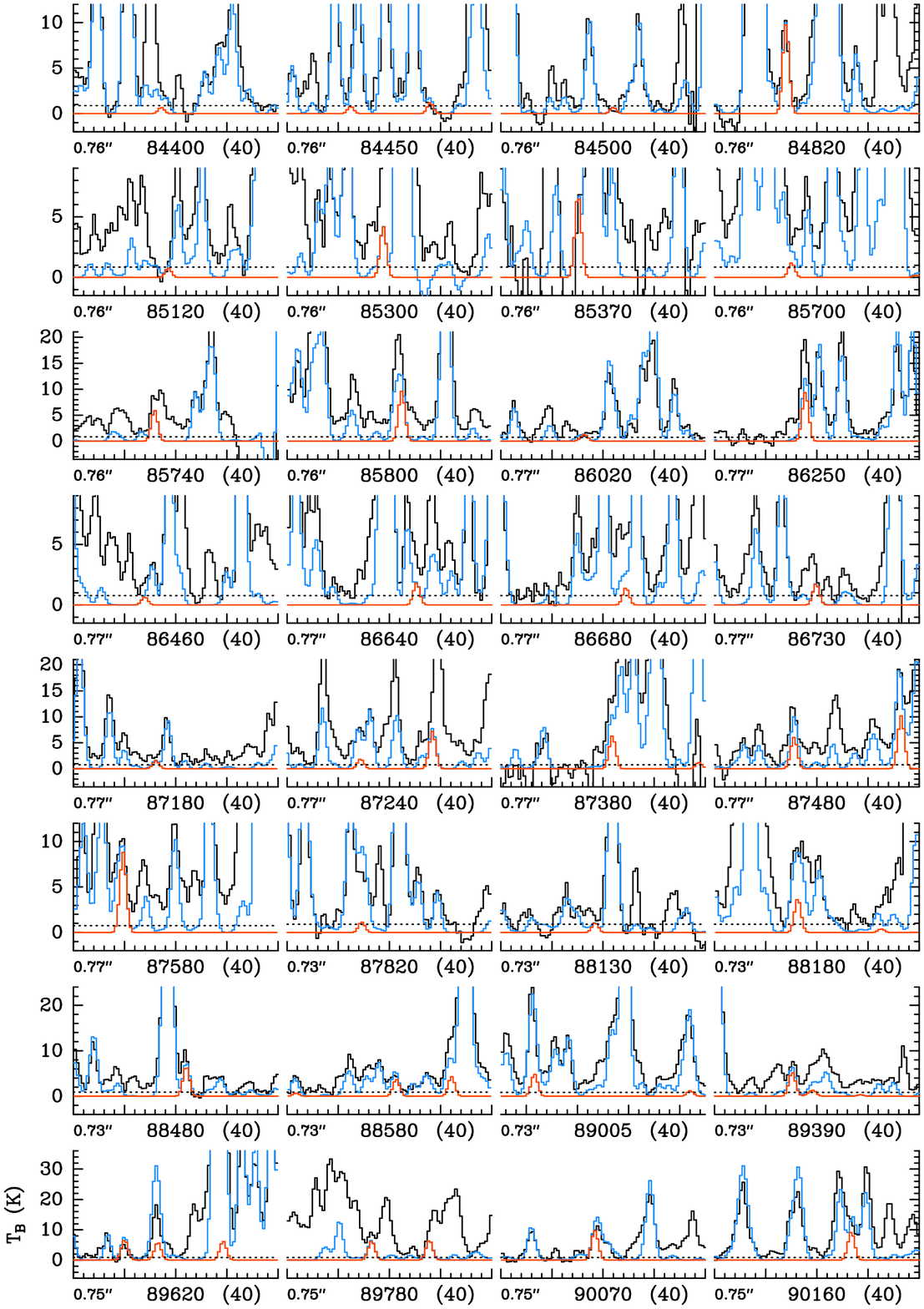}}}
\caption{Same as Fig.~\ref{f:spec_nh2conh2_ve0} for CH$_3$C(O)NH$_2$, 
$\varv_{\rm t} = 2$.
}
\label{f:spec_ch3conh2_ve2}
\end{figure*}

\clearpage
\begin{figure*}
\addtocounter{figure}{-1}
\centerline{\resizebox{0.82\hsize}{!}{\includegraphics[angle=0]{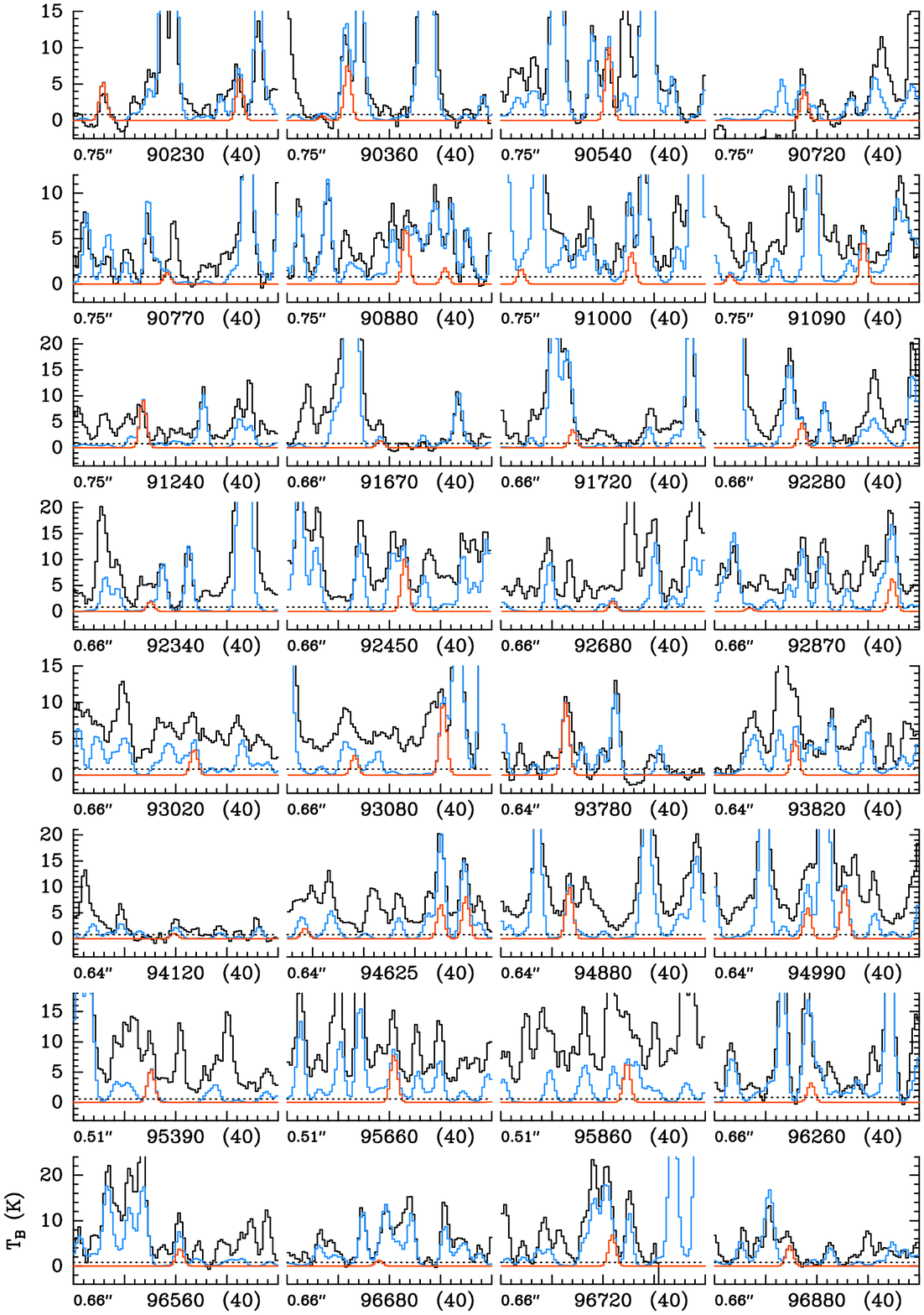}}}
\caption{continued.
}
\end{figure*}

\clearpage
\begin{figure*}
\addtocounter{figure}{-1}
\centerline{\resizebox{0.82\hsize}{!}{\includegraphics[angle=0]{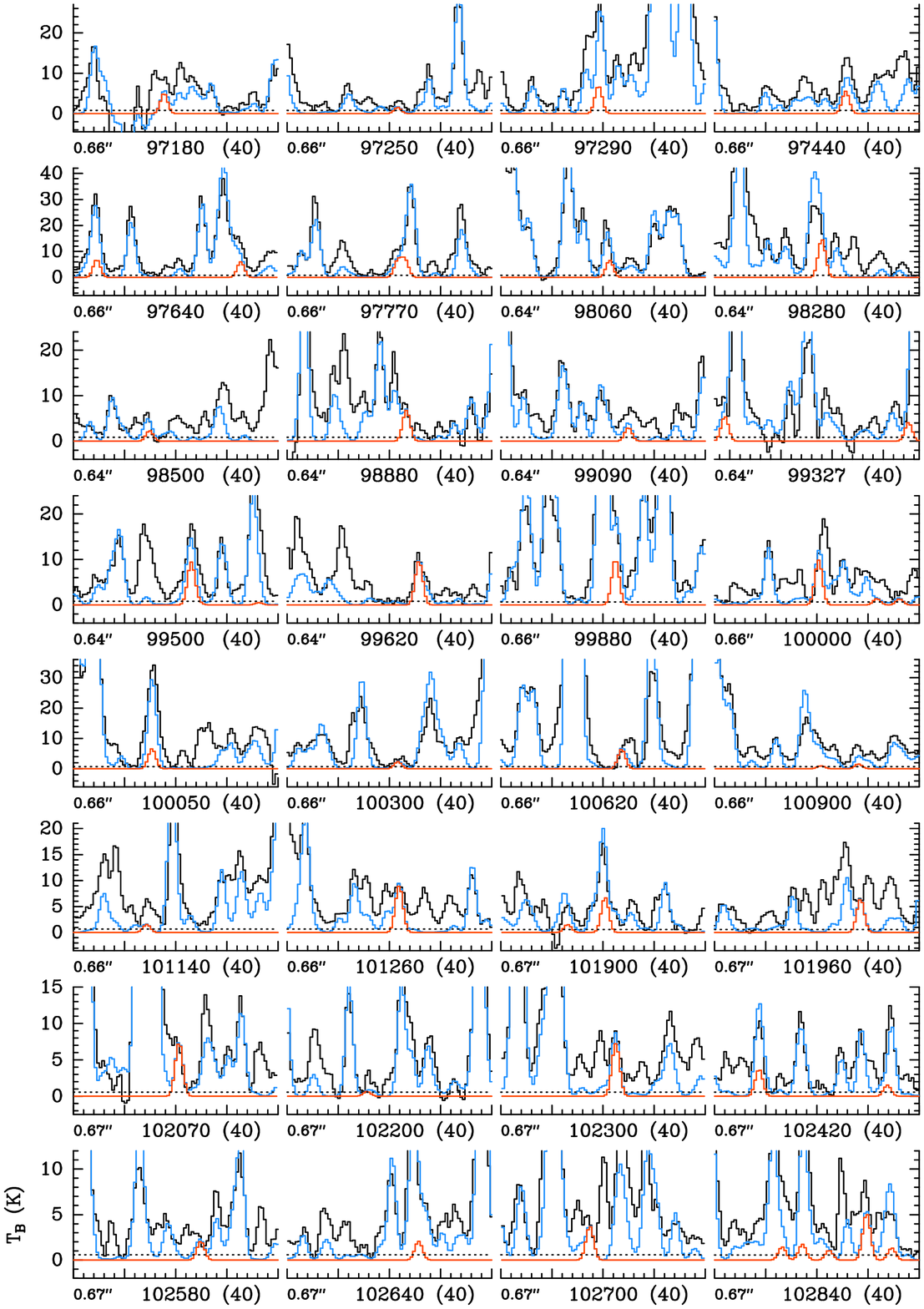}}}
\caption{continued.
}
\end{figure*}

\clearpage
\begin{figure*}
\addtocounter{figure}{-1}
\centerline{\resizebox{0.82\hsize}{!}{\includegraphics[angle=0]{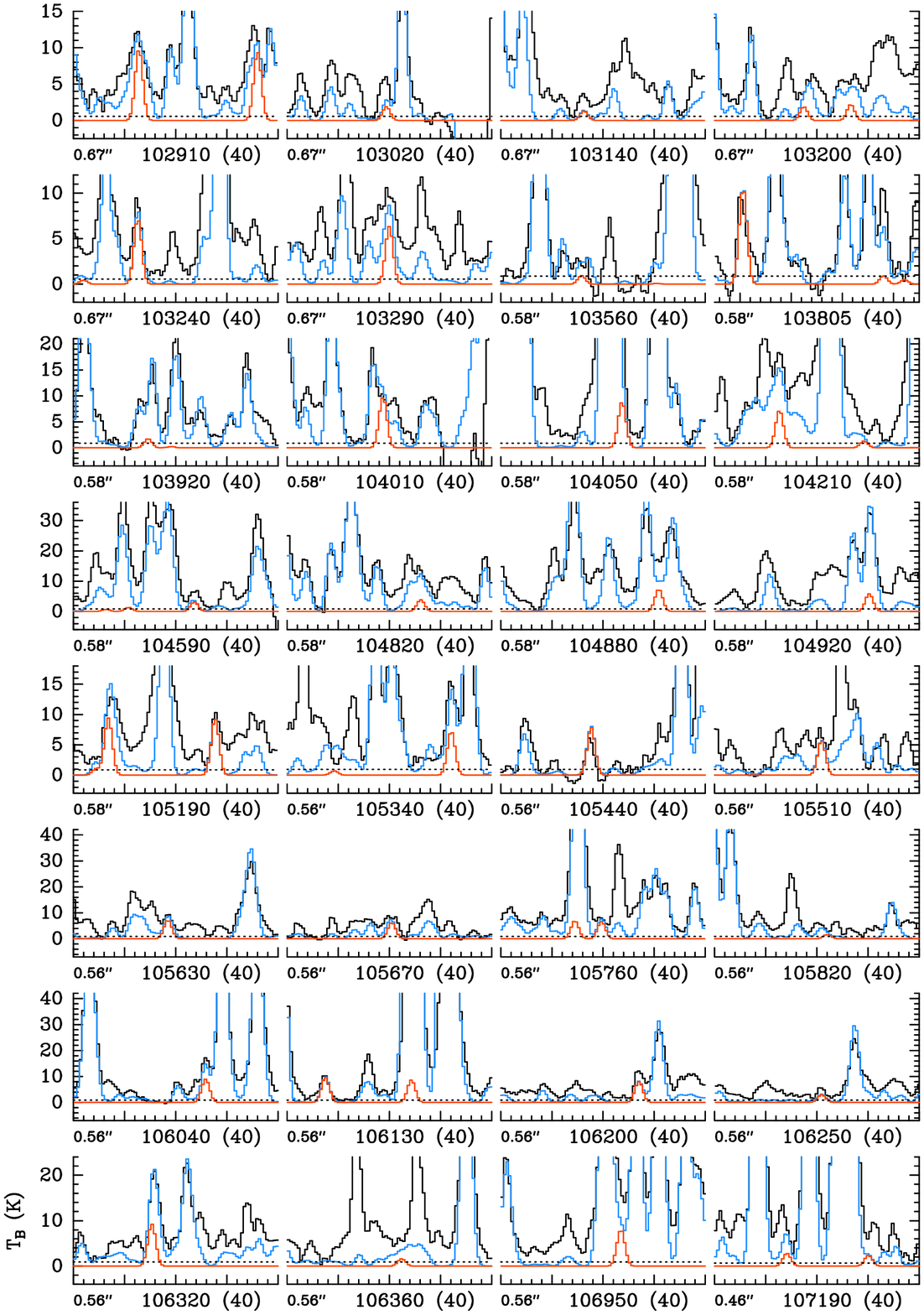}}}
\caption{continued.
}
\end{figure*}

\clearpage
\begin{figure*}
\addtocounter{figure}{-1}
\centerline{\resizebox{0.82\hsize}{!}{\includegraphics[angle=0]{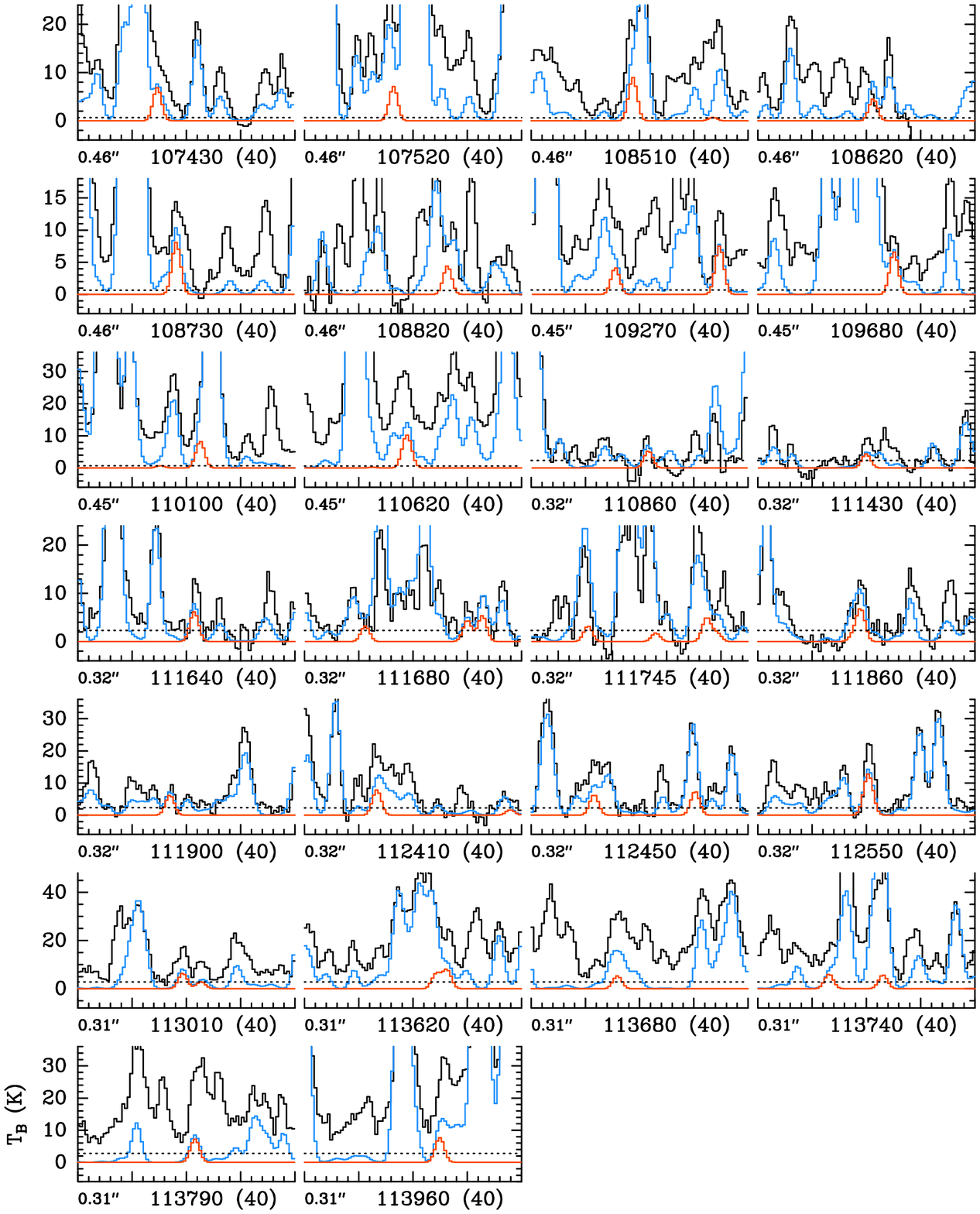}}}
\caption{continued.
}
\end{figure*}

\clearpage
\begin{figure*}
\centerline{\resizebox{0.82\hsize}{!}{\includegraphics[angle=0]{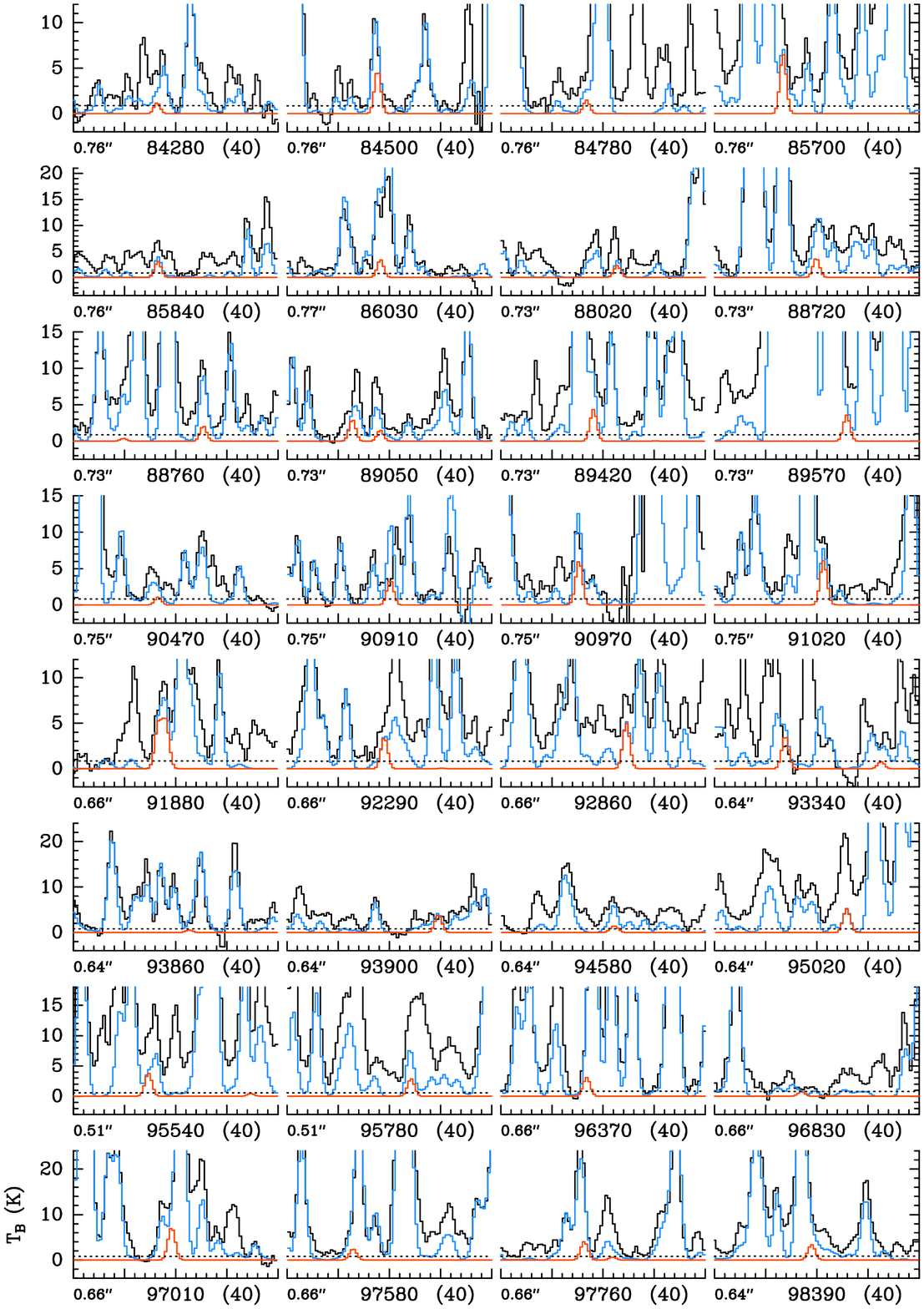}}}
\caption{Same as Fig.~\ref{f:spec_nh2conh2_ve0} for CH$_3$C(O)NH$_2$, 
$\Delta\varv_{\rm t} \neq 0$.
}
\label{f:spec_ch3conh2_cv}
\end{figure*}

\clearpage
\begin{figure*}
\addtocounter{figure}{-1}
\centerline{\resizebox{0.82\hsize}{!}{\includegraphics[angle=0]{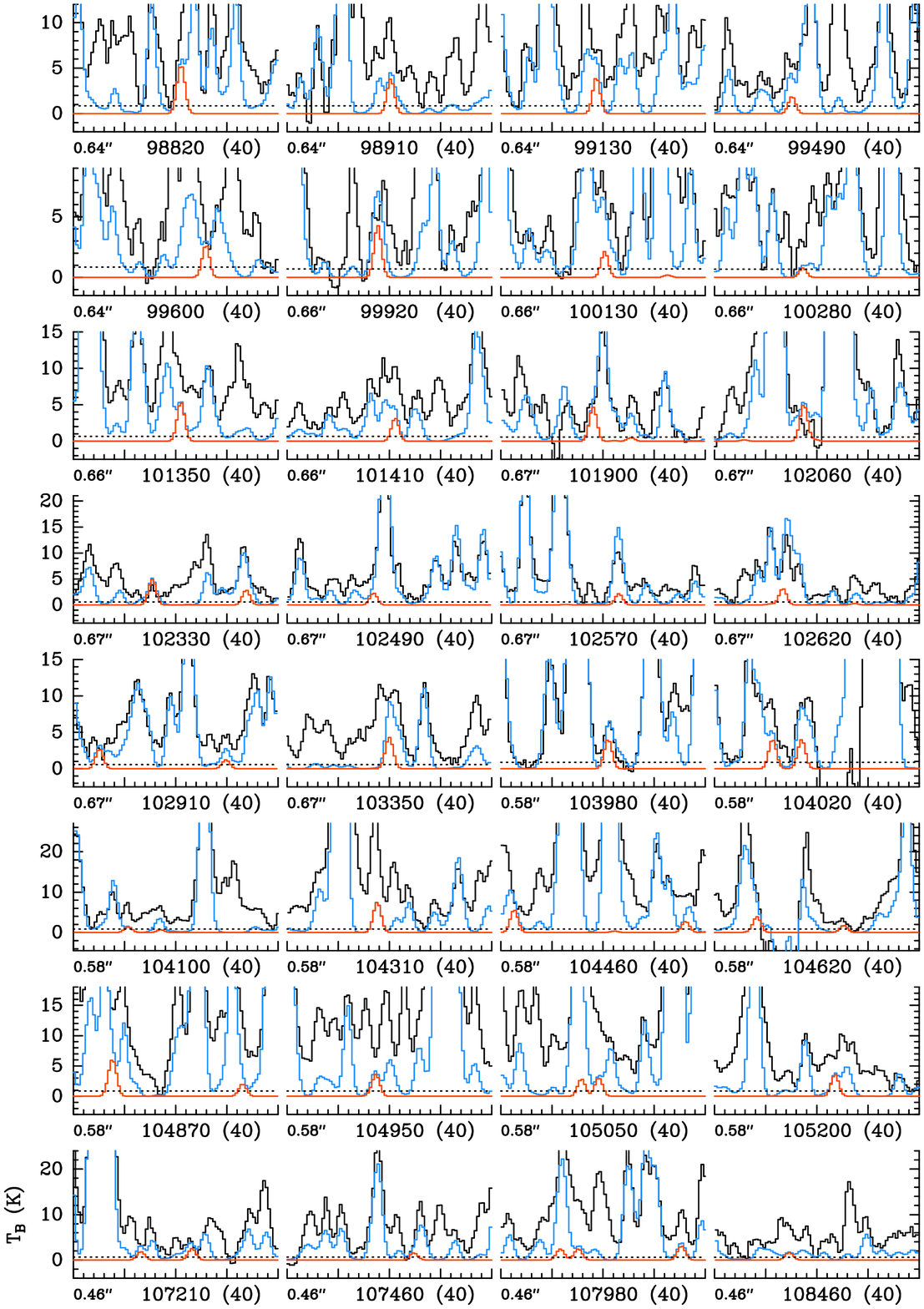}}}
\caption{continued.
}
\end{figure*}

\clearpage
\begin{figure*}
\addtocounter{figure}{-1}
\centerline{\resizebox{0.82\hsize}{!}{\includegraphics[angle=0]{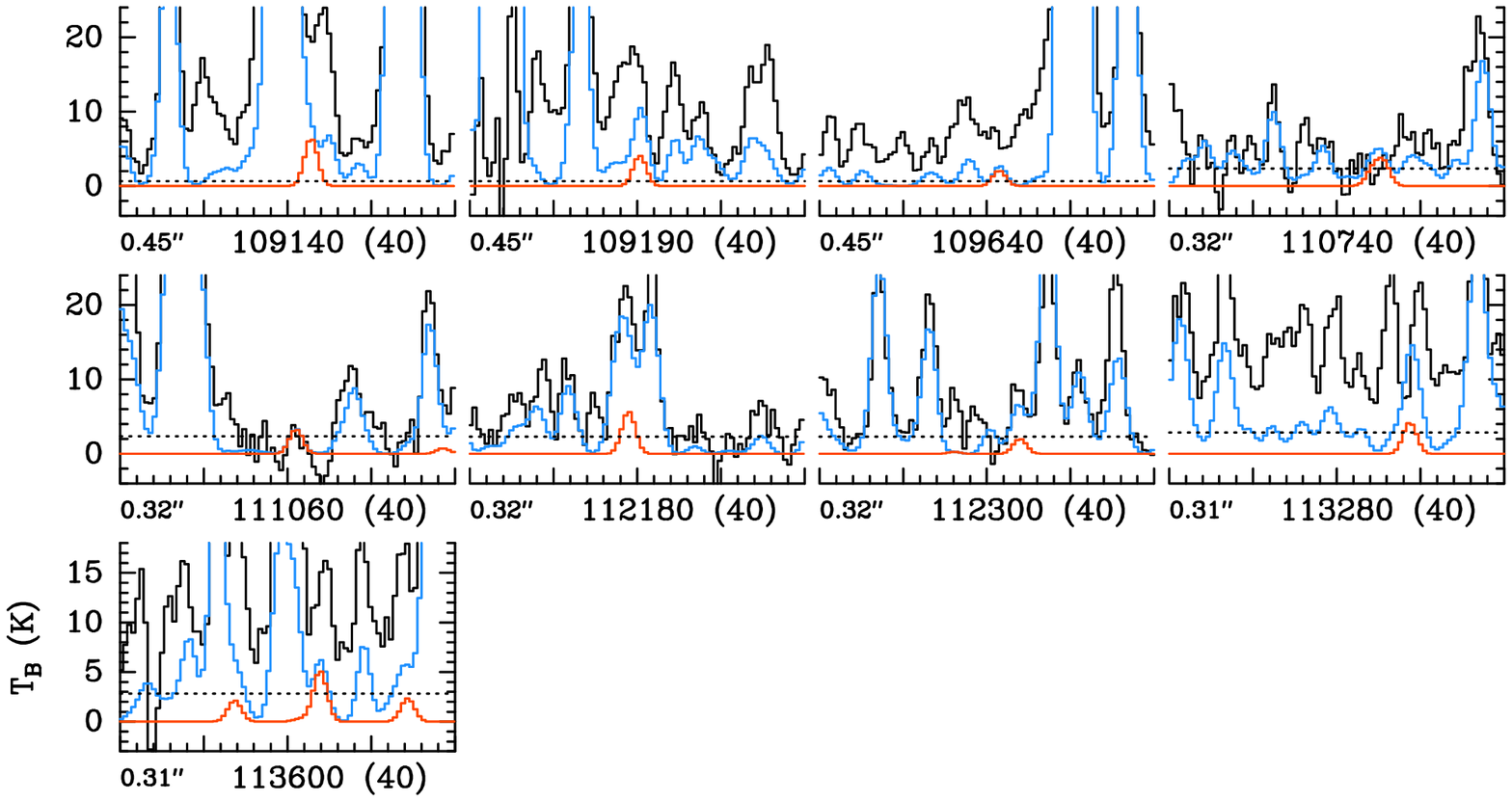}}}
\caption{continued.
}
\end{figure*}

\clearpage
\section{Complementary figures: Population diagrams}
\label{a:popdiag}

\begin{figure}
\centerline{\resizebox{0.95\hsize}{!}{\includegraphics[angle=0]{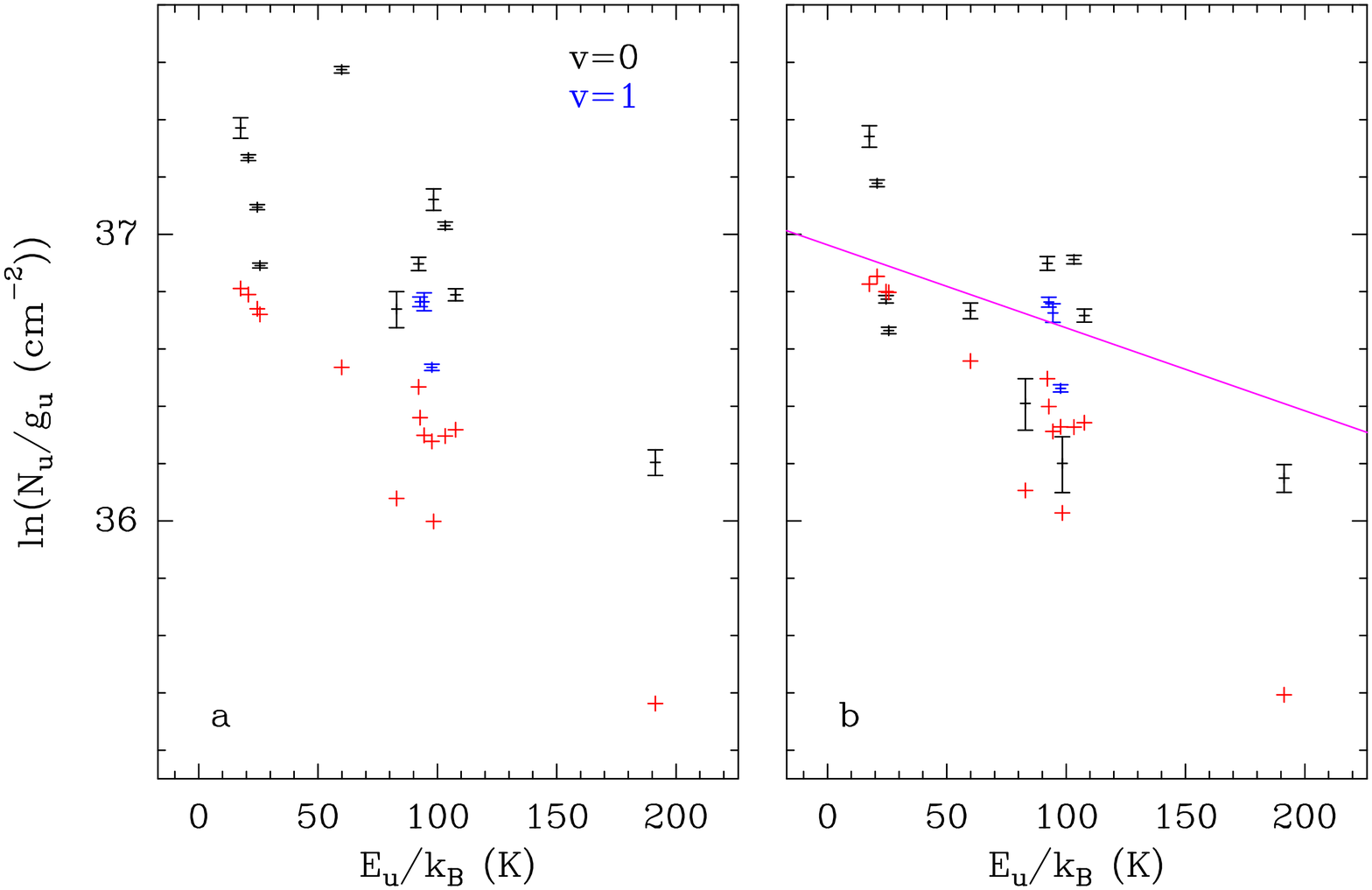}}}
\caption{Population diagram of NH$_2$C(O)NH$_2$ toward Sgr~B2(N1S). The 
observed datapoints are shown in various colors (but not red) as indicated in 
the upper right corner of panel \textbf{a} while the synthetic populations are 
shown in red. No correction is applied in panel \textbf{a}. 
In panel \textbf{b}, the optical depth correction has been applied to both the 
observed and synthetic populations and the contamination by all other 
species included in the full model has been removed from the observed 
datapoints. The purple line is a linear fit to the observed populations (in 
linear-logarithmic space).
}
\label{f:popdiag_nh2conh2}
\end{figure}

\begin{figure}
\centerline{\resizebox{0.95\hsize}{!}{\includegraphics[angle=0]{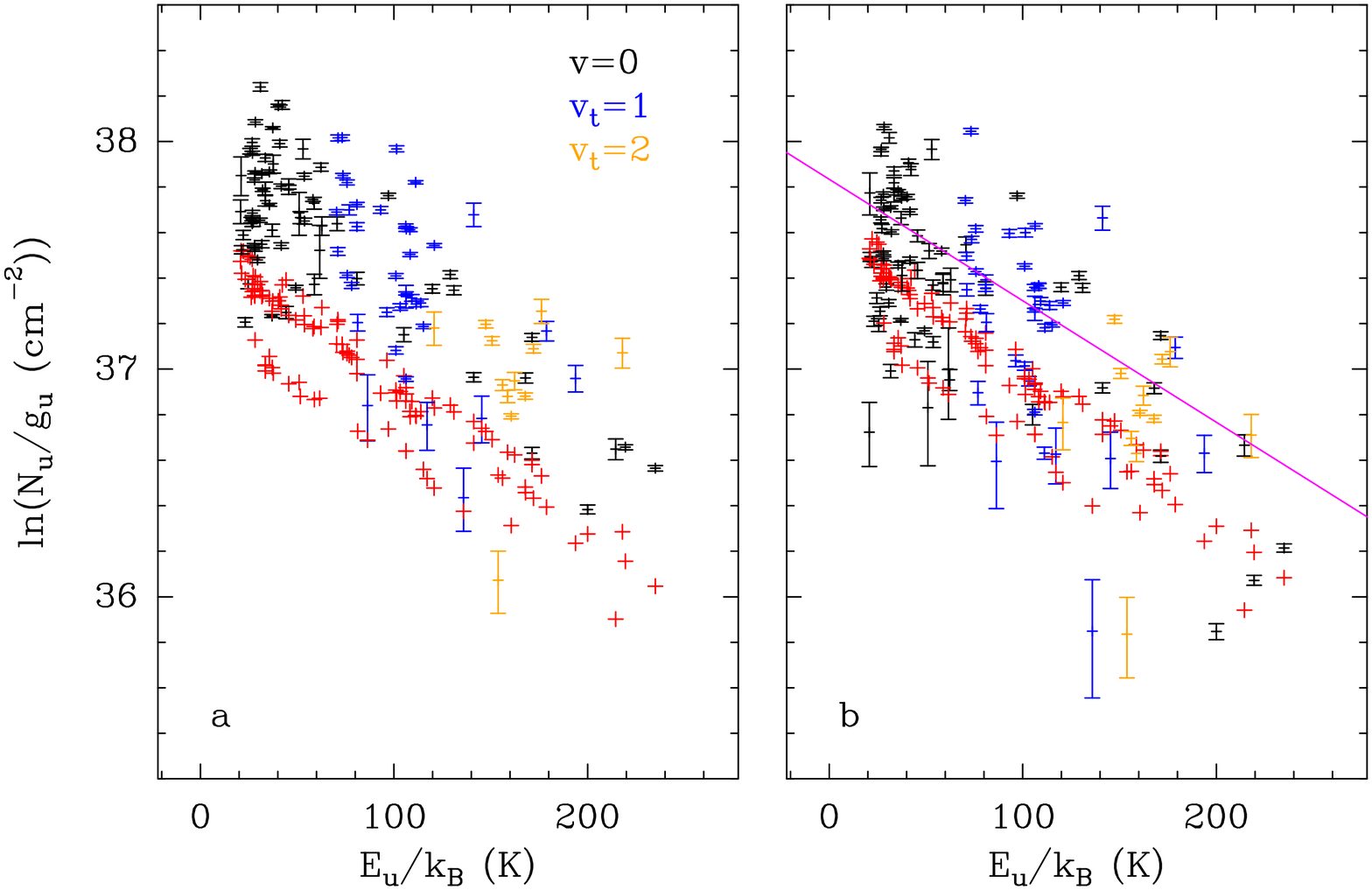}}}
\caption{Same as Fig.~\ref{f:popdiag_nh2conh2} for CH$_3$NHCHO.
}
\label{f:popdiag_ch3nhcho}
\end{figure}

\begin{figure}
\centerline{\resizebox{0.95\hsize}{!}{\includegraphics[angle=0]{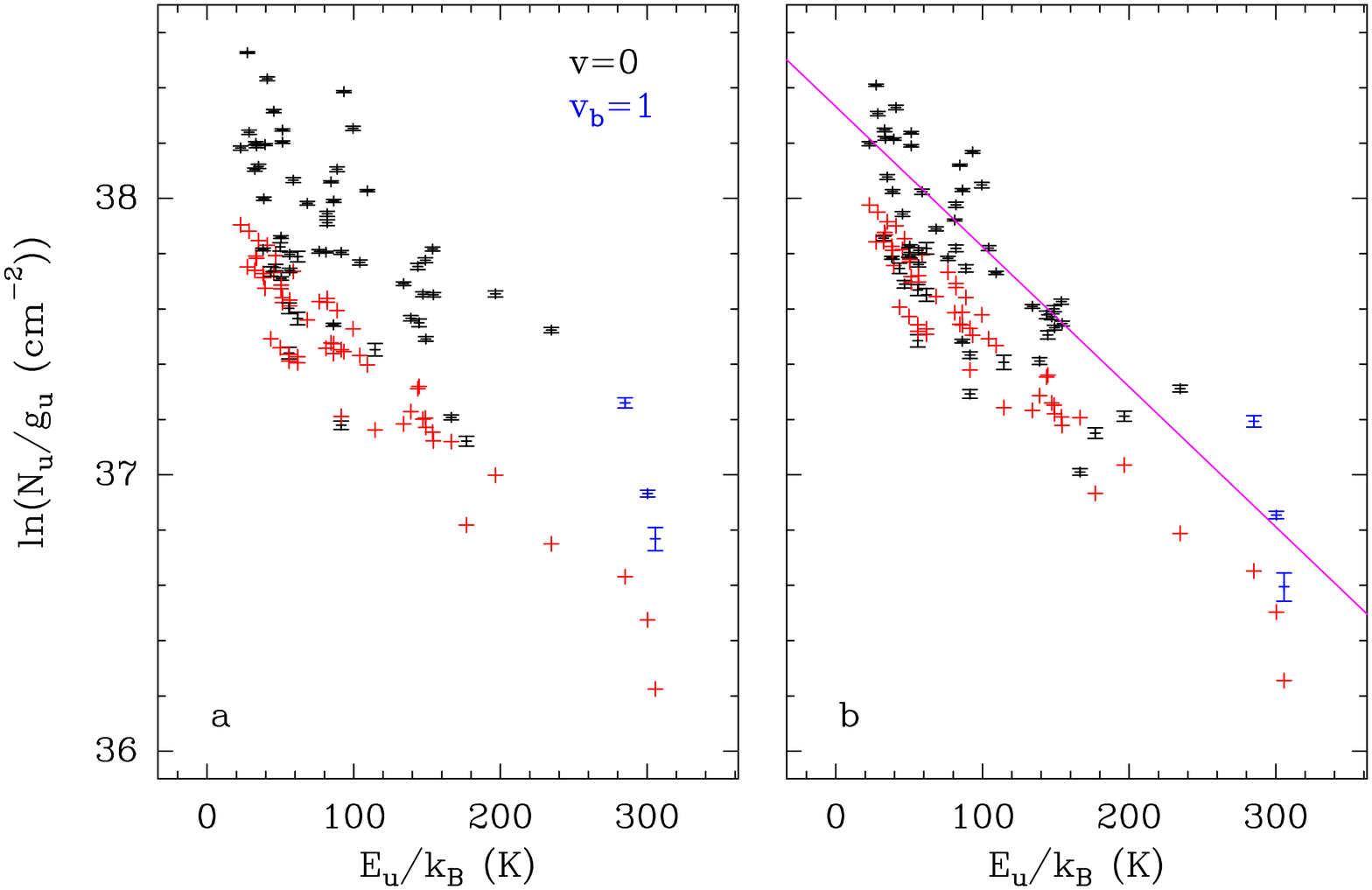}}}
\caption{Same as Fig.~\ref{f:popdiag_nh2conh2} for CH$_3$NCO.
}
\label{f:popdiag_ch3nco}
\end{figure}

\begin{figure}
\centerline{\resizebox{0.95\hsize}{!}{\includegraphics[angle=0]{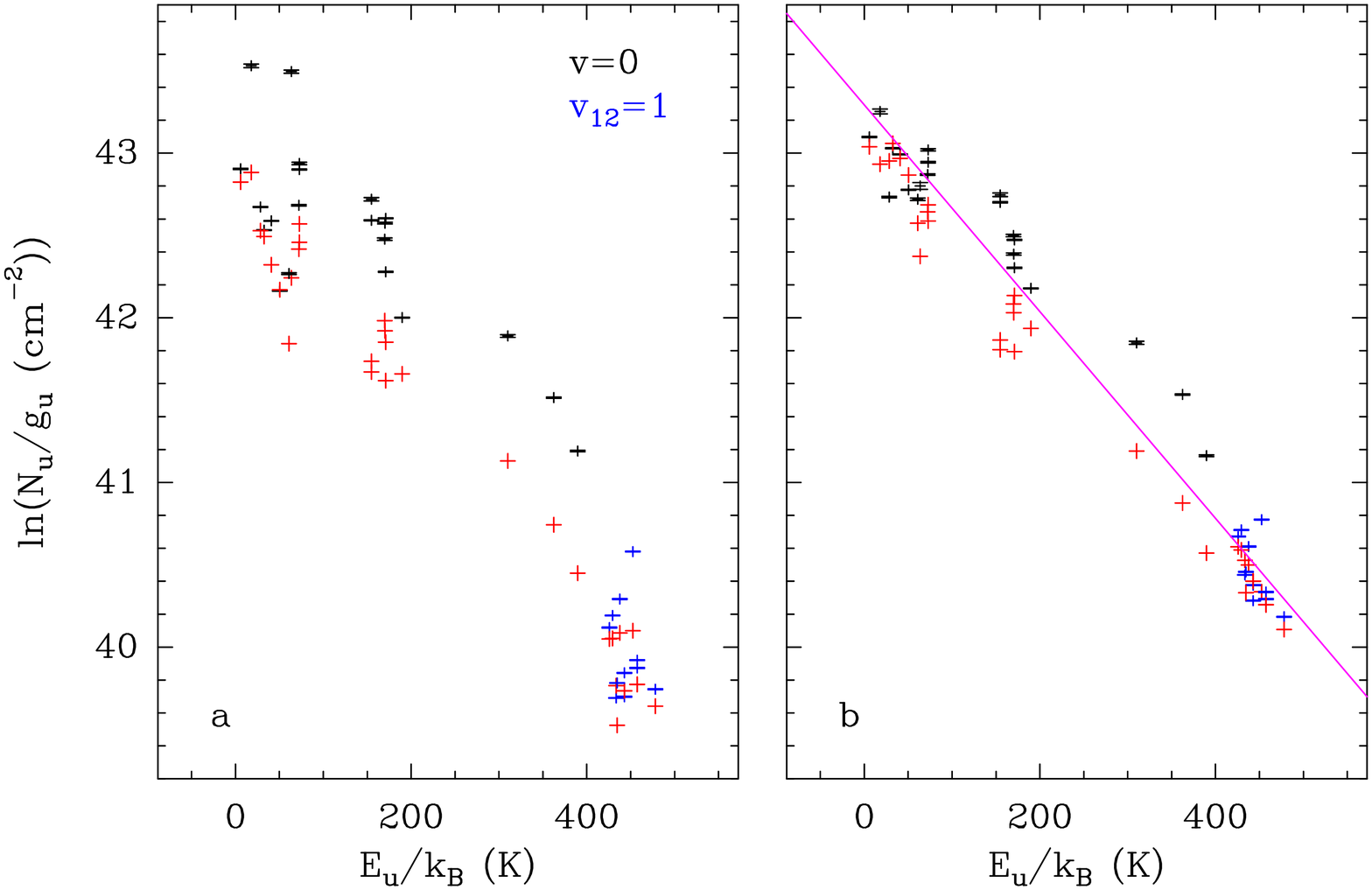}}}
\caption{Same as Fig.~\ref{f:popdiag_nh2conh2} for NH$_2$CHO.
}
\label{f:popdiag_nh2cho}
\end{figure}

\begin{figure}
\centerline{\resizebox{0.95\hsize}{!}{\includegraphics[angle=0]{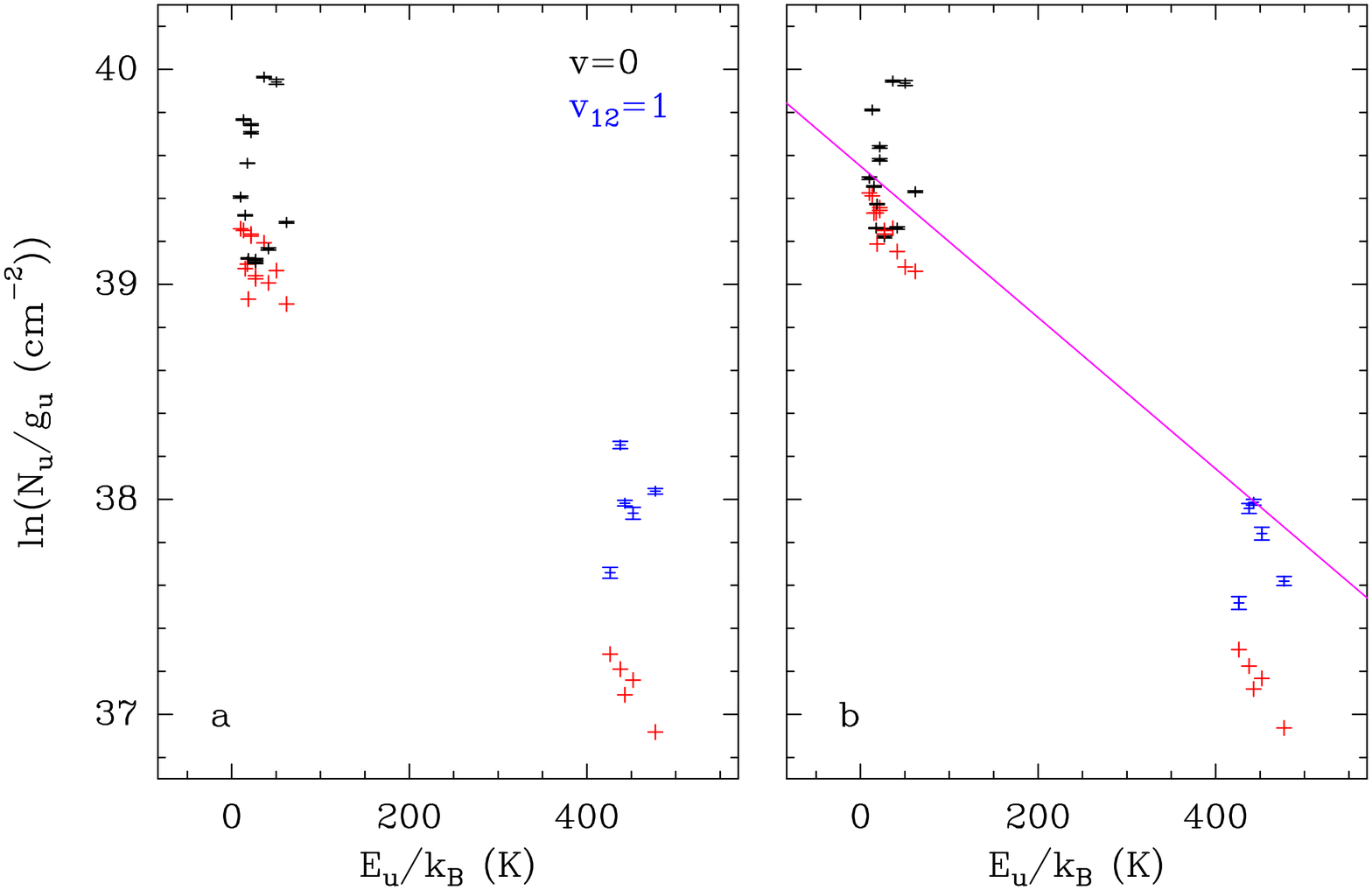}}}
\caption{Same as Fig.~\ref{f:popdiag_nh2conh2} for NH$_2$$^{13}$CHO.
}
\label{f:popdiag_nh2cho_13c}
\end{figure}

\begin{figure}
\centerline{\resizebox{0.95\hsize}{!}{\includegraphics[angle=0]{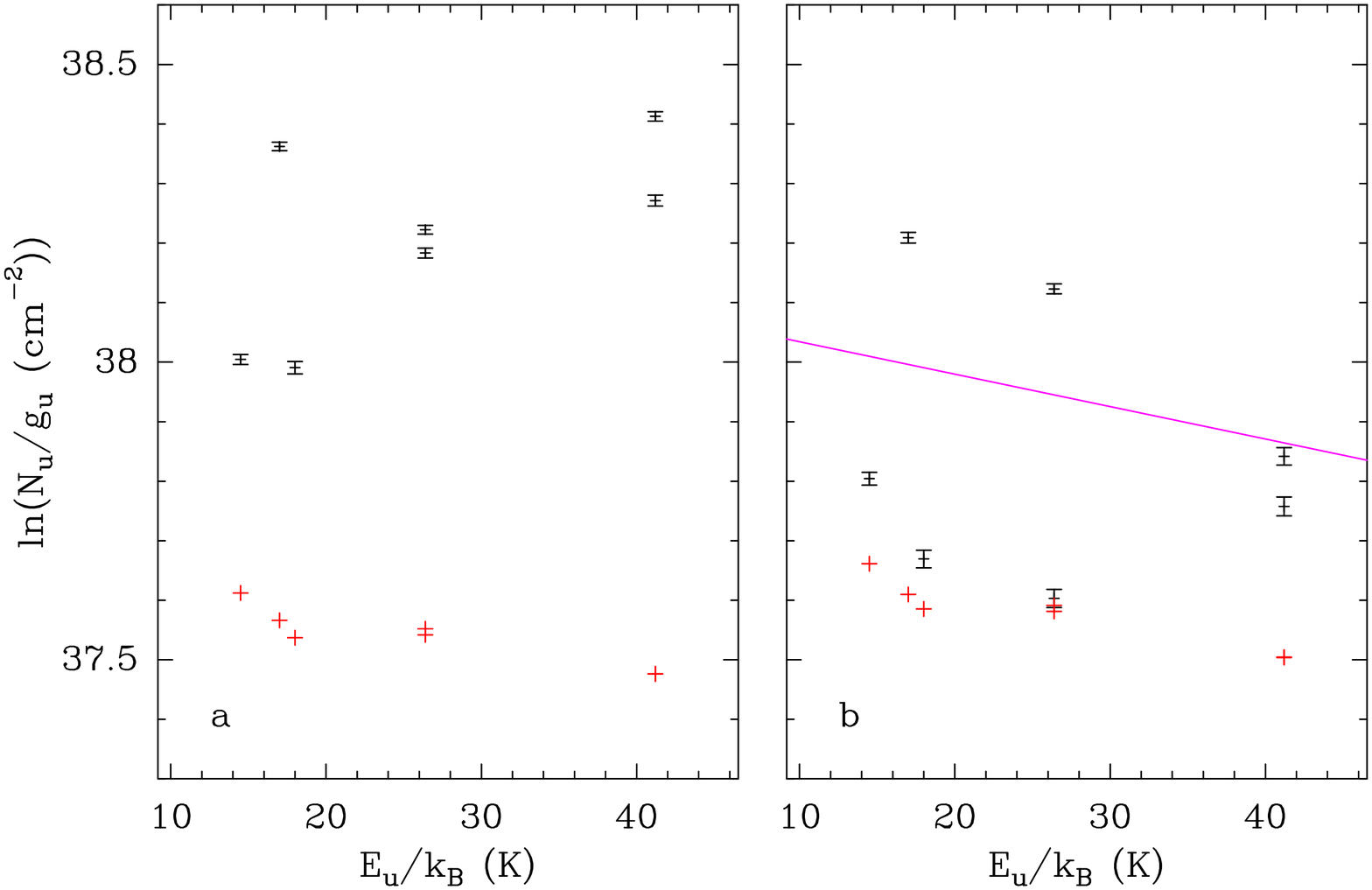}}}
\caption{Same as Fig.~\ref{f:popdiag_nh2conh2} for NH$_2$CH$^{18}$O.
}
\label{f:popdiag_nh2cho_18o}
\end{figure}

\begin{figure}
\centerline{\resizebox{0.95\hsize}{!}{\includegraphics[angle=0]{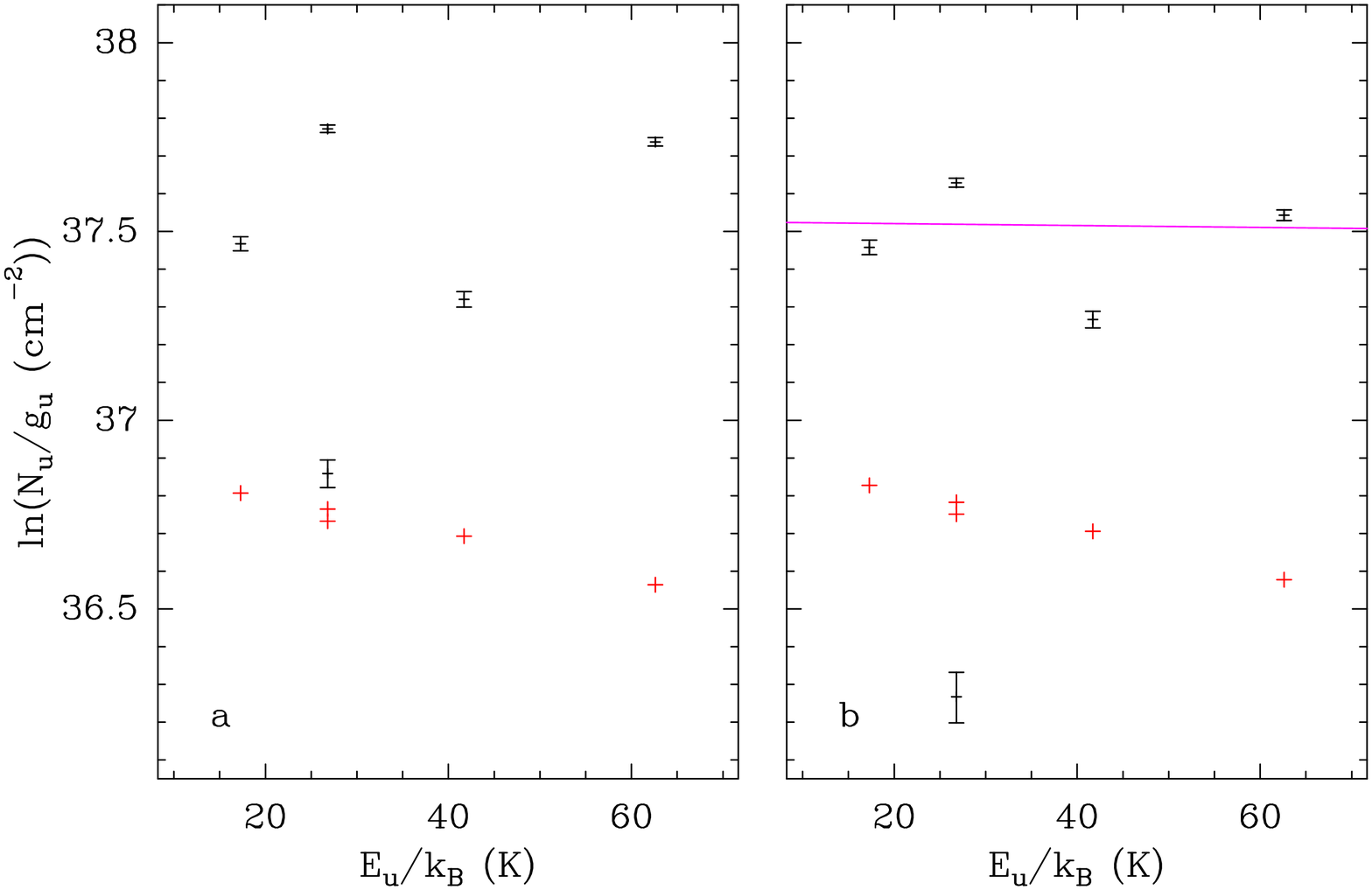}}}
\caption{Same as Fig.~\ref{f:popdiag_nh2conh2} for $^{15}$NH$_2$CHO.
}
\label{f:popdiag_nh2cho_15n}
\end{figure}

\begin{figure}
\centerline{\resizebox{0.95\hsize}{!}{\includegraphics[angle=0]{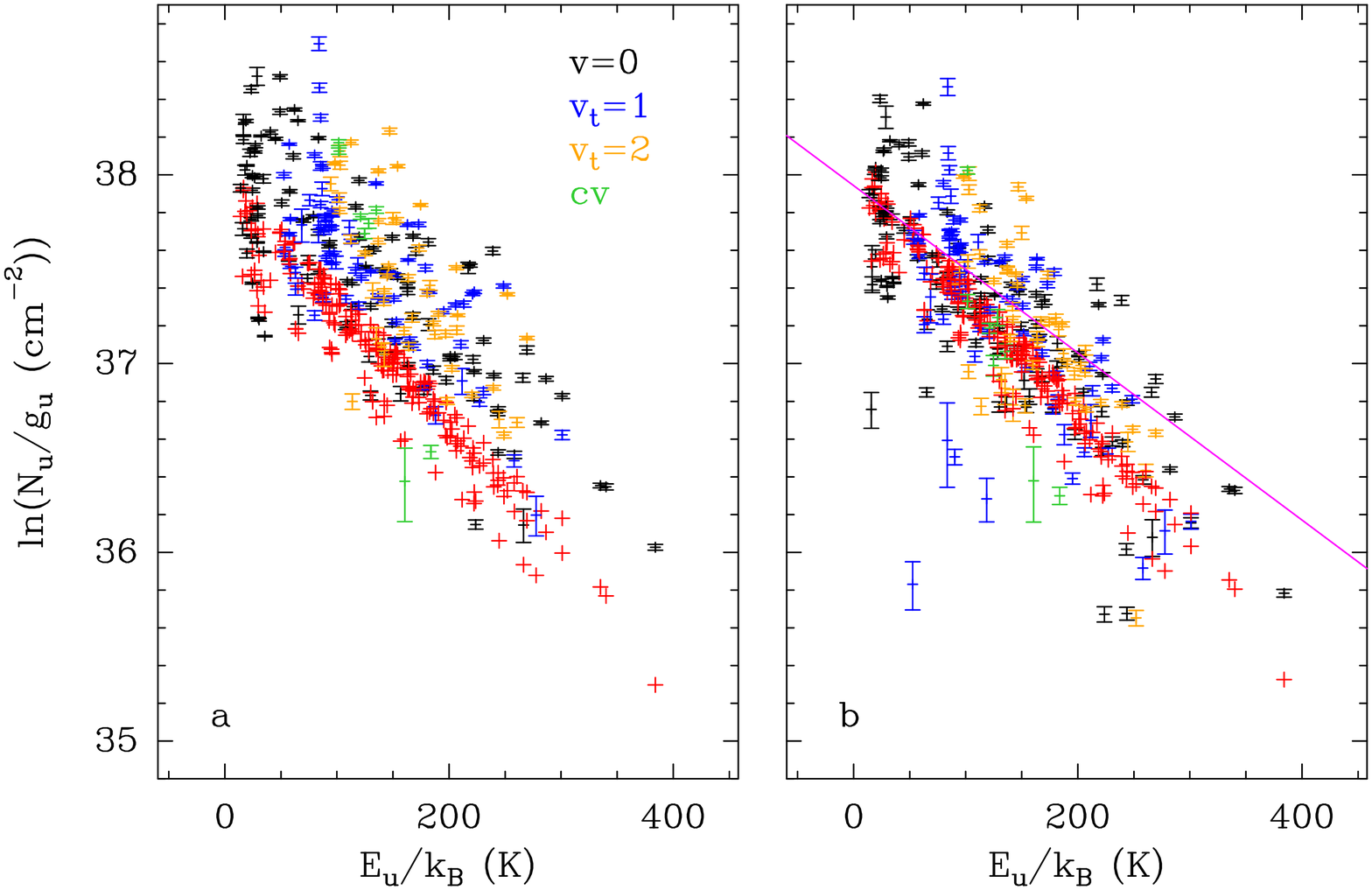}}}
\caption{Same as Fig.~\ref{f:popdiag_nh2conh2} for CH$_3$C(O)NH$_2$.
}
\label{f:popdiag_ch3conh2}
\end{figure}

Figures~\ref{f:popdiag_nh2conh2}--\ref{f:popdiag_ch3conh2} show the population 
diagrams of NH$_2$C(O)NH$_2$, CH$_3$NHCHO, CH$_3$NCO, NH$_2$CHO and its 
$^{13}$C, $^{18}$O, and $^{15}$N isotopologs, and CH$_3$C(O)NH$_2$ toward 
Sgr~B2(N1S).

\clearpage

\end{appendix}

\end{document}